\documentclass[12pt]{article}
\begin{document}

\author{C. Bizdadea\thanks{%
e-mail address: bizdadea@central.ucv.ro}, E. M. Cioroianu\thanks{%
e-mail address: manache@central.ucv.ro}, S. O. Saliu\thanks{%
e-mail address: osaliu@central.ucv.ro}, S. C. S\u {a}raru\thanks{%
e-mail address: scsararu@central.ucv.ro} \\
Faculty of Physics, University of Craiova\\
13 A. I. Cuza Str., Craiova 200585, Romania}
\title{Couplings of a collection of BF models to matter theories}
\maketitle

\begin{abstract}
The couplings of a collection of BF models to matter theories are addressed
in the framework of the antifield-BRST deformation procedure. The general
theory is exemplified in the case where the matter fields are a set of Dirac
spinors and respectively a collection of real scalar fields.

\end{abstract}

\section{Introduction}

The power of the BRST formalism was strongly increased by its cohomological
development, which allowed, among others, a useful investigation of many
interesting aspects related to the perturbative renormalization problem~\cite
{4a,4d,5}, anomaly-tracking mechanism~\cite{5,6}, simultaneous study of
local and rigid invariances of a given theory~\cite{7}, as well as to the
reformulation of the construction of consistent interactions in gauge
theories~\cite{7a} in terms of the deformation theory~\cite{8a,8b}, or,
actually, in terms of the deformation of the solution to the master equation.

The main aim of this paper is to construct all consistent Lagrangian
interactions in four spacetime dimensions that can be added to a ``free''
model that describes a generic matter theory uncoupled to a collection of
abelian BF models~\cite{12} by means of deforming the solution to the master
equation with the help of specific cohomological techniques. The field
sector of the four-dimensional BF model consists in a collection of scalar
fields, two sets of vector fields and a system of two-forms. Interacting
topological field theories of BF-type are important in view of their
relationship with Poisson Sigma Models, which are known to explain
interesting aspects of two-dimensional gravity, including the study of
classical solutions~\cite{psm1}. Other interesting aspects envisaging BF
models can be found in~\cite{bfaspects} from a Lagrangian perspective and
also in~\cite{BFhamnoPT} from a Hamiltonian point of view. The results
presented here extend our former work~\cite{BFmat} on the Lagrangian
couplings between a sole BF model and matter fields.

The couplings are obtained on the grounds of smoothness, locality,
(background) Lorentz invariance and Poincar\'{e} invariance. In addition, we
require the conservation of the number of derivatives on each field in order
to prevent any changes in the differential order of the field equations with
respect to the ``free'' model. The entire Lagrangian formulation of the
interacting theory is obtained from the computation of the deformed solution
to the master equation, order by order in the coupling constant $g$. The
existence of consistent couplings of order $g$ between the matter fields and
the BF ones is ensured under the supplementary, reasonable hypothesis that
the matter theory is invariant under some (non-trivial) bosonic global
symmetries, which produce some (non-trivially) conserved currents $%
j_{a}^{\mu }$. It is essential that the number of rigid symmetries is equal
to the number of BF fields from the collection. Based on the derivative
order assumption, we argue that the generators of the rigid symmetries
cannot involve the derivatives of the matter fields and consequently we take
them to be linear in these fields, with some coefficients that are the
elements of a set of constant matrices $T_{a}$. The consistency of the
deformation procedure at order $g^{2}$ requires the existence of some
antisymmetric functions $W_{ab}$ that depend only on the undifferentiated
scalar fields, which have the meaning of the components of the two-tensor on
a Poisson manifold with the scalar fields viewed as the local coordinates on
the target space, and holds in two situations. In the first case (type I
solution) all the matrices $T_{a}$ commute and in the second one (type II
solution) their commutators close according to a Lie algebra $L\left(
\mathcal{G}\right) $ with the structure constants $-\bar{f}_{bc}^{a}$. Type
II solutions also restrict the components of the two-tensor $W_{ab}$ to be
polynomials of order one in the scalar fields, with the coefficients of the
linear terms precisely $\bar{f}_{bc}^{a}$. The deformation procedure stops
at order one if the matter currents $j_{a}^{\mu }$ include no derivatives
and if they either remain invariant under the gauge version of the rigid
symmetries in the first case or transform under the gauge version according
to the adjoint representation of $L\left( \mathcal{G}\right) $ in the second
case. Otherwise, there appear deformations of order $g^{2}$ and possibly of
higher orders. It turns out that our procedure deforms everything, namely,
the Lagrangian action, its gauge transformations and also the accompanying
gauge algebra.

The paper is organized into six sections. Section 2 introduces the model to
be considered and construct its ``free'' Lagrangian BRST symmetry. Section 3
briefly reviews the procedure of adding consistent interactions in gauge
theories based on the deformation of the solution to the master equation. In
Section 4 we construct the Lagrangian interactions for the starting ``free''
system in four dimensions by solving the deformation equations with the help
of standard cohomological techniques. Section 5 applies the theoretical part
of the paper to the case where the role of the matter fields is played by a
set of Dirac fields and respectively by a collection of real scalar fields
and Section 6 ends the paper with the main conclusions.

\section{``Free'' BRST symmetry}

We begin with a ``free'' theory in four spacetime dimensions, described by
the sum between a collection of BF-like Lagrangian actions and a matter
action
\begin{eqnarray}
S_{0}\left[ A_{\mu }^{a},H_{\mu }^{a},\varphi _{a},B_{a}^{\mu \nu
},y^{i}\right] &=&\int d^{4}x\left( H_{\mu }^{a}\partial ^{\mu }\varphi _{a}+%
\frac{1}{2}B_{a}^{\mu \nu }\partial _{[\mu }A_{\nu ]}^{a}+\mathcal{L}%
_{0}\left( \left[ y^{i}\right] \right) \right)  \nonumber \\
&\equiv &S_{0}^{\mathrm{BF}}\left[ A_{\mu }^{a},H_{\mu }^{a},\varphi
_{a},B_{a}^{\mu \nu }\right] +S_{0}^{\mathrm{matt}}\left[ y^{i}\right] ,
\label{bfn1}
\end{eqnarray}
where the BF field spectrum contains a set of two-forms $\left\{ B_{a}^{\mu
\nu }\right\} $, two systems of one-forms $\left\{ A_{\mu }^{a},H_{\mu
}^{a}\right\} $, and a collection of scalar fields $\left\{ \varphi
_{a}\right\} $. The discrete index $a$ is an integer valued from $1$ to $N$,
and the number of matter fields is denoted by $I$ ($i=\overline{1,I}$). Here
and in the sequel the notation $f\left( \left[ q\right] \right) $ signifies
that $f$ depends on $q$ and its spacetime derivatives up to a finite order,
and $\left[ \mu \nu \cdots \right] $ (or possibly $\left[ ab\cdots \right] $%
) means full antisymmetrization with respect to the indices between brackets
such that all the independent terms appear only once and are not multiplied
by additional numerical factors. We assume that the Lagrangian density $%
\mathcal{L}_{0}$ is however no more than second-order in the derivatives of
the matter fields $y^{i}$ and that it displays no non-trivial gauge
symmetries. In what follows $\varepsilon _{i}$ denotes the Grassmann parity
of the matter field $y^{i}$. Action (\ref{bfn1}) is found invariant under
the gauge transformations (generating set)
\begin{eqnarray}
\delta _{\epsilon }A_{\mu }^{a} &=&\partial _{\mu }\epsilon ^{a}\equiv
\left( R_{\mu }^{(A)a}\right) _{b}\epsilon ^{b},\;\delta _{\epsilon }H_{\mu
}^{a}=2\partial ^{\nu }\epsilon _{\mu \nu }^{a}\equiv \left( R_{\mu
}^{(H)a}\right) _{b}^{\nu \rho }\epsilon _{\nu \rho }^{b},  \label{bfn2a} \\
\delta _{\epsilon }\varphi _{a} &=&0,\;\delta _{\epsilon }B_{a}^{\mu \nu
}=-3\partial _{\rho }\epsilon _{a}^{\mu \nu \rho }\equiv \left(
R_{a}^{(B)\mu \nu }\right) _{\rho \lambda \sigma }^{b}\epsilon _{b}^{\rho
\lambda \sigma },\;\delta _{\epsilon }y^{i}=0,  \label{bfn2}
\end{eqnarray}
that are abelian and off-shell, second-stage reducible. The gauge parameters
are all bosonic, with $\epsilon _{\mu \nu }^{a}$ and $\epsilon _{a}^{\mu \nu
\rho }$ completely antisymmetric in their Lorentz indices. The redundancy of
the gauge generators of the fields $H_{\mu }^{a}$ and $B_{a}^{\mu \nu }$
decouples at the level of the ``free'' model, as it can be seen from the
reducibility relations
\begin{eqnarray}
\left( R_{\mu }^{(H)a}\right) _{b}^{\nu \rho }\left( Z_{\nu \rho
}^{(1)b}\right) _{c}^{\lambda \sigma \tau } &=&0,\;\left( R_{a}^{(B)\mu \nu
}\right) _{\rho \lambda \sigma }^{b}\left( Z_{b}^{(1)\rho \lambda \sigma
}\right) _{\tau \kappa \varsigma \xi }^{c}=0,  \label{bfn3a} \\
\left( Z_{\nu \rho }^{(1)b}\right) _{c}^{\lambda \sigma \tau }\left(
Z_{\lambda \sigma \tau }^{(2)c}\right) _{d}^{\kappa \varsigma \xi \theta }
&=&0,  \label{bfn3b}
\end{eqnarray}
where the reducibility functions respectively take the form
\begin{eqnarray}
\left( Z_{\nu \rho }^{(1)b}\right) _{c}^{\lambda \sigma \tau } &=&-\frac{1}{2%
}\delta _{c}^{b}\delta _{\nu }^{[\lambda }\delta _{\rho }^{\sigma }\partial
^{\tau ]},  \label{bfn3} \\
\left( Z_{b}^{(1)\rho \lambda \sigma }\right) _{\tau \kappa \varsigma \xi
}^{c} &=&\frac{1}{6}\delta _{b}^{c}\delta _{[\tau }^{\rho }\delta _{\kappa
}^{\lambda }\delta _{\varsigma }^{\sigma }\partial _{\xi ]},  \label{bfn5} \\
\left( Z_{\lambda \sigma \tau }^{(2)c}\right) _{d}^{\kappa \varsigma \xi
\theta } &=&\frac{1}{6}\delta _{d}^{c}\delta _{\lambda }^{[\kappa }\delta
_{\sigma }^{\varsigma }\delta _{\tau }^{\xi }\partial ^{\theta ]}.
\label{bfn4}
\end{eqnarray}
All these properties can be synthesized by the statement that the ``free''
model under discussion is a so-called ``normal'' gauge theory, of Cauchy
order equal to four. In particular, the matter action $S_{0}^{\mathrm{matt}%
}\left[ y^{i}\right] $ is assumed to define a theory of Cauchy order equal
to one, while the BF model alone, with the action $S_{0}^{\mathrm{BF}}\left[
A_{\mu }^{a},H_{\mu }^{a},\varphi _{a},B_{a}^{\mu \nu }\right] $, is
described by a linear gauge theory of Cauchy order equal to four.

With the purpose of constructing all consistent deformations of this theory
in mind, we initially determine its BRST symmetry. The BRST algebra is
generated by the field, ghost and antifield spectra
\begin{eqnarray}
\Phi ^{\alpha _{0}} &=&\left( A_{\mu }^{a},H_{\mu }^{a},\varphi
_{a},B_{a}^{\mu \nu },y^{i}\right) ,\;\Phi _{\alpha _{0}}^{*}=\left(
A_{a}^{*\mu },H_{a}^{*\mu },\varphi ^{*a},B_{\mu \nu }^{*a},y_{i}^{*}\right)
,  \label{bfn6} \\
\eta ^{\alpha _{1}} &=&\left( \eta ^{a},C_{\mu \nu }^{a},\eta _{a}^{\mu \nu
\rho }\right) ,\;\eta _{\alpha _{1}}^{*}=\left( \eta _{a}^{*},C_{a}^{*\mu
\nu },\eta _{\mu \nu \rho }^{*a}\right) ,  \label{bfn7} \\
\eta ^{\alpha _{2}} &=&\left( C_{\mu \nu \rho }^{a},\eta _{a}^{\mu \nu \rho
\lambda }\right) ,\;\eta _{\alpha _{2}}^{*}=\left( C_{a}^{*\mu \nu \rho
},\eta _{\mu \nu \rho \lambda }^{*a}\right) ,  \label{bfn8} \\
\eta ^{\alpha _{3}} &=&C_{\mu \nu \rho \lambda }^{a},\;\eta _{\alpha
_{3}}^{*}=C_{a}^{*\mu \nu \rho \lambda },  \label{bfn9}
\end{eqnarray}
where the fermionic ghosts $\eta ^{\alpha _{1}}$ and $\eta ^{\alpha _{3}}$%
correspond to the gauge parameters, respectively, to the second-order
reducibility relations, while the bosonic ghosts $\eta ^{\alpha _{2}}$ are
due to the first-order redundancy of the generating set. The star variables
denote the antifields and exhibit statistics opposite to that of the
corresponding fields/ghosts.

Since both the gauge generators and the reducibility functions are
field-independent, it follows that the BRST differential reduces to
\begin{equation}
s=\delta +\gamma ,  \label{decs}
\end{equation}
where $\delta $ is the Koszul-Tate differential, and $\gamma $ means the
exterior longitudinal derivative. The Koszul-Tate differential is graded in
terms of the antighost number ($\mathrm{agh}$, $\mathrm{agh}\left( \delta
\right) =-1$) and enforces a resolution of the algebra of smooth functions
defined on the stationary surface of field equations for action (\ref{bfn1}%
), $C^{\infty }\left( \Sigma \right) $, $\Sigma :\delta S_{0}/\delta \Phi
^{\alpha _{0}}=0$. The exterior longitudinal derivative is graded in terms
of the pure ghost number ($\mathrm{pgh}$, $\mathrm{pgh}\left( \gamma \right)
=1$) and is correlated with the original gauge symmetry via its cohomology
in pure ghost number zero computed in $C^{\infty }\left( \Sigma \right) $,
which is isomorphic to the algebra of physical observables for this ``free''
theory. The degrees of the generators (\ref{bfn6})--(\ref{bfn9}) from the
BRST complex are valued like
\begin{eqnarray}
\mathrm{pgh}(\Phi ^{\alpha _{0}}) &=&\mathrm{pgh}(\Phi _{\alpha _{0}}^{*})=%
\mathrm{pgh}(\eta _{\alpha _{1}}^{*})=0,\;\mathrm{pgh}(\eta ^{\alpha
_{1}})=1,  \label{bfn10} \\
\mathrm{agh}(\Phi ^{\alpha _{0}}) &=&\mathrm{agh}(\eta ^{\alpha _{1}})=0,\;%
\mathrm{agh}(\Phi _{\alpha _{0}}^{*})=1,\;\mathrm{agh}(\eta _{\alpha
_{1}}^{*})=2,  \label{bfn11} \\
\mathrm{pgh}(\eta ^{\alpha _{2}}) &=&2,\;\mathrm{pgh}(\eta ^{\alpha
_{3}})=3,\;\mathrm{pgh}(\eta _{\alpha _{2}}^{*})=\mathrm{pgh}(\eta _{\alpha
_{3}}^{*})=0,  \label{bfn12} \\
\mathrm{agh}(\eta ^{\alpha _{2}}) &=&\mathrm{agh}(\eta ^{\alpha _{3}})=0,\;%
\mathrm{agh}(\eta _{\alpha _{2}}^{*})=3,\;\mathrm{agh}(\eta _{\alpha
_{3}}^{*})=4,  \label{bfn12a}
\end{eqnarray}
and the actions of $\delta $ and $\gamma $ on them are given by
\begin{equation}
\delta \Phi ^{\alpha _{0}}=\delta \eta ^{\alpha _{1}}=\delta \eta ^{\alpha
_{2}}=\delta \eta ^{\alpha _{3}}=0,  \label{bfn13}
\end{equation}
\begin{equation}
\delta y_{i}^{*}=-\frac{\delta ^{L}\mathcal{L}_{0}}{\delta y^{i}},\;\delta
A_{a}^{*\mu }=\partial _{\nu }B_{a}^{\nu \mu },\;\delta H_{a}^{*\mu
}=-\partial ^{\mu }\varphi _{a},\;\delta \varphi ^{*a}=\partial ^{\mu
}H_{\mu }^{a},  \label{bfn14}
\end{equation}
\begin{equation}
\delta B_{\mu \nu }^{*a}=-\frac{1}{2}\partial _{[\mu }A_{\nu ]}^{a},\;\delta
\eta _{a}^{*}=-\partial _{\mu }A_{a}^{*\mu },\;\delta C_{a}^{*\mu \nu
}=\partial ^{[\mu }H_{a}^{*\nu ]},  \label{bfn15}
\end{equation}
\begin{equation}
\delta \eta _{\mu \nu \rho }^{*a}=\partial _{[\mu }B_{\nu \rho
]}^{*a},\;\delta C_{a}^{*\mu \nu \rho }=-\partial ^{[\mu }C_{a}^{*\nu \rho
]},  \label{bfn16}
\end{equation}
\begin{equation}
\delta \eta _{\mu \nu \rho \lambda }^{*a}=-\partial _{[\mu }\eta _{\nu \rho
\lambda ]}^{*a},\;\delta C_{a}^{*\mu \nu \rho \lambda }=\partial ^{[\mu
}C_{a}^{*\nu \rho \lambda ]},  \label{bfn17}
\end{equation}
\begin{equation}
\gamma \Phi _{\alpha _{0}}^{*}=\gamma \eta _{\alpha _{1}}^{*}=\gamma \eta
_{\alpha _{2}}^{*}=\gamma \eta _{\alpha _{3}}^{*}=0,  \label{bfn18}
\end{equation}
\begin{equation}
\gamma A_{\mu }^{a}=\partial _{\mu }\eta ^{a},\;\gamma H_{\mu
}^{a}=2\partial ^{\nu }C_{\mu \nu }^{a},\;\gamma \varphi _{a}=0,\;\gamma
B_{a}^{\mu \nu }=-3\partial _{\rho }\eta _{a}^{\mu \nu \rho },  \label{bfn19}
\end{equation}
\begin{equation}
\gamma y^{i}=0,\;\gamma \eta ^{a}=0,\;\gamma C_{\mu \nu }^{a}=-3\partial
^{\rho }C_{\mu \nu \rho }^{a},\;\gamma \eta _{a}^{\mu \nu \rho }=4\partial
_{\lambda }\eta _{a}^{\mu \nu \rho \lambda },  \label{bfn20}
\end{equation}
\begin{equation}
\gamma C_{\mu \nu \rho }^{a}=4\partial ^{\lambda }C_{\mu \nu \rho \lambda
}^{a},\;\gamma \eta _{a}^{\mu \nu \rho \lambda }=0,\;\gamma C_{\mu \nu \rho
\lambda }^{a}=0.  \label{bfn21}
\end{equation}

The overall degree from the BRST complex is named ghost number ($\mathrm{gh}$%
) and is defined like the difference between the pure ghost number and the
antighost number, such that $\mathrm{gh}\left( \delta \right) =\mathrm{gh}%
\left( \gamma \right) =\mathrm{gh}\left( s\right) =1$. The BRST symmetry
admits a canonical action $s\cdot =\left( \cdot ,S\right) $, where its
canonical generator ($\mathrm{gh}\left( S\right) =0$, $\varepsilon \left(
S\right) =0$) satisfies the classical master equation
\begin{equation}
\left( S,S\right) =0.  \label{masteq}
\end{equation}
The antibracket $\left( ,\right) $ is obtained by decreeing the
fields/ghosts respectively conjugated with the corresponding antifields. In
the case of the ``free'' theory under discussion, the solution to the master
equation takes the form
\begin{eqnarray}
S &=&S_{0}+\int d^{4}x\left( A_{a}^{*\mu }\partial _{\mu }\eta
^{a}+2H_{a}^{*\mu }\partial ^{\nu }C_{\mu \nu }^{a}-3B_{\mu \nu
}^{*a}\partial _{\rho }\eta _{a}^{\mu \nu \rho }\right.  \nonumber \\
&&-\left. 3C_{a}^{*\mu \nu }\partial ^{\rho }C_{\mu \nu \rho }^{a}+4\eta
_{\mu \nu \rho }^{*a}\partial _{\lambda }\eta _{a}^{\mu \nu \rho \lambda
}+4C_{a}^{*\mu \nu \rho }\partial ^{\lambda }C_{\mu \nu \rho \lambda
}^{a}\right) ,  \label{bfn22}
\end{eqnarray}
and it contains pieces of antighost number ranging from zero to three.

\section{Basic equations of the deformation procedure}

We consider the problem of constructing the consistent interactions that can
be added to the ``free'' Lagrangian action (\ref{bfn1}), $S_{0}\left[ \Phi
^{\alpha _{0}}\right] $ (where $\Phi ^{\alpha _{0}}$ means the original
field spectrum in (\ref{bfn6})), invariant under the gauge transformations (%
\ref{bfn2a})--(\ref{bfn2}), written in a compact form like
\begin{equation}
\delta _{\epsilon }\Phi ^{\alpha _{0}}=R_{\;\;\alpha _{1}}^{\alpha
_{0}}\epsilon ^{\alpha _{1}},\;\frac{\delta S_{0}}{\delta \Phi ^{\alpha _{0}}%
}R_{\;\;\alpha _{1}}^{\alpha _{0}}=0,  \label{bfn2.1}
\end{equation}
such that the couplings preserve both the field spectrum and the original
number of independent gauge symmetries. This issue is addressed by means of
reformulating the problem of constructing consistent interactions like a
deformation problem of the solution (\ref{bfn22}) to the master equation
corresponding to the ``free'' theory~\cite{8a}. Such a reformulation is
possible due to the fact that the solution to the master equation contains
all the information on the gauge structure of the theory. If a consistent
interacting gauge theory can be constructed, then the solution $S$ to the
master equation associated with the ``free'' theory, $\left( S,S\right) =0$,
can be deformed into a solution $\bar{S}$,
\begin{eqnarray}
&&S\rightarrow \bar{S}=S+gS_{1}+g^{2}S_{2}+\cdots =  \nonumber \\
&&S+g\int d^{4}x\,a+g^{2}\int d^{4}x\,b+\cdots ,  \label{bfn2.2}
\end{eqnarray}
of the master equation for the deformed theory
\begin{equation}
\left( \bar{S},\bar{S}\right) =0,  \label{bfn2.3}
\end{equation}
such that both the field/ghost and antifield spectra of the initial theory, (%
\ref{bfn6})--(\ref{bfn9}), are preserved. The equation (\ref{bfn2.3})
splits, according to the various orders in the coupling constant
(deformation parameter) $g$, into (\ref{masteq}) and
\begin{eqnarray}
2\left( S_{1},S\right) &=&0,  \label{bfn2.5} \\
2\left( S_{2},S\right) +\left( S_{1},S_{1}\right) &=&0,  \label{bfn2.6} \\
\left( S_{3},S\right) +\left( S_{1},S_{2}\right) &=&0,  \label{bfn2.7} \\
&&\vdots  \nonumber
\end{eqnarray}

The equation (\ref{masteq}) is fulfilled by hypothesis. The next one, (\ref
{bfn2.5}), requires that the first-order deformation of the solution to the
master equation, $S_{1}$, is a cocycle of the ``free'' BRST differential $s$%
. However, only cohomologically non-trivial solutions to (\ref{bfn2.5})
should be taken into account, since the BRST-exact ones can be eliminated by
a (in general non-linear) field redefinition. This means that $S_{1}$
pertains to the ghost number zero cohomological space of $s$, $H^{0}\left(
s\right) $, which is generically non-empty due to its isomorphism to the
space of physical observables of the ``free'' theory. It has been shown in~%
\cite{8a} on behalf of the triviality of the antibracket map in the
cohomology of the BRST differential that there are no obstructions in
finding solutions to the remaining equations ((\ref{bfn2.6})--(\ref{bfn2.7}%
), etc.). Unfortunately, the resulting interactions may be non-local, and
there might even appear obstructions if one insists on their locality. The
analysis of these obstructions can be done with the help of cohomological
techniques. As it will be seen below, all the consistent interactions in the
case of the model under study turn out to be local.

\section{Deformation of the ``free'' solution to the master equation}

Here, we compute the consistent Lagrangian interactions that can be added to
the ``free'' model analyzed so far, which describes a generic matter theory
plus a topological BF-type model in four spacetime dimensions. This is
achieved by solving the deformation equations (\ref{bfn2.5})--(\ref{bfn2.7}%
), etc., with the help of some specific cohomological techniques. Our aim is
to determine the complete deformed solution to the master equation, which is
consistent to all orders in the coupling constant. For obvious reasons, we
consider only smooth, local, (background) Lorentz invariant and, moreover,
Poincar\'{e} invariant quantities (i.e. we do not allow explicit dependence
on the spacetime coordinates). The smoothness of the deformations refers to
the fact that the deformed solution (\ref{bfn2.2}) to the master equation is
smooth in the coupling constant $g$ and reduces to the original solution (%
\ref{bfn22}) in the free limit $g=0$. In addition, we require the
conservation of the number of derivatives on each field (this condition is
frequently met in the literature; for instance, see~\cite{gen2,multi,lingr}%
). The last requirement will be brought in only after the derivation of the
general form of the first-order deformation, during the consistency
procedure.

\subsection{First-order deformation}

\subsubsection{Standard material: $H\left( \gamma \right) $ and $H\left(
\delta |d\right) $\label{stmat}}

Initially, we approach the first-order deformation of the solution to the
master equation, described by the equation (\ref{bfn2.5}). Using the
notations from (\ref{bfn2.2}), its local form is
\begin{equation}
sa=\partial _{\mu }m^{\mu },  \label{bfa24}
\end{equation}
for some local current $m^{\mu }$, so it requires that $a$ is a $s$-cocycle
modulo the exterior spacetime differential $d$. In order to analyze the
above equation, we develop $a$ according to the antighost number
\begin{equation}
a=\sum_{k=0}^{J}a_{k},\;\mathrm{agh}\left( a_{k}\right) =k,\;\mathrm{gh}%
\left( a_{k}\right) =0,\;\varepsilon \left( a_{k}\right) =0,  \label{bfa25}
\end{equation}
and assume, without loss of generality, that $a$ stops at some finite value $%
J$ of the antighost number. This can be shown, for instance, like in~\cite
{gen2} (Section 3), under the sole assumption that the interacting
Lagrangian at the first order in the coupling constant, $a_{0}$, has a
finite, but otherwise arbitrary derivative order. By taking into account the
decomposition (\ref{decs}) of the BRST differential, the equation (\ref
{bfa24}) is equivalent to a tower of local equations, corresponding to the
various decreasing values of the antighost number
\begin{eqnarray}
\gamma a_{J} &=&\partial _{\mu }\stackrel{(J)}{m}^{\mu },  \label{r64} \\
\delta a_{J}+\gamma a_{J-1} &=&\partial _{\mu }\stackrel{(J-1)}{m}^{\mu },
\label{r65} \\
\delta a_{k}+\gamma a_{k-1} &=&\partial _{\mu }\stackrel{(k-1)}{m}^{\mu
},\;1\leq k\leq J-1,  \label{r66}
\end{eqnarray}
where $\left( \stackrel{(k)}{m}^{\mu }\right) _{k=\overline{0,J}}$ are some
local currents, with $\mathrm{agh}\left( \stackrel{(k)}{m}^{\mu }\right) =k$%
. It can be proved that the equation (\ref{r64}) can be replaced at strictly
positive values of the antighost number by
\begin{equation}
\gamma a_{J}=0,\;J>0.  \label{bfa26}
\end{equation}
The proof of this result is standard material and can be found for instance
in~\cite{gen2,multi,lingr,hepth04}. Then, in order to solve the equation (%
\ref{bfa24}) (equivalent with (\ref{bfa26}) and (\ref{r65})--(\ref{r66})) we
clearly need to compute the cohomology of $\gamma $, $H\left( \gamma \right)
$. On behalf of (\ref{bfn18})--(\ref{bfn21}) it is simple to see that $%
H\left( \gamma \right) $ is spanned by
\begin{eqnarray}
\omega ^{\Delta } &=&\left( F_{\mu \nu }^{a}=\partial _{\left[ \mu \right.
}A_{\left. \nu \right] }^{a},\partial ^{\mu }H_{\mu }^{a},\varphi
_{a},\partial _{\nu }B_{a}^{\mu \nu },y^{i}\right) ,  \label{omega} \\
\chi ^{*} &=&\left( \Phi _{\alpha _{0}}^{*},\eta _{\alpha _{1}}^{*},\eta
_{\alpha _{2}}^{*},\eta _{\alpha _{3}}^{*}\right) ,  \label{antif}
\end{eqnarray}
and by their spacetime derivatives, as well as by the undifferentiated
ghosts
\begin{equation}
\eta ^{A_{1}}=\left( \eta ^{a},\eta _{a}^{\mu \nu \rho \lambda },C_{\mu \nu
\rho \lambda }^{a}\right) .  \label{ghosts}
\end{equation}
(The derivatives of the ghosts $\eta ^{A_{1}}$ are removed from $H\left(
\gamma \right) $ since they are $\gamma $-exact, in agreement with the first
relation in (\ref{bfn19}), the last formula in (\ref{bfn20}) and
respectively the first definition from (\ref{bfn21}).) If we denote by $%
e^{M}\left( \eta ^{A_{1}}\right) $ the elements with the pure ghost number
equal to $M$ of a basis in the space of polynomials in the ghosts $\eta
^{A_{1}}$, it follows that the general, local solution to the equation (\ref
{bfa26}) takes the form (up to trivial, $\gamma $-exact contributions)
\begin{equation}
a_{J}=\mu _{J}\left( \left[ \omega ^{\Delta }\right] ,\left[ \chi
^{*}\right] \right) e^{J}\left( \eta ^{A_{1}}\right) ,  \label{bfa27}
\end{equation}
where $\mathrm{agh}\left( \mu _{J}\right) =J$ and $\mathrm{pgh}\left(
e^{J}\right) =J$. The objects $\mu _{J}$ (obviously non-trivial in $%
H^{0}\left( \gamma \right) $) were taken to have a bounded number of
derivatives and a finite antighost number, so they are polynomials in the
antifields $\chi ^{*}$, in their derivatives, in all the quantities $\omega
^{\Delta }$ excepting the undifferentiated scalar fields $\varphi _{a}$ and
the undifferentiated bosonic matter fields $y^{i}$ (if any), as well as in
all of their derivatives. However, the $\mu _{J}$'s may in principle have
coefficients that are infinite series in $\varphi _{a}$ and in all commuting
$y^{i}$. Due to their $\gamma $-closeness and (partial) polynomial
character, $\mu _{J}$ will be called ``invariant polynomials''. A useful
property is that the cohomology of $d$ in the space of invariant polynomials
is trivial in form degree strictly less than four and in strictly positive
antighost number (for a general proof, see~\cite{dubplb}). This further
leads to the conclusion that there is no non-trivial descent for $H\left(
\gamma |d\right) $ in strictly positive antighost number, or, to put it
otherwise, that the equation (\ref{r64}) can always be replaced with (\ref
{bfa26}) for $J>0$. The proof to the last result can be done like in~\cite
{multi,lingr,hepth04}.

Replacing (\ref{bfa27}) in (\ref{r65}) we remark that a necessary (but not
sufficient) condition for the existence of (non-trivial) solutions $a_{J-1}$
is that the invariant polynomials $\mu _{J}$ from (\ref{bfa27}) are
(non-trivial) objects from the local cohomology of the Koszul-Tate
differential $H\left( \delta |d\right) $ in antighost number $J>0$ and in
pure ghost number equal to zero, $\mu _{J}\in H_{J}\left( \delta |d\right) $%
, i.e.
\begin{equation}
\delta \mu _{J}=\partial _{\mu }\stackrel{\left( J-1\right) }{j}^{\mu },\;%
\mathrm{agh}\left( \stackrel{\left( J-1\right) }{j}^{\mu }\right) =J-1,\;%
\mathrm{pgh}\left( \stackrel{\left( J-1\right) }{j}^{\mu }\right) =0.
\label{r80}
\end{equation}

Consequently, we need to investigate some of the main properties of the
local cohomology of the Koszul-Tate differential in pure ghost number zero
and in strictly positive antighost numbers\footnote{%
We recall that the local cohomology $H\left( \delta |d\right) $ is
completely trivial in both strictly positive antighost \textit{and} pure
ghost numbers (for instance, see~\cite{gen1}, Theorem 5.4 and~\cite{commun1}%
).} in order to fully determine the component $a_{J}$ of highest antighost
number from the first-order deformation. To this end we observe that the
form (\ref{bfn1}) of the ``free'' Lagrangian action together with the
definitions (\ref{bfn13})--(\ref{bfn17}) enable us to analyze $H_{k}\left(
\delta |d\right) $ in terms of the local cohomologies $H_{k}^{\mathrm{matt}%
}\left( \delta |d\right) $ and $H_{k}^{\mathrm{BF}}\left( \delta |d\right) $%
, where the last local cohomologies in antighost number $k$ refer to the
Koszul-Tate operator that acts non-trivially only in the matter sector,
respectively, only in the BF one\footnote{%
Indeed, we can decompose $\delta $ like $\delta =\delta ^{\mathrm{matt}%
}+\delta ^{\mathrm{BF}}$, where $\delta ^{\mathrm{matt}}\left( \mathrm{%
matter\;variables}\right) =\delta \left( \mathrm{matter\;variables}\right) $
and $\delta ^{\mathrm{matt}}\left( \mathrm{BF\;variables}\right) =0$,
respectively, $\delta ^{\mathrm{BF}}\left( \mathrm{matter\;variables}\right)
=0$ and $\delta ^{\mathrm{BF}}\left( \mathrm{BF\;variables}\right) =\delta
\left( \mathrm{BF\;variables}\right) $. According to this decomposition, $H^{%
\mathrm{matt}}\left( \delta |d\right) $ and $H^{\mathrm{BF}}\left( \delta
|d\right) $ must be understood only like some more suggestive notations for $%
H\left( \delta ^{\mathrm{matt}}|d\right) $ and $H\left( \delta ^{\mathrm{BF}%
}|d\right) $ respectively.}. In the light of the general results from~\cite
{gen1}, according to which the local cohomology of the Koszul-Tate
differential in pure ghost number zero for a given gauge theory is trivial
in antighost numbers strictly greater than the Cauchy order of this theory,
combined with the fact that the separate Cauchy orders of the matter theory
and of the BF model are equal to one and respectively to four, we can state
that
\begin{eqnarray}
H_{k}^{\mathrm{matt}}\left( \delta |d\right) &=&0,\;k>1,  \label{hmat} \\
H_{k}^{\mathrm{BF}}\left( \delta |d\right) &=&0,\;k>4.  \label{hbf}
\end{eqnarray}
By means of the above results it follows that
\begin{equation}
H_{k}\left( \delta |d\right) =0,\;k>4  \label{hbfx}
\end{equation}
for the overall ``free'' theory (\ref{bfn1}) and, moreover, that $%
H_{k}\left( \delta |d\right) =H_{k}^{\mathrm{BF}}\left( \delta |d\right) $
for $k=2,3,4$. As for $H_{1}\left( \delta |d\right) $, this is the only case
where the general representative of the local cohomology of the Koszul-Tate
differential involves, and possibly mixes, the field/ghost and antifield
spectra of both BF and matter sectors. It is quite reasonable to assume that
if the invariant polynomial $\mu _{k}$, with $\mathrm{agh}\left( \mu
_{k}\right) =k\geq 4$, is trivial in $H_{k}\left( \delta |d\right) $, then
it can be taken to be trivial also in $H_{k}^{\mathrm{inv}}\left( \delta
|d\right) $%
\begin{equation}
\left( \mu _{k}=\delta b_{k+1}+\partial _{\mu }\stackrel{(k)}{c}^{\mu },\;%
\mathrm{agh}\left( \mu _{k}\right) =k\geq 4\right) \Rightarrow \mu
_{k}=\delta \beta _{k+1}+\partial _{\mu }\stackrel{(k)}{\gamma }^{\mu },
\label{r81d}
\end{equation}
where $\beta _{k+1}$ and $\stackrel{(k)}{\gamma }^{\mu }$ are invariant
polynomials. [An element of $H_{k}^{\mathrm{inv}}\left( \delta |d\right) $
is defined via an equation similar to (\ref{r80}) for $J\rightarrow k$, but
with the corresponding current an invariant polynomial.] This assumption is
based on what happens in many gauge theories. For instance, see~\cite
{gen2,multi,lingr,hepth04}. The results (\ref{hbfx})--(\ref{r81d}) yield the
conclusion that all the local cohomology of the Koszul-Tate differential in
the space of invariant polynomials in antighost numbers strictly greater
than four is indeed trivial
\begin{equation}
H_{k}^{\mathrm{inv}}\left( \delta |d\right) =0,\;k>4.  \label{r81c}
\end{equation}
The previous results on $H\left( \delta |d\right) $ and $H^{\mathrm{inv}%
}\left( \delta |d\right) $ in strictly positive antighost numbers are
important because they control the obstructions to removing the antifields
from the first-order deformation. This statement is also standard material
and can be shown like in~\cite{gen2,multi,lingr,hepth04}. Its proof is
mainly based on the formulas (\ref{hbfx})--(\ref{r81c}) and relies on the
fact that we can successively eliminate all the pieces of antighost number
strictly greater that four from the non-integrated density of the
first-order deformation by adding \textit{only} trivial terms. As a
consequence, we can safely take the first-order deformation to stop at
antighost number four ($J=4$ in the expansion (\ref{bfa25}))
\begin{equation}
a=a_{0}+a_{1}+a_{2}+a_{3}+a_{4},  \label{bfn24}
\end{equation}
where $a_{4}$ is of the form (\ref{bfa27}) and $\mu _{4}$ is a non-trivial
element from $H_{4}^{\mathrm{inv}}\left( \delta |d\right) =H_{4}^{\mathrm{%
invBF}}\left( \delta |d\right) $.

\subsubsection{Computation of the first-order deformation}

After some computation, we infer that the most general, non-trivial
representative of $H_{4}^{\mathrm{invBF}}(\delta |d)$ can be taken of the
type
\begin{eqnarray}
&&\left( \mu _{4}\right) ^{\mu \nu \rho \lambda }=\left( \frac{\partial U}{%
\partial \varphi _{e}}C_{e}^{*\mu \nu \rho \lambda }+\frac{\partial ^{2}U}{%
\partial \varphi _{e}\partial \varphi _{f}}\left( H_{e}^{*[\mu }C_{f}^{*\nu
\rho \lambda ]}+C_{e}^{*[\mu \nu }C_{f}^{*\rho \lambda ]}\right) \right.
\nonumber \\
&&\left. +\frac{\partial ^{3}U}{\partial \varphi _{e}\partial \varphi
_{f}\partial \varphi _{g}}H_{e}^{*[\mu }H_{f}^{*\nu }C_{g}^{*\rho \lambda ]}+%
\frac{\partial ^{4}U}{\partial \varphi _{e}\partial \varphi _{f}\partial
\varphi _{g}\partial \varphi _{h}}H_{e}^{*\mu }H_{f}^{*\nu }H_{g}^{*\rho
}H_{h}^{*\lambda }\right) ,  \label{bfn25}
\end{eqnarray}
where $U=U(\varphi _{a})$ is an arbitrary function depending only on the
\textit{undifferentiated} scalar fields. We omit the proof of this result,
which is rather tedious and not illuminating, but observe that it is in
perfect agreement with the similar one given in~\cite{BFmat} in the absence
of BF collection indices and also with that resulting from the Hamiltonian
analysis of action (\ref{bfn1}) in $n$ dimensions and reported in~\cite
{BFhamnoPT}. On the other hand, the elements of pure ghost number equal to
four of a basis in the space of the polynomials in the ghosts (\ref{ghosts})
is
\begin{equation}
\eta ^{a}C_{\mu \nu \rho \lambda }^{b},\;\eta ^{a}\eta ^{b}\eta _{c\mu \nu
\rho \lambda },\;\eta ^{a}\eta ^{b}\eta ^{c}\eta ^{d},\;\eta _{a\mu \nu \rho
\lambda }\eta _{b\mu ^{\prime }\nu ^{\prime }\rho ^{\prime }\lambda ^{\prime
}}.  \label{bfn28}
\end{equation}
Thus, the last representative from the expansion (\ref{bfn24}) is provided
by directly ``gluing'' (\ref{bfn25}) to the elements from (\ref{bfn28}) by
means of some appropriate functions of the scalar fields, and hence will be
expressed by
\begin{eqnarray}
&&a_{4}=\left( \frac{\partial W_{ab}}{\partial \varphi _{c}}C_{c}^{*\mu \nu
\rho \lambda }+\frac{\partial ^{2}W_{ab}}{\partial \varphi _{c}\partial
\varphi _{d}}\left( H_{c}^{*[\mu }C_{d}^{*\nu \rho \lambda ]}+C_{c}^{*[\mu
\nu }C_{d}^{*\rho \lambda ]}\right) \right.  \nonumber \\
&&\left. +\frac{\partial ^{3}W_{ab}}{\partial \varphi _{c}\partial \varphi
_{d}\partial \varphi _{e}}H_{c}^{*[\mu }H_{d}^{*\nu }C_{e}^{*\rho \lambda ]}+%
\frac{\partial ^{4}W_{ab}}{\partial \varphi _{c}\partial \varphi
_{d}\partial \varphi _{e}\partial \varphi _{f}}H_{c}^{*\mu }H_{d}^{*\nu
}H_{e}^{*\rho }H_{f}^{*\lambda }\right) \eta ^{a}C_{\mu \nu \rho \lambda
}^{b}  \nonumber \\
&&-\frac{1}{4}\left( \frac{\partial M_{ab}^{c}}{\partial \varphi _{d}}%
C_{d}^{*\mu \nu \rho \lambda }+\frac{\partial ^{2}M_{ab}^{c}}{\partial
\varphi _{d}\partial \varphi _{e}}\left( H_{d}^{*[\mu }C_{e}^{*\nu \rho
\lambda ]}+C_{d}^{*[\mu \nu }C_{e}^{*\rho \lambda ]}\right) +\frac{\partial
^{3}M_{ab}^{c}}{\partial \varphi _{d}\partial \varphi _{e}\partial \varphi
_{f}}\times \right.  \nonumber \\
&&\left. \times H_{d}^{*[\mu }H_{e}^{*\nu }C_{f}^{*\rho \lambda ]}+\frac{%
\partial ^{4}M_{ab}^{c}}{\partial \varphi _{d}\partial \varphi _{e}\partial
\varphi _{f}\partial \varphi _{g}}H_{d}^{*\mu }H_{e}^{*\nu }H_{f}^{*\rho
}H_{g}^{*\lambda }\right) \eta ^{a}\eta ^{b}\eta _{c\mu \nu \rho \lambda }
\nonumber \\
&&+\frac{1}{2}\varepsilon _{\mu \nu \rho \lambda }\left( \left( \frac{%
\partial M^{ab}}{\partial \varphi _{c}}C_{c}^{*\mu \nu \rho \lambda }+\frac{%
\partial ^{2}M^{ab}}{\partial \varphi _{c}\partial \varphi _{d}}\left(
H_{c}^{*[\mu }C_{d}^{*\nu \rho \lambda ]}+C_{c}^{*[\mu \nu }C_{d}^{*\rho
\lambda ]}\right) \right. \right.  \nonumber \\
&&\left. +\frac{\partial ^{3}M^{ab}}{\partial \varphi _{c}\partial \varphi
_{d}\partial \varphi _{e}}H_{c}^{*[\mu }H_{d}^{*\nu }C_{e}^{*\rho \lambda ]}+%
\frac{\partial ^{4}M^{ab}}{\partial \varphi _{c}\partial \varphi
_{d}\partial \varphi _{e}\partial \varphi _{f}}H_{c}^{*\mu }H_{d}^{*\nu
}H_{e}^{*\rho }H_{f}^{*\lambda }\right) \eta _{a}^{\sigma \tau \kappa
\varsigma }\eta _{b\sigma \tau \kappa \varsigma }  \nonumber \\
&&-\frac{1}{2\cdot \left( 4!\right) ^{2}}\left( \frac{\partial M_{mnpq}}{%
\partial \varphi _{c}}C_{c}^{*\mu \nu \rho \lambda }+\frac{\partial
^{2}M_{mnpq}}{\partial \varphi _{c}\partial \varphi _{d}}\left( H_{c}^{*[\mu
}C_{d}^{*\nu \rho \lambda ]}\right. \right.  \nonumber \\
&&\left. +C_{c}^{*[\mu \nu }C_{d}^{*\rho \lambda ]}\right) +\frac{\partial
^{3}M_{mnpq}}{\partial \varphi _{c}\partial \varphi _{d}\partial \varphi _{e}%
}H_{c}^{*[\mu }H_{d}^{*\nu }C_{e}^{*\rho \lambda ]}  \nonumber \\
&&\left. \left. +\frac{\partial ^{4}M_{mnpq}}{\partial \varphi _{c}\partial
\varphi _{d}\partial \varphi _{e}\partial \varphi _{f}}H_{c}^{*\mu
}H_{d}^{*\nu }H_{e}^{*\rho }H_{f}^{*\lambda }\right) \eta ^{m}\eta ^{n}\eta
^{p}\eta ^{q}\right) .  \label{bfn27}
\end{eqnarray}
In the above the functions $W_{ab}$, $M_{ab}^{c}$, $M^{ab}$ and $M_{mnpq}$
depend only on the undifferentiated scalar fields. Meantime, $M_{ab}^{c}$
together with $M_{mnpq}$ are antisymmetric in their lower indices due to the
anticommutation among the ghosts $\eta ^{a}$, while $M^{ab}$ are symmetric
as a consequence of the commutation among the ghosts for ghosts $\eta
_{a}^{\mu \nu \rho \lambda }$. The factors $-1/4$, $+1/2$ and respectively $%
-1/2\cdot \left( 4!\right) ^{2}$ in front of the last three pieces were
added for further convenience.

By computing the action of $\delta $ on $a_{4}$ and by taking into account
the relations (\ref{bfn18})--(\ref{bfn21}), it follows that the solution $%
a_{3}$ of the equation (\ref{r65}) for $J=4$ is precisely given by
\begin{eqnarray}
&&a_{3}=\left( \frac{\partial W_{ab}}{\partial \varphi _{c}}C_{c}^{*\mu \nu
\rho }+\frac{\partial ^{2}W_{ab}}{\partial \varphi _{c}\partial \varphi _{d}}%
H_{c}^{*[\mu }C_{d}^{*\nu \rho ]}+\frac{\partial ^{3}W_{ab}}{\partial
\varphi _{c}\partial \varphi _{d}\partial \varphi _{e}}H_{c}^{*\mu
}H_{d}^{*\nu }H_{e}^{*\rho }\right) \times  \nonumber \\
&&\times \left( -\eta ^{a}C_{\mu \nu \rho }^{b}+4A^{a\lambda }C_{\mu \nu
\rho \lambda }^{b}\right) +2\left( 6\left( \frac{\partial W_{ab}}{\partial
\varphi _{c}}C_{c}^{*\mu \nu }+\frac{\partial ^{2}W_{ab}}{\partial \varphi
_{c}\partial \varphi _{d}}H_{c}^{*\mu }H_{d}^{*\nu }\right) B^{*a\rho
\lambda }\right.  \nonumber \\
&&\left. +4\frac{\partial W_{ab}}{\partial \varphi _{c}}H_{c}^{*\mu }\eta
^{*a\nu \rho \lambda }+W_{ab}\eta ^{*a\mu \nu \rho \lambda }\right) C_{\mu
\nu \rho \lambda }^{b}+\frac{1}{2}\left( \frac{\partial M_{ab}^{c}}{\partial
\varphi _{d}}C_{d}^{*\mu \nu \rho }\right.  \nonumber \\
&&\left. +\frac{\partial ^{2}M_{ab}^{c}}{\partial \varphi _{d}\partial
\varphi _{e}}H_{d}^{*[\mu }C_{e}^{*\nu \rho ]}+\frac{\partial ^{3}M_{ab}^{c}%
}{\partial \varphi _{d}\partial \varphi _{e}\partial \varphi _{f}}%
H_{d}^{*\mu }H_{e}^{*\nu }H_{f}^{*\rho }\right) \left( \frac{1}{2}\eta
^{a}\eta ^{b}\eta _{c\mu \nu \rho }\right.  \nonumber \\
&&\left. -4A^{a\lambda }\eta ^{b}\eta _{c\mu \nu \rho \lambda }\right)
-\left( 6\left( \frac{\partial M_{ab}^{c}}{\partial \varphi _{d}}C_{d}^{*\mu
\nu }+\frac{\partial ^{2}M_{ab}^{c}}{\partial \varphi _{d}\partial \varphi
_{e}}H_{d}^{*\mu }H_{e}^{*\nu }\right) B^{*a\rho \lambda }\right.  \nonumber
\\
&&\left. +4\frac{\partial M_{ab}^{c}}{\partial \varphi _{d}}H_{d}^{*\mu
}\eta ^{*a\nu \rho \lambda }+M_{ab}^{c}\eta ^{*a\mu \nu \rho \lambda
}\right) \eta ^{b}\eta _{c\mu \nu \rho \lambda }-\varepsilon _{\mu \nu \rho
\lambda }\left( \frac{\partial M^{ab}}{\partial \varphi _{c}}C_{c}^{*\sigma
\tau \kappa }\right.  \nonumber \\
&&\left. +\frac{\partial ^{2}M^{ab}}{\partial \varphi _{c}\partial \varphi
_{d}}H_{c}^{*[\sigma }C_{d}^{*\tau \kappa ]}+\frac{\partial ^{3}M^{ab}}{%
\partial \varphi _{c}\partial \varphi _{d}\partial \varphi _{e}}%
H_{c}^{*\sigma }H_{d}^{*\tau }H_{e}^{*\kappa }\right) \eta _{a\sigma \tau
\kappa }\eta _{b}^{\mu \nu \rho \lambda }  \nonumber \\
&&-\frac{\varepsilon _{\mu \nu \rho \lambda }}{\left( 4!\right) ^{2}}\left(
4\left( \frac{\partial M_{mnpq}}{\partial \varphi _{c}}C_{c}^{*\mu \nu \rho
}+\frac{\partial ^{2}M_{mnpq}}{\partial \varphi _{c}\partial \varphi _{d}}%
H_{c}^{*[\mu }C_{d}^{*\nu \rho ]}\right. \right.  \nonumber \\
&&\left. +\frac{\partial ^{3}M_{mnpq}}{\partial \varphi _{c}\partial \varphi
_{d}\partial \varphi _{e}}H_{c}^{*\mu }H_{d}^{*\nu }H_{e}^{*\rho }\right)
A^{m\lambda }+12\left( \frac{\partial M_{mnpq}}{\partial \varphi _{c}}%
C_{c}^{*\mu \nu }\right.  \nonumber \\
&&\left. +\frac{\partial ^{2}M_{mnpq}}{\partial \varphi _{c}\partial \varphi
_{d}}H_{c}^{*\mu }H_{d}^{*\nu }\right) B^{*m\rho \lambda }+8\frac{\partial
M_{mnpq}}{\partial \varphi _{c}}H_{c}^{*\mu }\eta ^{*m\nu \rho \lambda }
\nonumber \\
&&\left. +2M_{mnpq}\eta ^{*m\mu \nu \rho \lambda }\right) \eta ^{n}\eta
^{p}\eta ^{q}.  \label{bfn30}
\end{eqnarray}
By means of the equation (\ref{r66}) for $k=3$%
\begin{equation}
\delta a_{3}+\gamma a_{2}=\partial _{\mu }\stackrel{(2)}{m}^{\mu },
\label{bfa34}
\end{equation}
the solution (\ref{bfn30}) and the definitions (\ref{bfn18})--(\ref{bfn21})
lead to
\begin{eqnarray}
a_{2} &=&\left( \frac{\partial W_{ab}}{\partial \varphi _{c}}C_{c}^{*\mu \nu
}+\frac{\partial ^{2}W_{ab}}{\partial \varphi _{c}\partial \varphi _{d}}%
H_{c}^{*\mu }H_{d}^{*\nu }\right) \left( \eta ^{a}C_{\mu \nu }^{b}-3A^{a\rho
}C_{\mu \nu \rho }^{b}\right)  \nonumber \\
&&-2\left( 3\frac{\partial W_{ab}}{\partial \varphi _{c}}H_{c}^{*\mu
}B^{*a\nu \rho }+W_{ab}\eta ^{*a\mu \nu \rho }\right) C_{\mu \nu \rho }^{b}
\nonumber \\
&&-\frac{1}{2}\left( \frac{\partial M_{ab}^{c}}{\partial \varphi _{d}}%
C_{d}^{*\mu \nu }+\frac{\partial ^{2}M_{ab}^{c}}{\partial \varphi
_{d}\partial \varphi _{e}}H_{d}^{*\mu }H_{e}^{*\nu }\right) \left( \frac{1}{2%
}\eta ^{a}\eta ^{b}B_{c\mu \nu }-3A^{a\rho }\eta ^{b}\eta _{c\mu
\nu \rho
}\right)  \nonumber \\
&&+\left( 3\frac{\partial M_{ab}^{c}}{\partial \varphi _{d}}H_{d}^{*\mu
}B^{*a\nu \rho }+M_{ab}^{c}\eta ^{*a\mu \nu \rho }\right) \eta ^{b}\eta
_{c\mu \nu \rho }  \nonumber \\
&&+\frac{1}{2}\left( -\frac{\partial M_{ab}^{c}}{\partial \varphi _{d}}%
H_{d}^{*\mu }A_{c\mu }^{*}+M_{ab}^{c}\eta _{c}^{*}\right) \eta ^{a}\eta ^{b}
\nonumber \\
&&+\left( 3\left( \frac{\partial M_{ab}^{c}}{\partial \varphi _{d}}%
C_{d}^{*\mu \nu }+\frac{\partial ^{2}M_{ab}^{c}}{\partial \varphi
_{d}\partial \varphi _{e}}H_{d}^{*\mu }H_{e}^{*\nu }\right) A^{a\rho }+12%
\frac{\partial M_{ab}^{c}}{\partial \varphi _{d}}H_{d}^{*\mu }B^{*a\nu \rho
}\right.  \nonumber \\
&&\left. +4M_{ab}^{c}\eta ^{*a\mu \nu \rho }\right) A^{b\lambda }\eta _{c\mu
\nu \rho \lambda }-6M_{ab}^{c}B_{\mu \nu }^{*a}B_{\rho \lambda }^{*b}\eta
_{c}^{\mu \nu \rho \lambda }  \nonumber \\
&&+\frac{9}{2}\varepsilon ^{\mu \nu \rho \lambda }\left( \frac{\partial
M^{ab}}{\partial \varphi _{d}}C_{d\mu \nu }^{*}+\frac{\partial ^{2}M^{ab}}{%
\partial \varphi _{d}\partial \varphi _{e}}H_{d\mu }^{*}H_{e\nu }^{*}\right)
\eta _{a\rho \sigma \tau }\eta _{b\lambda }^{\;\;\;\sigma \tau }  \nonumber
\\
&&+\varepsilon _{\mu \nu \rho \lambda }\left( \left( \frac{\partial M^{ab}}{%
\partial \varphi _{d}}C_{d}^{*\sigma \tau }+\frac{\partial ^{2}M^{ab}}{%
\partial \varphi _{d}\partial \varphi _{e}}H_{d}^{*\sigma }H_{e}^{*\tau
}\right) B_{b\sigma \tau }\right.  \nonumber \\
&&\left. +2\frac{\partial M^{ab}}{\partial \varphi _{d}}H_{d}^{*\sigma
}A_{b\sigma }^{*}-2M^{ab}\eta _{b}^{*}\right) \eta _{a}^{\mu \nu \rho
\lambda }  \nonumber \\
&&+\frac{3\varepsilon _{\mu \nu \rho \lambda }}{\left( 4!\right) ^{2}}\left(
6\left( \frac{\partial M_{mnpq}}{\partial \varphi _{d}}C_{d}^{*\mu \nu }+%
\frac{\partial ^{2}M_{mnpq}}{\partial \varphi _{d}\partial \varphi _{e}}%
H_{d}^{*\mu }H_{e}^{*\nu }\right) A^{m\rho }A^{n\lambda }\right.  \nonumber
\\
&&+24\frac{\partial M_{mnpq}}{\partial \varphi _{d}}H_{d}^{*\mu }B^{*m\nu
\rho }A^{n\lambda }+8M_{mnpq}\eta ^{*m\mu \nu \rho }A^{n\lambda }  \nonumber
\\
&&\left. -12M_{mnpq}B^{*m\mu \nu }B^{*n\rho \lambda }\right) \eta ^{p}\eta
^{q}.  \label{bfn31}
\end{eqnarray}
Next, we investigate the equation (\ref{r66}) for $k=2$%
\begin{equation}
\delta a_{2}+\gamma a_{1}=\partial _{\mu }\stackrel{(1)}{m}^{\mu },
\label{bfa36}
\end{equation}
which, combined with (\ref{bfn31}), provide $a_{1}$ like
\begin{eqnarray}
a_{1} &=&\frac{\partial W_{ab}}{\partial \varphi _{c}}H_{c}^{*\mu }\left(
-\eta ^{a}H_{\mu }^{b}+2A^{a\nu }C_{\mu \nu }^{b}\right) +W_{ab}\left(
2B_{\mu \nu }^{*a}C^{b\mu \nu }-\varphi ^{*a}\eta ^{b}\right)  \nonumber \\
&&-\frac{\partial M_{ab}^{c}}{\partial \varphi _{d}}H_{d}^{*\mu }A^{a\nu
}\left( \eta ^{b}B_{c\mu \nu }+\frac{3}{2}A^{b\rho }\eta _{c\mu \nu \rho
}\right)  \nonumber \\
&&-M_{ab}^{c}\left( B_{\mu \nu }^{*a}\eta ^{b}B_{c}^{\mu \nu }+A_{\mu
}^{a}\eta ^{b}A_{c}^{*\mu }+3B_{\mu \nu }^{*a}A_{\rho }^{b}\eta _{c}^{\mu
\nu \rho }\right)  \nonumber \\
&&+2\varepsilon _{\nu \rho \sigma \lambda }\left( \frac{\partial M^{ab}}{%
\partial \varphi _{c}}H_{c\mu }^{*}B_{a}^{\mu \nu }-M^{ab}A_{a}^{*\nu
}\right) \eta _{b}^{\rho \sigma \lambda }  \nonumber \\
&&+\frac{\varepsilon ^{\mu \nu \rho \lambda }}{4!}\left( \frac{\partial
M_{mnpq}}{\partial \varphi _{c}}H_{c\mu }^{*}A_{\nu }^{m}+3M_{mnpq}B_{\mu
\nu }^{*m}\right) A_{\rho }^{n}A_{\lambda }^{p}\eta ^{q}+\overline{a}_{1},
\label{bfn32}
\end{eqnarray}
where
\begin{eqnarray}
\bar{a}_{1} &=&\left( B_{\mu \nu }^{*a}T_{ab}^{\mu \nu }\left( \left[ \omega
^{\Delta }\right] \right) +A_{\mu }^{*a}\tilde{T}_{ab}^{\mu }\left( \left[
\omega ^{\Delta }\right] \right) +\varphi ^{*a}T_{ab}\left( \left[ \omega
^{\Delta }\right] \right) \right.  \nonumber \\
&&\left. +H_{a}^{*\mu }T_{\mu b}^{a}\left( \left[ \omega ^{\Delta }\right]
\right) +y_{i}^{*}\bar{T}_{b}^{i}\left( \left[ \omega ^{\Delta }\right]
\right) \right) \eta ^{b}\equiv \lambda _{b}\left( \left[ \omega ^{\Delta
}\right] \right) \eta ^{b}  \label{uvw}
\end{eqnarray}
and $\omega ^{\Delta }$ is explained in (\ref{omega}). In order to produce a
bosonic $\bar{a}_{1}$, as required by the standard rules of the BRST
formalism, the gauge invariant functions $T_{ab}^{\mu \nu }$, $\tilde{T}%
_{ab}^{\mu }$, $T_{ab}$ and $T_{\mu b}^{a}$ must be bosonic, while the
Grassmann parity of $\bar{T}_{b}^{i}$ should be equal to $\varepsilon _{i}$
for each $b=\overline{1,N}$. The term (\ref{uvw}) added in the right-hand
side of (\ref{bfn32}) appears like the general solution to the `homogeneous'
equation $\gamma \bar{a}_{1}=0$ and takes into account the fact that both
the BF and the matter theories participate in the local cohomology of the
Koszul-Tate differential in antighost number one. Its form is given by the
general solution (\ref{bfa27}) for $J=1$. Such terms correspond to $\bar{a}%
_{2}=0$ and thus they do not modify either the gauge algebra or the
reducibility functions, but only the gauge transformations of the
interacting theory. We emphasize that the solutions $a_{3}$ and $a_{2}$
obtained previously also include the general ones, corresponding to the
`homogeneous' equations $\gamma \bar{a}_{3}=0$ and $\gamma \bar{a}_{2}=0$.
In order to simplify the exposition we avoided the discussion regarding the
selection procedure of these solutions such as to comply with obtaining some
consistent $a_{2}$ and $a_{1}$. It is however interesting to note that this
procedure allows no new functions of the scalar fields beside $W_{ab}$, $%
M_{ab}^{c}$, $M^{ab}$ and $M_{abcd}$ to enter $a_{3}$ or $a_{2}$.

In order to solve the equation (\ref{r66}) at antighost number zero
\begin{equation}
\delta a_{1}+\gamma a_{0}=\partial ^{\mu }\stackrel{(0)}{m}_{\mu },
\label{bfagh0}
\end{equation}
whose solution is nothing but the deformed Lagrangian at order one in $g$,
from (\ref{bfn32}) we observe that
\begin{eqnarray}
\delta a_{1} &=&-\gamma \left[ -W_{ab}A^{a\mu }H_{\mu }^{b}+\frac{1}{2}%
M_{ab}^{c}A_{\mu }^{a}A_{\nu }^{b}B_{c}^{\mu \nu }\right.  \nonumber \\
&&\left. +\frac{1}{2}\varepsilon ^{\mu \nu \rho \lambda }\left(
M^{ab}B_{a\mu \nu }B_{b\rho \lambda }-\frac{1}{2\cdot 4!}M_{abcd}A_{\mu
}^{a}A_{\nu }^{b}A_{\rho }^{c}A_{\lambda }^{d}\right) \right]  \nonumber \\
&&+\partial ^{\mu }\left( W_{ab}\left( -\eta ^{a}H_{\mu }^{b}+2A^{a\nu
}C_{\mu \nu }^{b}\right) -M_{ab}^{c}A^{a\nu }\left( \eta ^{b}B_{c\mu \nu }+%
\frac{3}{2}A^{b\rho }\eta _{c\mu \nu \rho }\right) \right.  \nonumber \\
&&\left. +2\varepsilon ^{\nu \rho \sigma \lambda }M^{ab}B_{a\mu \nu }\eta
_{b\rho \sigma \lambda }+\frac{1}{4!}\varepsilon _{\mu \nu \rho \lambda
}M_{abcd}A^{a\nu }A^{b\rho }A^{c\lambda }\eta ^{d}\right) +\delta \overline{a%
}_{1}.  \label{bfn33a}
\end{eqnarray}
Thus, the consistency of the deformation procedure at order one in the
coupling constant requires that $\delta \bar{a}_{1}$ must independently be $%
\gamma $-exact modulo $d$%
\begin{equation}
\delta \bar{a}_{1}+\gamma \bar{a}_{0}=\partial _{\mu }\bar{j}^{\mu }.
\label{bfnbar0}
\end{equation}
At the first sight it seems that $\bar{a}_{1}$ is not an essential
ingredient of the deformation procedure since in its absence the equation (%
\ref{bfagh0}) would still allow solutions for $a_{0}$, as it can be observed
from (\ref{bfn33a}) in which we set $\bar{a}_{1}=0$. However, it is
important to note that in its absence there are \textit{no couplings} of the
matter fields to the BF sector. Indeed, the results (\ref{hbfx}) and $%
H_{k}\left( \delta |d\right) =H_{k}^{\mathrm{BF}}\left( \delta |d\right) $
for $k=2,3,4$ imply that the earliest step where the matter generators may
be brought in during the deformation process is given by the solutions of
the `homogeneous' equation $\gamma \bar{a}_{1}=0$ at antighost number one.
Since the scope of this paper is to analyze the structure of possible
interactions between the matter and the BF fields, in what follows we focus
on the conditions that should be satisfied such that $\bar{a}_{1}$ indeed
furnishes a consistent $\bar{a}_{0}$.

More precisely, we determine the allowed form of the functions $T_{ab}^{\mu
\nu }$, $\tilde{T}_{ab}^{\mu }$, $T_{ab}$, $T_{\mu b}^{a}$ and $\bar{T}%
_{b}^{i}$ in (\ref{uvw}) such that (\ref{bfnbar0}) is obeyed. Recalling the
definitions (\ref{bfn13})--(\ref{bfn17}), it follows that
\begin{equation}
\delta \bar{a}_{1}=-\left( \delta \lambda _{b}\right) \eta ^{b}.  \label{xx1}
\end{equation}
In the meantime, from the definitions (\ref{bfn19}) and the first two
relations in (\ref{bfn20}) we read that the equation (\ref{bfnbar0})
possesses solutions if and only if there exist some bosonic, $\gamma $%
-invariant currents $\bar{j}_{b}^{\mu }$ with both pure ghost and antighost
numbers equal to zero
\begin{equation}
\gamma \bar{j}_{b}^{\mu }=0,\;\mathrm{pgh}\left( \bar{j}_{b}^{\mu }\right)
=0=\mathrm{agh}\left( \bar{j}_{b}^{\mu }\right) ,\;\varepsilon \left( \bar{j}%
_{b}^{\mu }\right) =0,  \label{bfnbar3}
\end{equation}
such that
\begin{equation}
-\delta \lambda _{b}=\partial _{\mu }\bar{j}_{b}^{\mu }.  \label{bfnbar2}
\end{equation}
Analyzing the expression of $\lambda _{b}$ from (\ref{uvw}), after some
computation we find that a necessary condition for (\ref{bfnbar2}) to hold
is
\begin{equation}
T_{ab}^{\mu \nu }=0,\;\tilde{T}_{ab}^{\mu }=0,\;T_{ab}=0.  \label{bfnbar8}
\end{equation}
Substituting the partial solutions (\ref{bfnbar8}) back into the equation (%
\ref{bfnbar2}), the latter becomes
\begin{equation}
\left( \partial ^{\mu }\varphi _{a}\right) T_{\mu b}^{a}\left( \left[ \omega
^{\Delta }\right] \right) +\left( -\right) ^{\epsilon _{i}}\frac{\delta ^{L}%
\mathcal{L}_{0}}{\delta y^{i}}\bar{T}_{b}^{i}\left( \left[ \omega ^{\Delta
}\right] \right) =\partial _{\mu }\bar{j}_{b}^{\mu }.  \label{xbfn38a}
\end{equation}
A practical manner of exhibiting solutions to (\ref{xbfn38a}), and hence, by
virtue of the above discussion, of introducing couplings between the BF and
matter theories, is to suppose that there exist some local functions $%
T_{a}^{i}$ involving only the matter fields and their derivatives, whose
Grassmann parities are $\varepsilon _{i}$, such that
\begin{equation}
\left( -\right) ^{\varepsilon _{i}}\frac{\delta ^{L}\mathcal{L}_{0}}{\delta
y^{i}}T_{a}^{i}\left( \left[ y^{i}\right] \right) =\partial _{\mu
}j_{a}^{\mu }\left( \left[ y^{i}\right] \right) .  \label{bfn39}
\end{equation}
Here, $j_{a}^{\mu }$ are some bosonic, local currents that depend only on
the purely matter field spectrum. The last relation is nothing but Noether's
theorem expressing the appearance of the on-shell conserved currents $%
j_{a}^{\mu }\left( \left[ y^{i}\right] \right) $ (on-shell means here on the
stationary surface of field equations for the purely matter theory) deriving
from the invariance of the Lagrangian action of the matter fields under the
rigid symmetries
\begin{equation}
\Delta y^{i}=T_{a}^{i}\left( \left[ y^{i}\right] \right) \xi ^{a},
\label{bfn40}
\end{equation}
with $\xi ^{a}$ some constant, bosonic parameters. From now on we work under
the hypothesis that the matter theory indeed displays such rigid symmetries.

The equation (\ref{bfn39}) may be rewritten in terms of the Koszul-Tate
differential like
\begin{equation}
\partial _{\mu }j_{a}^{\mu }=\delta \left( -y_{i}^{*}T_{a}^{i}\right) \equiv
\delta \sigma _{a},  \label{bfa41}
\end{equation}
and it correlates the rigid symmetries (\ref{bfn40}) to certain
cohomological classes from the space $H_{1}\left( \delta |d\right) $.
Explicitly, it shows that some global symmetries (materialized in the
conserved currents $j_{a}^{\mu }$) define some elements $\sigma _{a}$ from $%
H_{1}\left( \delta |d\right) $, i.e., some elements of antighost number
equal to one that are $\delta $-closed modulo $d$. A global symmetry is said
to be trivial if the corresponding $\sigma _{a}$ are in a trivial class of $%
H_{1}\left( \delta |d\right) $, hence if they are $\delta $-exact modulo $d$%
\begin{equation}
\sigma _{a}=\delta \rho _{a}+\partial _{\mu }c_{a}^{\mu },\;\mathrm{agh}%
\left( \rho _{a}\right) =2,\;\mathrm{agh}\left( c_{a}^{\mu }\right) =1.
\label{bfn42}
\end{equation}
The currents associated with a trivial global symmetry are trivial (see the
first reference from \cite{8a}), and we assume that this is not the case
here.

Inserting (\ref{bfn39}) in (\ref{xbfn38a}), we find that the left-hand side
of (\ref{xbfn38a}) reduces to a total derivative if and only if
\begin{equation}
\bar{T}_{b}^{i}=T_{a}^{i}\left( \left[ y^{i}\right] \right) U_{b}^{a}\left(
\varphi \right) ,\;T_{\mu b}^{a}=j_{c\mu }\left( \left[ y^{i}\right] \right)
\frac{\partial U_{b}^{c}\left( \varphi \right) }{\partial \varphi _{a}},
\label{xbfny42}
\end{equation}
where $U_{b}^{a}\left( \varphi \right) $ are some arbitrary functions of the
undifferentiated scalar fields. In this case, the bar current from (\ref
{xbfn38a}) is related to the purely matter one via
\begin{equation}
\bar{j}_{b}^{\mu }=j_{a}^{\mu }\left( \left[ y^{i}\right] \right)
U_{b}^{a}\left( \varphi \right) .  \label{barunbarcurr}
\end{equation}
Using the solutions (\ref{bfnbar8}) and (\ref{xbfny42}) in (\ref{uvw}), we
completely determine $\bar{a}_{1}$ under the form
\begin{equation}
\bar{a}_{1}=y_{i}^{*}T_{a}^{i}\left( \left[ y^{i}\right] \right)
U_{b}^{a}\left( \varphi \right) \eta ^{b}+H_{a}^{*\mu }j_{c\mu }\left(
\left[ y^{i}\right] \right) \frac{\partial U_{b}^{c}\left( \varphi \right) }{%
\partial \varphi _{a}}\eta ^{b}.  \label{xyz}
\end{equation}
With $\bar{a}_{1}$ at hand, from (\ref{uvw}) we find that the solution to (%
\ref{bfnbar0}) is
\begin{equation}
\bar{a}_{0}=j_{a}^{\mu }\left( \left[ y^{i}\right] \right) U_{b}^{a}\left(
\varphi \right) A_{\mu }^{b},  \label{solabar0}
\end{equation}
which, correlated with (\ref{bfn33a}), enables us to write the full
antighost number zero component in the first-order deformation like
\begin{eqnarray}
a_{0} &=&j_{a}^{\mu }\left( \left[ y^{i}\right] \right) U_{b}^{a}\left(
\varphi \right) A_{\mu }^{b}-W_{ab}\left( \varphi \right) A^{a\mu }H_{\mu
}^{b}+\frac{1}{2}M_{ab}^{c}\left( \varphi \right) A_{\mu }^{a}A_{\nu
}^{b}B_{c}^{\mu \nu }  \nonumber \\
&&+\frac{1}{2}\varepsilon ^{\mu \nu \rho \lambda }\left( M^{ab}\left(
\varphi \right) B_{a\mu \nu }B_{b\rho \lambda }-\frac{1}{2\cdot 4!}%
M_{abcd}\left( \varphi \right) A_{\mu }^{a}A_{\nu }^{b}A_{\rho
}^{c}A_{\lambda }^{d}\right) .  \label{bfn43}
\end{eqnarray}

The basic conclusion of the above discussion is that the appearance of
consistent couplings (at order one in the deformation parameter) of the
matter fields to the BF ones is obtained under the hypothesis that the
matter theory is invariant under some (non-trivial) bosonic global
transformations of the type (\ref{bfn40}), that result, via Noether's
theorem (\ref{bfn39}), in the (non-trivial) conserved currents $j_{a}^{\mu }$%
. It is essential that the number of rigid symmetries is equal to the number
of BF fields from the collection ($N$).

Putting together the formulas (\ref{bfn27}), (\ref{bfn30}), (\ref{bfn31}), (%
\ref{bfn32}), (\ref{xyz}) and (\ref{bfn43}), we conclude that the
first-order deformation of the solution to the master equation for the model
under study can be written in the form
\begin{eqnarray}
&&S_{1}=\int d^{4}x\left( \left( \frac{\partial W_{ab}}{\partial \varphi _{c}%
}C_{c}^{*\mu \nu \rho \lambda }+\frac{\partial ^{2}W_{ab}}{\partial \varphi
_{c}\partial \varphi _{d}}\left( H_{c}^{*[\mu }C_{d}^{*\nu \rho \lambda
]}+C_{c}^{*[\mu \nu }C_{d}^{*\rho \lambda ]}\right) \right. \right.
\nonumber \\
&&\left. +\frac{\partial ^{3}W_{ab}}{\partial \varphi _{c}\partial \varphi
_{d}\partial \varphi _{e}}H_{c}^{*[\mu }H_{d}^{*\nu }C_{e}^{*\rho \lambda ]}+%
\frac{\partial ^{4}W_{ab}}{\partial \varphi _{c}\partial \varphi
_{d}\partial \varphi _{e}\partial \varphi _{f}}H_{c}^{*\mu }H_{d}^{*\nu
}H_{e}^{*\rho }H_{f}^{*\lambda }\right) \eta ^{a}C_{\mu \nu \rho \lambda
}^{b}  \nonumber \\
&&-\frac{1}{4}\left( \frac{\partial M_{ab}^{c}}{\partial \varphi _{d}}%
C_{d}^{*\mu \nu \rho \lambda }+\frac{\partial ^{2}M_{ab}^{c}}{\partial
\varphi _{d}\partial \varphi _{e}}\left( H_{d}^{*[\mu }C_{e}^{*\nu \rho
\lambda ]}+C_{d}^{*[\mu \nu }C_{e}^{*\rho \lambda ]}\right) \right.
\nonumber \\
&&\left. +\frac{\partial ^{3}M_{ab}^{c}}{\partial \varphi _{d}\partial
\varphi _{e}\partial \varphi _{f}}H_{d}^{*[\mu }H_{e}^{*\nu }C_{f}^{*\rho
\lambda ]}+\frac{\partial ^{4}M_{ab}^{c}}{\partial \varphi _{d}\partial
\varphi _{e}\partial \varphi _{f}\partial \varphi _{g}}H_{d}^{*\mu
}H_{e}^{*\nu }H_{f}^{*\rho }H_{g}^{*\lambda }\right) \eta ^{a}\eta ^{b}\eta
_{c\mu \nu \rho \lambda }  \nonumber \\
&&+\frac{1}{2}\varepsilon _{\mu \nu \rho \lambda }\left( \left( \frac{%
\partial M^{ab}}{\partial \varphi _{c}}C_{c}^{*\mu \nu \rho \lambda }+\frac{%
\partial ^{2}M^{ab}}{\partial \varphi _{c}\partial \varphi _{d}}\left(
H_{c}^{*[\mu }C_{d}^{*\nu \rho \lambda ]}+C_{c}^{*[\mu \nu }C_{d}^{*\rho
\lambda ]}\right) \right. \right.  \nonumber \\
&&\left. +\frac{\partial ^{3}M^{ab}}{\partial \varphi _{c}\partial \varphi
_{d}\partial \varphi _{e}}H_{c}^{*[\mu }H_{d}^{*\nu }C_{e}^{*\rho \lambda ]}+%
\frac{\partial ^{4}M^{ab}}{\partial \varphi _{c}\partial \varphi
_{d}\partial \varphi _{e}\partial \varphi _{f}}H_{c}^{*\mu
}H_{d}^{*\nu }H_{e}^{*\rho }H_{f}^{*\lambda }\right) \eta
_{a}^{\sigma \tau \kappa \varsigma
}\eta _{b\sigma \tau \kappa \varsigma }  \nonumber \\
&&-\frac{1}{2\cdot \left( 4!\right) ^{2}}\left( \frac{\partial M_{mnpq}}{%
\partial \varphi _{c}}C_{c}^{*\mu \nu \rho \lambda }+\frac{\partial
^{2}M_{mnpq}}{\partial \varphi _{c}\partial \varphi _{d}}\left( H_{c}^{*[\mu
}C_{d}^{*\nu \rho \lambda ]}+C_{c}^{*[\mu \nu }C_{d}^{*\rho \lambda
]}\right) \right.  \nonumber \\
&&+\frac{\partial ^{3}M_{mnpq}}{\partial \varphi _{c}\partial \varphi
_{d}\partial \varphi _{e}}H_{c}^{*[\mu }H_{d}^{*\nu }C_{e}^{*\rho \lambda ]}
\nonumber \\
&&\left. \left. +\frac{\partial ^{4}M_{mnpq}}{\partial \varphi _{c}\partial
\varphi _{d}\partial \varphi _{e}\partial \varphi _{f}}H_{c}^{*\mu
}H_{d}^{*\nu }H_{e}^{*\rho }H_{f}^{*\lambda }\right) \eta ^{m}\eta ^{n}\eta
^{p}\eta ^{q}\right)  \nonumber \\
&&+\left( \frac{\partial W_{ab}}{\partial \varphi _{c}}C_{c}^{*\mu \nu \rho
}+\frac{\partial ^{2}W_{ab}}{\partial \varphi _{c}\partial \varphi _{d}}%
H_{c}^{*[\mu }C_{d}^{*\nu \rho ]}+\frac{\partial ^{3}W_{ab}}{\partial
\varphi _{c}\partial \varphi _{d}\partial \varphi _{e}}H_{c}^{*\mu
}H_{d}^{*\nu }H_{e}^{*\rho }\right) \times  \nonumber \\
&&\times \left( -\eta ^{a}C_{\mu \nu \rho }^{b}+4A^{a\lambda }C_{\mu \nu
\rho \lambda }^{b}\right) +2\left( 6\left( \frac{\partial W_{ab}}{\partial
\varphi _{c}}C_{c}^{*\mu \nu }+\frac{\partial ^{2}W_{ab}}{\partial \varphi
_{c}\partial \varphi _{d}}H_{c}^{*\mu }H_{d}^{*\nu }\right) B^{*a\rho
\lambda }\right.  \nonumber \\
&&\left. +4\frac{\partial W_{ab}}{\partial \varphi _{c}}H_{c}^{*\mu }\eta
^{*a\nu \rho \lambda }+W_{ab}\eta ^{*a\mu \nu \rho \lambda }\right) C_{\mu
\nu \rho \lambda }^{b}+\frac{1}{2}\left( \frac{\partial M_{ab}^{c}}{\partial
\varphi _{d}}C_{d}^{*\mu \nu \rho }\right.  \nonumber \\
&&\left. +\frac{\partial ^{2}M_{ab}^{c}}{\partial \varphi _{d}\partial
\varphi _{e}}H_{d}^{*[\mu }C_{e}^{*\nu \rho ]}+\frac{\partial ^{3}M_{ab}^{c}%
}{\partial \varphi _{d}\partial \varphi _{e}\partial \varphi _{f}}%
H_{d}^{*\mu }H_{e}^{*\nu }H_{f}^{*\rho }\right) \left( \frac{1}{2}\eta
^{a}\eta ^{b}\eta _{c\mu \nu \rho }\right.  \nonumber \\
&&\left. -4A^{a\lambda }\eta ^{b}\eta _{c\mu \nu \rho \lambda }\right)
-\left( 6\left( \frac{\partial M_{ab}^{c}}{\partial \varphi _{d}}C_{d}^{*\mu
\nu }+\frac{\partial ^{2}M_{ab}^{c}}{\partial \varphi _{d}\partial \varphi
_{e}}H_{d}^{*\mu }H_{e}^{*\nu }\right) B^{*a\rho \lambda }\right.  \nonumber
\\
&&\left. +4\frac{\partial M_{ab}^{c}}{\partial \varphi _{d}}H_{d}^{*\mu
}\eta ^{*a\nu \rho \lambda }+M_{ab}^{c}\eta ^{*a\mu \nu \rho \lambda
}\right) \eta ^{b}\eta _{c\mu \nu \rho \lambda }-\varepsilon _{\mu \nu \rho
\lambda }\left( \frac{\partial M^{ab}}{\partial \varphi _{c}}C_{c}^{*\sigma
\tau \kappa }\right.  \nonumber \\
&&\left. +\frac{\partial ^{2}M^{ab}}{\partial \varphi _{c}\partial \varphi
_{d}}H_{c}^{*[\sigma }C_{d}^{*\tau \kappa ]}+\frac{\partial ^{3}M^{ab}}{%
\partial \varphi _{c}\partial \varphi _{d}\partial \varphi _{e}}%
H_{c}^{*\sigma }H_{d}^{*\tau }H_{e}^{*\kappa }\right) \eta _{a\sigma \tau
\kappa }\eta _{b}^{\mu \nu \rho \lambda }  \nonumber \\
&&-\frac{\varepsilon _{\mu \nu \rho \lambda }}{\left( 4!\right) ^{2}}\left(
4\left( \frac{\partial M_{mnpq}}{\partial \varphi _{c}}C_{c}^{*\mu \nu \rho
}+\frac{\partial ^{2}M_{mnpq}}{\partial \varphi _{c}\partial \varphi _{d}}%
H_{c}^{*[\mu }C_{d}^{*\nu \rho ]}\right. \right.  \nonumber \\
&&\left. +\frac{\partial ^{3}M_{mnpq}}{\partial \varphi _{c}\partial \varphi
_{d}\partial \varphi _{e}}H_{c}^{*\mu }H_{d}^{*\nu }H_{e}^{*\rho }\right)
A^{m\lambda }+12\left( \frac{\partial M_{mnpq}}{\partial \varphi _{c}}%
C_{c}^{*\mu \nu }\right.  \nonumber \\
&&\left. +\frac{\partial ^{2}M_{mnpq}}{\partial \varphi _{c}\partial \varphi
_{d}}H_{c}^{*\mu }H_{d}^{*\nu }\right) B^{*m\rho \lambda }  \nonumber \\
&&\left. +8\frac{\partial M_{mnpq}}{\partial \varphi _{c}}H_{c}^{*\mu }\eta
^{*m\nu \rho \lambda }+2M_{mnpq}\eta ^{*m\mu \nu \rho \lambda }\right) \eta
^{n}\eta ^{p}\eta ^{q}  \nonumber \\
&&+\left( \frac{\partial W_{ab}}{\partial \varphi _{c}}C_{c}^{*\mu \nu }+%
\frac{\partial ^{2}W_{ab}}{\partial \varphi _{c}\partial \varphi _{d}}%
H_{c}^{*\mu }H_{d}^{*\nu }\right) \left( \eta ^{a}C_{\mu \nu }^{b}-3A^{a\rho
}C_{\mu \nu \rho }^{b}\right)  \nonumber \\
&&-2\left( 3\frac{\partial W_{ab}}{\partial \varphi _{c}}H_{c}^{*\mu
}B^{*a\nu \rho }+W_{ab}\eta ^{*a\mu \nu \rho }\right) C_{\mu \nu \rho }^{b}
\nonumber \\
&&-\frac{1}{2}\left( \frac{\partial M_{ab}^{c}}{\partial \varphi _{d}}%
C_{d}^{*\mu \nu }+\frac{\partial ^{2}M_{ab}^{c}}{\partial \varphi
_{d}\partial \varphi _{e}}H_{d}^{*\mu }H_{e}^{*\nu }\right) \left( \frac{1}{2%
}\eta ^{a}\eta ^{b}B_{c\mu \nu }-3A^{a\rho }\eta ^{b}\eta _{c\mu \nu \rho
}\right)  \nonumber \\
&&+\left( 3\frac{\partial M_{ab}^{c}}{\partial \varphi _{d}}H_{d}^{*\mu
}B^{*a\nu \rho }+M_{ab}^{c}\eta ^{*a\mu \nu \rho }\right) \eta ^{b}\eta
_{c\mu \nu \rho }  \nonumber \\
&&+\frac{1}{2}\left( -\frac{\partial M_{ab}^{c}}{\partial \varphi _{d}}%
H_{d}^{*\mu }A_{c\mu }^{*}+M_{ab}^{c}\eta _{c}^{*}\right) \eta ^{a}\eta ^{b}
\nonumber \\
&&+\left( 3\left( \frac{\partial M_{ab}^{c}}{\partial \varphi _{d}}%
C_{d}^{*\mu \nu }+\frac{\partial ^{2}M_{ab}^{c}}{\partial \varphi
_{d}\partial \varphi _{e}}H_{d}^{*\mu }H_{e}^{*\nu }\right) A^{a\rho }+12%
\frac{\partial M_{ab}^{c}}{\partial \varphi _{d}}H_{d}^{*\mu }B^{*a\nu \rho
}\right.  \nonumber \\
&&\left. +4M_{ab}^{c}\eta ^{*a\mu \nu \rho }\right) A^{b\lambda }\eta _{c\mu
\nu \rho \lambda }-6M_{ab}^{c}B_{\mu \nu }^{*a}B_{\rho \lambda }^{*b}\eta
_{c}^{\mu \nu \rho \lambda }  \nonumber \\
&&+\frac{9}{2}\varepsilon ^{\mu \nu \rho \lambda }\left( \frac{\partial
M^{ab}}{\partial \varphi _{d}}C_{d\mu \nu }^{*}+\frac{\partial ^{2}M^{ab}}{%
\partial \varphi _{d}\partial \varphi _{e}}H_{d\mu }^{*}H_{e\nu }^{*}\right)
\eta _{a\rho \sigma \tau }\eta _{b\lambda }^{\;\;\;\sigma \tau }  \nonumber
\\
&&+\varepsilon _{\mu \nu \rho \lambda }\left( \left( \frac{\partial M^{ab}}{%
\partial \varphi _{d}}C_{d}^{*\sigma \tau }+\frac{\partial ^{2}M^{ab}}{%
\partial \varphi _{d}\partial \varphi _{e}}H_{d}^{*\sigma }H_{e}^{*\tau
}\right) B_{b\sigma \tau }\right.  \nonumber \\
&&\left. +2\frac{\partial M^{ab}}{\partial \varphi _{d}}H_{d}^{*\sigma
}A_{b\sigma }^{*}-2M^{ab}\eta _{b}^{*}\right) \eta _{a}^{\mu \nu \rho
\lambda }  \nonumber \\
&&+\frac{3\varepsilon _{\mu \nu \rho \lambda }}{\left( 4!\right) ^{2}}\left(
6\left( \frac{\partial M_{mnpq}}{\partial \varphi _{d}}C_{d}^{*\mu \nu }+%
\frac{\partial ^{2}M_{mnpq}}{\partial \varphi _{d}\partial \varphi _{e}}%
H_{d}^{*\mu }H_{e}^{*\nu }\right) A^{m\rho }A^{n\lambda }\right.  \nonumber
\\
&&+24\frac{\partial M_{mnpq}}{\partial \varphi _{d}}H_{d}^{*\mu }B^{*m\nu
\rho }A^{n\lambda }  \nonumber \\
&&\left. +8M_{mnpq}\eta ^{*m\mu \nu \rho }A^{n\lambda }-12M_{mnpq}B^{*m\mu
\nu }B^{*n\rho \lambda }\right) \eta ^{p}\eta ^{q}  \nonumber \\
&&+\frac{\partial W_{ab}}{\partial \varphi _{c}}H_{c}^{*\mu }\left( -\eta
^{a}H_{\mu }^{b}+2A^{a\nu }C_{\mu \nu }^{b}\right) +W_{ab}\left( 2B_{\mu \nu
}^{*a}C^{b\mu \nu }-\varphi ^{*a}\eta ^{b}\right)  \nonumber \\
&&-\frac{\partial M_{ab}^{c}}{\partial \varphi _{d}}H_{d}^{*\mu }A^{a\nu
}\left( \eta ^{b}B_{c\mu \nu }+\frac{3}{2}A^{b\rho }\eta _{c\mu \nu \rho
}\right)  \nonumber \\
&&-M_{ab}^{c}\left( B_{\mu \nu }^{*a}\eta ^{b}B_{c}^{\mu \nu }+A_{\mu
}^{a}\eta ^{b}A_{c}^{*\mu }+3B_{\mu \nu }^{*a}A_{\rho }^{b}\eta _{c}^{\mu
\nu \rho }\right)  \nonumber \\
&&+2\varepsilon _{\nu \rho \sigma \tau }\left( \frac{\partial M^{ab}}{%
\partial \varphi _{c}}H_{c\mu }^{*}B_{a}^{\mu \nu }-M^{ab}A_{a}^{*\nu
}\right) \eta _{b}^{\rho \sigma \tau }  \nonumber \\
&&+\frac{\varepsilon ^{\mu \nu \rho \lambda }}{4!}\left( \frac{\partial
M_{mnpq}}{\partial \varphi _{c}}H_{c\mu }^{*}A_{\nu }^{m}+3M_{mnpq}B_{\mu
\nu }^{*m}\right) A_{\rho }^{n}A_{\lambda }^{p}\eta
^{q}+y_{i}^{*}T_{a}^{i}U_{b}^{a}\eta ^{b}  \nonumber \\
&&+H_{m}^{*\mu }j_{a\mu }\frac{\partial U_{b}^{a}}{\partial \varphi _{m}}%
\eta ^{b}-W_{ab}A^{a\mu }H_{\mu }^{b}+\frac{1}{2}M_{ab}^{c}A_{\mu
}^{a}A_{\nu }^{b}B_{c}^{\mu \nu }+j_{a}^{\mu }U_{b}^{a}A_{\mu }^{b}
\nonumber \\
&&\left. +\frac{\varepsilon ^{\mu \nu \rho \lambda }}{2}\left( M^{ab}B_{a\mu
\nu }B_{b\rho \lambda }-\frac{1}{2\cdot 4!}M_{abcd}A_{\mu }^{a}A_{\nu
}^{b}A_{\rho }^{c}A_{\lambda }^{d}\right) \right) .  \label{bfn33}
\end{eqnarray}
It is by construction a $s$-cocycle of ghost number zero, such that $%
S+gS_{1} $ is solution to the master equation (\ref{bfn2.3}) up to order $g$.

\subsection{Higher-order deformations}

It is clear from (\ref{bfn33}) that $S_{1}$ strongly depends on the
structure of the matter theory. The only (reasonable) assumptions made so
far on the ``free'' matter theory are that its Lagrangian density $\mathcal{L%
}_{0}\left( \left[ y^{i}\right] \right) $ is at most second-order
in the derivatives of $y^{i}$ and that it separately describes a
``normal'' theory of Cauchy order equal to one. Until now we did
not restrict in any way the derivative order of the interacting
Lagrangian density, given at the first order in the coupling
constant by $a_{0}$ like in (\ref{bfn43}). However, as announced
in the beginning of this section, we will ask that the
interactions preserve the differential order with respect to the
``free'' field equations. Given the first-order differential
behaviour of the BF field equations resulting from the ``free''
action (\ref{bfn1}), it follows that each term in $a_{0}$ must be
restricted to have at most one spacetime derivative. The
quantities in $a_{0}$ that may contain spacetime derivatives of
the fields are proportional to $j_{a}^{\mu }\left( \left[
y^{i}\right] \right) $, and hence we must ask that the conserved
matter currents have no more than one derivative. In view of
Noether's theorem (\ref{bfn39}) it is
then sufficient to take the generators $T_{a}^{i}$ of the rigid symmetries (%
\ref{bfn40}) to be polynomials in the undifferentiated matter fields (or
even infinite series in the subset of commuting such fields). In order to
fix the ideas and manifestly ensure that the Grassmann parity of $T_{a}^{i}$
is $\varepsilon _{i}$ from now on we consider the case where the generators
of the rigid symmetries (\ref{bfn40}) are linear in the matter fields, i.e.
\begin{equation}
T_{a}^{i}=\left( T_{a}\right) _{\;j}^{i}y^{j},  \label{ec8}
\end{equation}
with $\left( T_{a}\right) _{\;j}^{i}$ denoting the components of some
constant matrices $T_{a}$. Consequently, we obtain that the derivative order
of $j_{a}^{\mu }$ is less than that of the matter Lagrangian density by one
unit, namely it may be either zero or one.

Next, we discuss the higher-order deformation equations (\ref{bfn2.6})--(\ref
{bfn2.7}), etc. Initially, we analyze under what conditions the first-order
deformation (\ref{bfn33}) is consistent at the second order in the coupling
constant, namely the equation (\ref{bfn2.6}) holds. We will see that these
conditions impose various restrictions on the functions entering (\ref{bfn33}%
), so on the one hand they fix the expression of $S_{1}$ itself and, on the
other hand, allow us to predict whether non-trivial second- and possibly
higher-order deformations of the solution to the master equation appear%
\footnote{%
Strictly speaking, we should have added to (\ref{bfn33}) also the solutions
to the equation $\gamma a_{0}^{\prime }=\partial _{\mu }\tau ^{\mu }$ at
antighost number zero. For the lack of simplicity we have omitted such
solutions since their consistency can be shown to enforce their triviality,
independently of the consistency equation for (\ref{bfn33}), which is
further discussed.}. The second-order deformation is governed by the
equation (\ref{bfn2.6}), which, if we maintain the notations from (\ref
{bfn2.2}) and consider that $\left( S_{1},S_{1}\right) =\int d^{4}x\,\Delta $%
, takes the local form
\begin{equation}
\Delta =-\frac{1}{2}sb+\partial ^{\mu }\theta _{\mu }.  \label{bfn33abc}
\end{equation}
At this point it is necessary to make some specifications. It is clear that
the expression of $\Delta $ depends, beside the BF sector, also on the
(derivative) structure of the corresponding matter currents $j_{a}^{\mu }$.
This is why we will approach distinctly the situation where the matter
currents display no derivatives from the case where the derivative order of
these currents is equal to one.

Assuming that the matter currents have no derivatives (or, equivalently,
that the matter Lagrangian density is first-order derivative), with the help
of (\ref{bfn33}) in which we set (\ref{ec8}) we infer that
\begin{eqnarray}
\Delta &=&K^{abc}t_{abc}+K_{d}^{abc}\frac{\partial t_{abc}}{\partial \varphi
_{d}}+K_{de}^{abc}\frac{\partial ^{2}t_{abc}}{\partial \varphi _{d}\partial
\varphi _{e}}+K_{def}^{abc}\frac{\partial ^{3}t_{abc}}{\partial \varphi
_{d}\partial \varphi _{e}\partial \varphi _{f}}  \nonumber \\
&&+K_{defg}^{abc}\frac{\partial ^{4}t_{abc}}{\partial \varphi _{d}\partial
\varphi _{e}\partial \varphi _{f}\partial \varphi _{g}}%
+U_{d}^{abc}t_{abc}^{d}+U_{d,e}^{abc}\frac{\partial t_{abc}^{d}}{\partial
\varphi _{e}}+U_{d,ef}^{abc}\frac{\partial ^{2}t_{abc}^{d}}{\partial \varphi
_{e}\partial \varphi _{f}}  \nonumber \\
&&+U_{d,efg}^{abc}\frac{\partial ^{3}t_{abc}^{d}}{\partial \varphi
_{e}\partial \varphi _{f}\partial \varphi _{g}}+U_{d,efgh}^{abc}\frac{%
\partial ^{4}t_{abc}^{d}}{\partial \varphi _{e}\partial \varphi _{f}\partial
\varphi _{g}\partial \varphi _{h}}+K^{abcdf}t_{abcdf}  \nonumber \\
&&+K_{e}^{abcdf}\frac{\partial t_{abcdf}}{\partial \varphi _{e}}%
+K_{eg}^{abcdf}\frac{\partial ^{2}t_{abcdf}}{\partial \varphi _{e}\partial
\varphi _{g}}+K_{egh}^{abcdf}\frac{\partial ^{3}t_{abcdf}}{\partial \varphi
_{e}\partial \varphi _{g}\partial \varphi _{h}}  \nonumber \\
&&+K_{eghl}^{abcdf}\frac{\partial ^{4}t_{abcdf}}{\partial \varphi
_{e}\partial \varphi _{g}\partial \varphi _{h}\partial \varphi _{l}}%
+K_{b}^{a}t_{a}^{b}+K_{b,c}^{a}\frac{\partial t_{a}^{b}}{\partial \varphi
_{c}}+K_{b,cd}^{a}\frac{\partial ^{2}t_{a}^{b}}{\partial \varphi
_{c}\partial \varphi _{d}}  \nonumber \\
&&+K_{b,cde}^{a}\frac{\partial ^{3}t_{a}^{b}}{\partial \varphi _{c}\partial
\varphi _{d}\partial \varphi _{e}}+K_{b,cdef}^{a}\frac{\partial ^{4}t_{a}^{b}%
}{\partial \varphi _{c}\partial \varphi _{d}\partial \varphi _{e}\partial
\varphi _{f}}+K_{ab}^{c}t_{c}^{ab}+K_{ab,d}^{c}\frac{\partial t_{c}^{ab}}{%
\partial \varphi _{d}}  \nonumber \\
&&+K_{ab,de}^{c}\frac{\partial ^{2}t_{c}^{ab}}{\partial \varphi _{d}\partial
\varphi _{e}}+K_{ab,def}^{c}\frac{\partial ^{3}t_{c}^{ab}}{\partial \varphi
_{d}\partial \varphi _{e}\partial \varphi _{f}}+K_{ab,defg}^{c}\frac{%
\partial ^{4}t_{c}^{ab}}{\partial \varphi _{d}\partial \varphi _{e}\partial
\varphi _{f}\partial \varphi _{g}}+\bar{\Delta}  \nonumber \\
&\equiv &\Pi +\bar{\Delta},  \label{bfn36}
\end{eqnarray}
where $\bar{\Delta}$ is responsible for the occurrence of the matter sector
and its expression is
\begin{eqnarray}
\bar{\Delta} &=&-4\varepsilon _{\mu \nu \rho \lambda }\left( \left(
H_{m}^{*\sigma }j_{a\sigma }\frac{\partial \left( U_{e}^{a}M^{eb}\right) }{%
\partial \varphi _{m}}+y_{i}^{*}\left( T_{a}\right)
_{\;j}^{i}y^{j}U_{e}^{a}M^{eb}\right) \eta _{b}^{\mu \nu \rho \lambda
}\right.  \nonumber \\
&&\left. +j_{a}^{\mu }U_{e}^{a}M^{eb}\eta _{b}^{\nu \rho \lambda }\right)
+y_{i}^{*}\left( \left( T_{a}\right) _{\;j}^{i}\left( U_{e}^{a}\frac{%
\partial W_{bc}}{\partial \varphi _{e}}\right. \right.  \nonumber \\
&&\left. +W_{eb}\frac{\partial U_{c}^{a}}{\partial \varphi _{e}}-W_{ec}\frac{%
\partial U_{b}^{a}}{\partial \varphi _{e}}\right) \left. +\left[
T_{d},T_{e}\right] _{\;j}^{i}U_{b}^{d}U_{c}^{e}\right) y^{j}\eta ^{b}\eta
^{c}  \nonumber \\
&&+H_{m\mu }^{*}\left( j_{a}^{\mu }\frac{\partial }{\partial \varphi _{m}}%
\left( U_{e}^{a}\frac{\partial W_{bc}}{\partial \varphi _{e}}+W_{eb}\frac{%
\partial U_{c}^{a}}{\partial \varphi _{e}}-W_{ec}\frac{\partial U_{b}^{a}}{%
\partial \varphi _{e}}\right) \right.  \nonumber \\
&&\left. +\frac{\delta ^{R}j_{a}^{\mu }}{\delta y^{i}}\left( T_{e}\right)
_{\;j}^{i}y^{j}\left( \frac{\partial U_{b}^{a}}{\partial \varphi _{m}}%
U_{c}^{e}-\frac{\partial U_{c}^{a}}{\partial \varphi _{m}}U_{b}^{e}\right)
\right) \eta ^{b}\eta ^{c}  \nonumber \\
&&+2\left( j_{a}^{\mu }\left( U_{e}^{a}\frac{\partial W_{bc}}{\partial
\varphi _{e}}+W_{eb}\frac{\partial U_{c}^{a}}{\partial \varphi _{e}}-W_{ec}%
\frac{\partial U_{b}^{a}}{\partial \varphi _{e}}\right) \right.  \nonumber \\
&&\left. +\frac{\delta ^{R}j_{a}^{\mu }}{\delta y^{i}}\left( T_{e}\right)
_{\;j}^{i}y^{j}U_{c}^{e}U_{b}^{a}\right) A_{\mu }^{b}\eta ^{c}.
\label{mattc}
\end{eqnarray}
The form of the $t$'s involved in (\ref{bfn36}) reads as
\begin{eqnarray}
t_{abc} &=&W_{ec}M_{ab}^{c}+W_{ea}\frac{\partial W_{bc}}{\partial \varphi
_{e}}+W_{eb}\frac{\partial W_{ca}}{\partial \varphi _{e}},  \label{bfn37} \\
t_{abc}^{d} &=&W_{e[a}\frac{\partial M_{bc]}^{d}}{\partial \varphi _{e}}%
+M_{e[a}^{d}M_{bc]}^{e}+M^{de}M_{eabc},  \label{bfn38} \\
t_{abcdf} &=&W_{e[a}\frac{\partial M_{bcdf]}}{\partial \varphi _{e}}%
+M_{e[abc}M_{df]}^{e},  \label{bfn38a} \\
t_{a}^{b} &=&M^{be}W_{ea},  \label{bfn38b} \\
t_{a}^{bc} &=&W_{ea}\frac{\partial M^{bc}}{\partial \varphi _{e}}%
+M_{ea}^{(b}M_{\left. {}\right. }^{c)e},  \label{bfn38c}
\end{eqnarray}
while the remaining objects, of the type $K$ and $U$, can be found in
Appendix \ref{notat}. The notation $\left( bc\cdots \right) $ signifies
complete symmetrization with respect to the indices between parentheses such
as to include all the independent terms only once and without normalization
factors. We recall that the expression of $\bar{\Delta}$ from (\ref{mattc})
was obtained under the assumption that the currents $j_{a}^{\mu }$ exhibit
\textit{no derivatives}.

In the complementary situation, where the matter currents do contain
derivatives (which is the same with assuming that the matter Lagrangian
density is second-order derivative), we find that the non-integrated density
of $\left( S_{1},S_{1}\right) =\int d^{4}x\,\Delta ^{\prime }$ can be
written as
\begin{equation}
\Delta ^{\prime }=\Delta +\Lambda \equiv \Pi +\bar{\Delta}+\Lambda ,
\label{mattc1}
\end{equation}
where $\Delta $ is given in (\ref{bfn36}) and $\Lambda $ contains
derivatives, involves both the BF and matter sectors and depends on the
concrete form of $j_{a}^{\mu }$. The general expression of $\Lambda $ is
nevertheless not illuminating in the sequel. Since $\Delta ^{\prime }$
includes $\Delta $, it is enough to analyze the consistency of the
first-order deformation using directly the former quantity. However, when
relevant differences between the cases where the currents do or do not
contain derivatives arise, they will be clearly emphasized. There are two
main types of terms in the right-hand side of (\ref{mattc1}): (i) the first
kind involves only the BF sector and was generically denoted by $\Pi $; (ii)
the second variety combines the matter and the BF spectra and is designated
by $\bar{\Delta}+\Lambda $, with $\bar{\Delta}$ given in (\ref{mattc}). Due
to their different nature, $\Pi $ and $\bar{\Delta}+\Lambda $ must
separately be $s$-boundaries modulo $d$, i.e. each of them has to be written
in a form similar to the right-hand side of (\ref{bfn33abc}).

None of the elements of the type (i) can be written like in the right-hand
side of (\ref{bfn33abc}) because none of them contains spacetime
derivatives, as does the action of $s$ on all fields, ghosts and antifields
from the BF sector. Thereby, $t_{abc}$, $t_{abc}^{d}$, $t_{abcdf}$, $%
t_{a}^{b}$ and $t_{a}^{bc}$ must vanish
\begin{equation}
t_{abc}=0,\ t_{abc}^{d}=0,\;t_{abcdf}=0,\;t_{a}^{b}=0,\;t_{a}^{bc}=0.
\label{sist}
\end{equation}
Using the expressions (\ref{bfn37})--(\ref{bfn38c}), we see that the
solution to the equations (\ref{sist}) is
\begin{equation}
M_{ab}^{c}=\frac{\partial W_{ab}}{\partial \varphi _{c}},\;M_{abcd}=f_{e[ab}%
\frac{\partial W_{cd]}}{\partial \varphi _{e}},\;M^{ab}=0,  \label{sist1}
\end{equation}
where, the functions $W_{ab}$ (which are antisymmetric now due to the first
relation in (\ref{sist1}) and to the established antisymmetry of $M_{ab}^{c}$%
) are restricted to satisfy the identity
\begin{equation}
W_{e[a}\frac{\partial W_{bc]}}{\partial \varphi _{e}}=0  \label{sist2}
\end{equation}
and $f_{abc}$ are arbitrary constants, completely antisymmetric in their
indices.

On behalf of (\ref{sist1}), we find that $\Delta ^{\prime }$ given by (\ref
{mattc1}) reduces only to components of the kind (ii)
\begin{eqnarray}
\Delta ^{\prime } &=&y_{i}^{*}\left( \left( T_{a}\right) _{\;j}^{i}\left(
U_{e}^{a}\frac{\partial W_{bc}}{\partial \varphi _{e}}+W_{eb}\frac{\partial
U_{c}^{a}}{\partial \varphi _{e}}-W_{ec}\frac{\partial U_{b}^{a}}{\partial
\varphi _{e}}\right) \right.  \nonumber \\
&&\left. +\left[ T_{d},T_{e}\right] _{\;j}^{i}U_{b}^{d}U_{c}^{e}\right)
y^{j}\eta ^{b}\eta ^{c}  \nonumber \\
&&+H_{m\mu }^{*}\left( j_{a}^{\mu }\frac{\partial }{\partial \varphi _{m}}%
\left( U_{e}^{a}\frac{\partial W_{bc}}{\partial \varphi _{e}}+W_{eb}\frac{%
\partial U_{c}^{a}}{\partial \varphi _{e}}-W_{ec}\frac{\partial U_{b}^{a}}{%
\partial \varphi _{e}}\right) \right.  \nonumber \\
&&\left. +\frac{\delta ^{R}j_{a}^{\mu }}{\delta y^{i}}\left( T_{e}\right)
_{\;j}^{i}y^{j}\left( \frac{\partial U_{b}^{a}}{\partial \varphi _{m}}%
U_{c}^{e}-\frac{\partial U_{c}^{a}}{\partial \varphi _{m}}U_{b}^{e}\right)
\right) \eta ^{b}\eta ^{c}  \nonumber \\
&&+2\left( j_{a}^{\mu }\left( U_{e}^{a}\frac{\partial W_{bc}}{\partial
\varphi _{e}}+W_{eb}\frac{\partial U_{c}^{a}}{\partial \varphi _{e}}-W_{ec}%
\frac{\partial U_{b}^{a}}{\partial \varphi _{e}}\right) \right.  \nonumber \\
&&\left. +\frac{\delta ^{R}j_{a}^{\mu }}{\delta y^{i}}\left( T_{e}\right)
_{\;j}^{i}y^{j}U_{c}^{e}U_{b}^{a}\right) A_{\mu }^{b}\eta ^{c}+\Lambda .
\label{sistx3}
\end{eqnarray}
We observe that the term from (\ref{sistx3}) proportional with $y_{i}^{*}$
cannot be written in a $s$-exact modulo $d$ form for the same reason like
before. In view of this, we impose that
\begin{equation}
\left( U_{e}^{a}\frac{\partial W_{bc}}{\partial \varphi _{e}}+W_{eb}\frac{%
\partial U_{c}^{a}}{\partial \varphi _{e}}-W_{ec}\frac{\partial U_{b}^{a}}{%
\partial \varphi _{e}}\right) \left( T_{a}\right) _{\;j}^{i}+\left[
T_{d},T_{e}\right] _{\;j}^{i}U_{b}^{d}U_{c}^{e}=0.  \label{ec9}
\end{equation}

There are two kinds of solutions to the system (\ref{ec9}). The first one
(to be called ``type \textbf{I} solution'') is
\begin{equation}
U_{b}^{a}=k^{ae}W_{eb},\;\left[ T_{a},T_{b}\right] =0  \label{sol14}
\end{equation}
and it does not impose further constraints on the functions $W_{ab}$, but
merely restricts the matrices $T_{a}$ to be commutative (the second relation
in (\ref{sol14})). In (\ref{sol14}) $k^{ae}$ are some constants. Due to the
identities (\ref{sist2}), the antisymmetric functions $W_{ab}$ can be viewed
like the two-tensor on a Poisson manifold with the dynamical scalar fields
taken like local coordinates on the target space. The second one (to be
named ``type \textbf{II} solution'') reads as
\begin{eqnarray}
W_{ab} &=&\bar{f}_{ab}^{c}\varphi _{c}+F_{ab},  \label{sol21} \\
M_{bc}^{a} &=&\bar{f}_{bc}^{a},  \label{sol22} \\
U_{b}^{a} &=&\delta _{b}^{a},  \label{sol23} \\
\left[ T_{a},T_{b}\right]  &=&-\bar{f}_{ab}^{c}T_{c},  \label{sol24}
\end{eqnarray}
and it restricts the functions $W_{ab}$ to be at most linear in the scalar
fields. In the above, $\bar{f}_{ab}^{c}$ and $F_{ab}$ are some constants,
antisymmetric in their lower indices. As a consequence of the identities (%
\ref{sist2}), we find that these constants are further subject to the
conditions
\begin{equation}
\bar{f}_{e[a}^{d}\bar{f}_{bc]}^{e}=0,\;F_{e[a}\bar{f}_{bc]}^{e}=0.
\label{sol25}
\end{equation}
We can thus interpret $\bar{f}_{ab}^{c}$ like the structure constants of a
certain Lie algebra $L\left( \mathcal{G}\right) $, where by $\mathcal{G}$ we
denoted the unique (since the Lie algebra is by hypothesis
finite-dimensional) connected, simply-connected Lie group having this
algebra like its Lie algebra. Then, according to (\ref{sol24}), the matrices
$T_{a}$ of elements $\left( T_{a}\right) _{\;k}^{i}$ can be viewed like a
basis of infinitesimal generators of an arbitrary linear representation of
dimension $I$ (the number of matter fields) of $L\left( \mathcal{G}\right) $%
. With these two types of solutions at hand, in the sequel we analyze the
existence of higher-order deformations of the solution to the master
equation.

\subsubsection{Type I solutions}

Substituting (\ref{sol14}) in (\ref{sistx3}) we obtain that
\begin{eqnarray}
&&\left( S_{1},S_{1}\right) =  \nonumber \\
&&\int d^{4}x\left( 2\frac{\delta ^{R}j_{a}^{\mu }}{\delta y^{i}}\left(
T_{b}\right) _{\;j}^{i}y^{j}k^{am}k^{bn}W_{nc}\eta ^{c}\left( W_{md}A_{\mu
}^{d}\right. \right.  \nonumber \\
&&\left. \left. +H_{e\mu }^{*}\frac{\partial W_{md}}{\partial \varphi _{e}}%
\eta ^{d}\right) +\Lambda _{\mathrm{I}}\right) ,  \label{bfny1}
\end{eqnarray}
where $\Lambda _{\mathrm{I}}$ means $\Lambda $ restricted to the type I
solutions. Two major situations met in practical applications deserve
special attention.

\textbf{I.a.} Initially, we consider the case where the matter currents are
invariant under the gauge version of the genuine rigid symmetries (\ref
{bfn40})
\begin{equation}
\frac{\delta ^{R}j_{a}^{\mu }}{\delta y^{i}}\left( T_{b}\right)
_{\;j}^{i}y^{j}=0.  \label{bfny2}
\end{equation}
It is obvious that if the matter currents $j_{a}^{\mu }$ contain
derivatives, then they cannot be invariant under the gauge version of the
rigid symmetries (\ref{bfn40}). As a consequence, the formula (\ref{bfny2})
might hold only if these currents involve no derivatives, in which situation
$\Lambda _{\mathrm{I}}$ vanishes. Then, from (\ref{bfny1}) we infer that $%
\left( S_{1},S_{1}\right) =0$, and so we can take $S_{2}=0$, and, in fact, $%
S_{k}=0$ for all $k\geq 2$. As a consequence, the deformed solution to the
master equation that is \textit{consistent to all orders in the coupling
constant} results in this case from (\ref{bfn33}) where we set (\ref{ec8}), (%
\ref{sist1}) and (\ref{sol14}), and reads as
\begin{eqnarray}
&&\bar{S}^{\left( \mathrm{I.a}\right) }=\int d^{4}x\left( H_{\mu }^{a}\hat{D}%
^{\mu }\varphi _{a}+\frac{1}{2}B_{a}^{\mu \nu }\hat{F}_{\mu \nu }^{a}-\frac{g%
}{4\cdot 4!}\varepsilon ^{\mu \nu \rho \lambda }f_{m\left[ ab\right. }\frac{%
\partial W_{\left. cd\right] }}{\partial \varphi _{m}}A_{\mu }^{a}A_{\nu
}^{b}A_{\rho }^{c}A_{\lambda }^{d}\right.  \nonumber \\
&&+\mathcal{L}_{0}\left( \left[ y^{i}\right] \right) +gj_{a}^{\mu
}k^{ae}W_{eb}A_{\mu }^{b}+H_{a}^{*\mu }\left( 2\left( \hat{D}^{\nu }\right)
_{\;\;b}^{a}C_{\mu \nu }^{b}+gj_{m\mu }k^{me}\frac{\partial W_{eb}}{\partial
\varphi _{a}}\eta ^{b}\right.  \nonumber \\
&&-g\frac{\partial W_{bc}}{\partial \varphi _{a}}\eta ^{b}H_{\mu }^{c}-g%
\frac{\partial ^{2}W_{bd}}{\partial \varphi _{a}\partial \varphi _{c}}%
A^{b\nu }\left( \eta ^{d}B_{c\mu \nu }+\frac{3}{2}A^{d\rho }\eta _{c\mu \nu
\rho }\right)  \nonumber \\
&&\left. +g\frac{\varepsilon _{\mu \nu \rho \lambda }}{4!}f_{c\left[
mn\right. }\frac{\partial ^{2}W_{\left. pq\right] }}{\partial \varphi
_{a}\partial \varphi _{c}}A^{m\nu }A^{n\rho }A^{p\lambda }\eta ^{q}\right)
-B_{\mu \nu }^{*a}\left( 3\left( \hat{D}_{\rho }\right) _{a}^{\;\;b}\eta
_{b}^{\mu \nu \rho }-2gW_{ab}C^{b\mu \nu }\right.  \nonumber \\
&&\left. +g\frac{\partial W_{ab}}{\partial \varphi _{c}}\eta ^{b}B_{c}^{\mu
\nu }-\frac{g}{8}\varepsilon ^{\mu \nu \rho \lambda }f_{d\left[ ab\right. }%
\frac{\partial W_{\left. pq\right] }}{\partial \varphi _{d}}A_{\rho
}^{b}A_{\lambda }^{p}\eta ^{q}\right)  \nonumber \\
&&+A_{a}^{*\mu }\left( \hat{D}_{\mu }\right) _{\;\;b}^{a}\eta ^{b}-g\varphi
^{*a}W_{ab}\eta ^{b}+gy_{i}^{*}\left( T_{a}\right)
_{\;j}^{i}y^{j}k^{ae}W_{eb}\eta ^{b}  \nonumber \\
&&-C_{a}^{*\mu \nu }\left( 3\left( \hat{D}^{\rho }\right) _{\;\;b}^{a}C_{\mu
\nu \rho }^{b}-g\frac{\partial W_{bc}}{\partial \varphi _{a}}\eta ^{b}C_{\mu
\nu }^{c}-g\frac{\partial ^{2}W_{bc}}{\partial \varphi _{a}\partial \varphi
_{d}}\left( 3A^{b\rho }A^{c\lambda }\eta _{d\mu \nu \rho \lambda }\right.
\right.  \nonumber \\
&&\left. \left. -\left( \frac{1}{4}\eta ^{b}B_{d\mu \nu }+\frac{3}{2}%
A^{b\rho }\eta _{d\mu \nu \rho }\right) \eta ^{c}\right) -\frac{3g}{4\cdot 4!%
}\varepsilon _{\mu \nu \rho \lambda }f_{d\left[ mn\right. }\frac{\partial
^{2}W_{\left. pq\right] }}{\partial \varphi _{a}\partial \varphi _{d}}%
A^{m\rho }A^{n\lambda }\eta ^{p}\eta ^{q}\right)  \nonumber \\
&&+\eta _{\mu \nu \rho }^{*a}\left( 4\left( \hat{D}_{\lambda }\right)
_{a}^{\;\;b}\eta _{b}^{\mu \nu \rho \lambda }-2gW_{ab}C^{b\mu \nu \rho }+g%
\frac{\partial W_{ab}}{\partial \varphi _{c}}\eta ^{b}\eta _{c}^{\mu \nu
\rho }\right.  \nonumber \\
&&\left. +\frac{g}{4!}\varepsilon ^{\mu \nu \rho \lambda }f_{m\left[
ab\right. }\frac{\partial W_{\left. pq\right] }}{\partial \varphi _{m}}%
A_{\lambda }^{b}\eta ^{p}\eta ^{q}\right) +\frac{g}{2}\frac{\partial W_{ab}}{%
\partial \varphi _{c}}\eta _{c}^{*}\eta ^{a}\eta ^{b}  \nonumber \\
&&+gH_{d}^{*\mu }H_{e}^{*\nu }\left( \frac{\partial ^{2}W_{ab}}{\partial
\varphi _{d}\partial \varphi _{e}}\left( \eta ^{a}C_{\mu \nu }^{b}-3A^{a\rho
}C_{\mu \nu \rho }^{b}\right) \right.  \nonumber \\
&&+\frac{\partial ^{3}W_{ab}}{\partial \varphi _{c}\partial \varphi
_{d}\partial \varphi _{e}}\left( 3A^{a\rho }A^{b\lambda }\eta _{c\mu \nu
\rho \lambda }-\frac{1}{4}\eta ^{a}\eta ^{b}B_{c\mu \nu }+\frac{3}{2}%
A^{a\rho }\eta ^{b}\eta _{c\mu \nu \rho }\right)  \nonumber \\
&&\left. +\frac{3}{4\cdot 4!}\varepsilon _{\mu \nu \rho \lambda }f_{a\left[
mn\right. }\frac{\partial ^{3}W_{\left. pq\right] }}{\partial \varphi
_{a}\partial \varphi _{d}\partial \varphi _{e}}A^{m\rho }A^{n\lambda }\eta
^{p}\eta ^{q}\right) -3gB_{\mu \nu }^{*a}B_{\rho \lambda }^{*b}\left( 2\frac{%
\partial W_{ab}}{\partial \varphi _{c}}\eta _{c}^{\mu \nu \rho \lambda
}\right.  \nonumber \\
&&\left. +\frac{1}{2\cdot 4!}\varepsilon ^{\mu \nu \rho \lambda }f_{c\left[
ab\right. }\frac{\partial W_{\left. pq\right] }}{\partial \varphi _{c}}\eta
^{p}\eta ^{q}\right) -\frac{g}{2}\frac{\partial ^{2}W_{ab}}{\partial \varphi
_{c}\partial \varphi _{d}}H_{d}^{*\mu }A_{c\mu }^{*}\eta ^{a}\eta ^{b}
\nonumber \\
&&+3gH_{d}^{*\mu }B^{*a\nu \rho }\left( -2\frac{\partial W_{ab}}{\partial
\varphi _{d}}C_{\mu \nu \rho }^{b}+\frac{\partial ^{2}W_{ab}}{\partial
\varphi _{c}\partial \varphi _{d}}\left( \eta ^{b}\eta _{c\mu \nu \rho
}+4A^{b\lambda }\eta _{c\mu \nu \rho \lambda }\right) \right.  \nonumber \\
&&\left. +\frac{1}{4!}\varepsilon _{\mu \nu \rho \lambda }f_{c\left[
ab\right. }\frac{\partial ^{2}W_{\left. pq\right] }}{\partial \varphi
_{c}\partial \varphi _{d}}A^{b\lambda }\eta ^{p}\eta ^{q}\right)
+C_{a}^{*\mu \nu \rho }\left( 4\left( \hat{D}^{\lambda }\right)
_{\;\;b}^{a}C_{\mu \nu \rho \lambda }^{b}\right.  \nonumber \\
&&-g\frac{\partial W_{bc}}{\partial \varphi _{a}}\eta ^{b}C_{\mu \nu \rho
}^{c}+g\frac{\partial ^{2}W_{bc}}{\partial \varphi _{a}\partial \varphi _{d}}%
\left( \frac{1}{4}\eta ^{b}\eta ^{c}\eta _{d\mu \nu \rho }-2A^{b\lambda
}\eta ^{c}\eta _{d\mu \nu \rho \lambda }\right)  \nonumber \\
&&\left. -\frac{g}{3!\cdot 4!}\varepsilon _{\mu \nu \rho \lambda }f_{c\left[
mn\right. }\frac{\partial ^{2}W_{\left. pq\right] }}{\partial \varphi
_{a}\partial \varphi _{c}}A^{m\lambda }\eta ^{n}\eta ^{p}\eta ^{q}\right)
+g\eta ^{*a\mu \nu \rho \lambda }\left( 2W_{ab}C_{\mu \nu \rho \lambda
}^{b}\right.  \nonumber \\
&&\left. -\frac{\partial W_{ab}}{\partial \varphi _{c}}\eta ^{b}\eta _{c\mu
\nu \rho \lambda }-\frac{2}{\left( 4!\right) ^{2}}\varepsilon _{\mu \nu \rho
\lambda }f_{b\left[ an\right. }\frac{\partial W_{\left. pq\right] }}{%
\partial \varphi _{b}}\eta ^{n}\eta ^{p}\eta ^{q}\right)  \nonumber \\
&&+g\left( \frac{\partial ^{2}W_{ab}}{\partial \varphi _{c}\partial \varphi
_{d}}H_{c}^{*[\mu }C_{d}^{*\nu \rho ]}+\frac{\partial ^{3}W_{ab}}{\partial
\varphi _{c}\partial \varphi _{d}\partial \varphi _{e}}H_{c}^{*\mu
}H_{d}^{*\nu }H_{e}^{*\rho }\right) \times  \nonumber \\
&&\times \left( -\eta ^{a}C_{\mu \nu \rho }^{b}+4A^{a\lambda }C_{\mu \nu
\rho \lambda }^{b}\right) +4g\left( 3\left( \frac{\partial W_{ab}}{\partial
\varphi _{c}}C_{c}^{*\mu \nu }+\frac{\partial ^{2}W_{ab}}{\partial \varphi
_{c}\partial \varphi _{d}}H_{c}^{*\mu }H_{d}^{*\nu }\right) B^{*a\rho
\lambda }\right.  \nonumber \\
&&\left. +2\frac{\partial W_{ab}}{\partial \varphi _{c}}H_{c}^{*\mu }\eta
^{*a\nu \rho \lambda }\right) C_{\mu \nu \rho \lambda }^{b}+\frac{1}{2}%
g\left( \frac{\partial ^{3}W_{ab}}{\partial \varphi _{c}\partial \varphi
_{d}\partial \varphi _{e}}H_{d}^{*[\mu }C_{e}^{*\nu \rho ]}\right.  \nonumber
\\
&&\left. +\frac{\partial ^{4}W_{ab}}{\partial \varphi _{c}\partial \varphi
_{d}\partial \varphi _{e}\partial \varphi _{f}}H_{d}^{*\mu }H_{e}^{*\nu
}H_{f}^{*\rho }\right) \left( \frac{1}{2}\eta ^{a}\eta ^{b}\eta _{c\mu \nu
\rho }-4A^{a\lambda }\eta ^{b}\eta _{c\mu \nu \rho \lambda }\right)
\nonumber \\
&&-2g\left( 3\left( \frac{\partial ^{2}W_{ab}}{\partial \varphi _{c}\partial
\varphi _{d}}C_{d}^{*\mu \nu }+\frac{\partial ^{3}W_{ab}}{\partial \varphi
_{c}\partial \varphi _{d}\partial \varphi _{e}}H_{d}^{*\mu }H_{e}^{*\nu
}\right) B^{*a\rho \lambda }\right.  \nonumber \\
&&\left. +2\frac{\partial ^{2}W_{ab}}{\partial \varphi _{c}\partial \varphi
_{d}}H_{d}^{*\mu }\eta ^{*a\nu \rho \lambda }\right) \eta ^{b}\eta _{c\mu
\nu \rho \lambda }  \nonumber \\
&&-\frac{g}{\left( 4!\right) ^{2}}\varepsilon _{\mu \nu \rho \lambda
}f_{a\left[ mn\right. }\left( 4\left( \frac{\partial ^{3}W_{\left. pq\right]
}}{\partial \varphi _{a}\partial \varphi _{c}\partial \varphi _{d}}%
H_{c}^{*[\mu }C_{d}^{*\nu \rho ]}\right. \right.  \nonumber \\
&&\left. +\frac{\partial ^{4}W_{\left. pq\right] }}{\partial \varphi
_{a}\partial \varphi _{c}\partial \varphi _{d}\partial \varphi _{e}}%
H_{c}^{*\mu }H_{d}^{*\nu }H_{e}^{*\rho }\right) A^{m\lambda }+12\left( \frac{%
\partial ^{2}W_{\left. pq\right] }}{\partial \varphi _{a}\partial \varphi
_{c}}C_{c}^{*\mu \nu }\right.  \nonumber \\
&&\left. \left. +\frac{\partial ^{3}W_{\left. pq\right] }}{\partial \varphi
_{a}\partial \varphi _{c}\partial \varphi _{d}}H_{c}^{*\mu }H_{d}^{*\nu
}\right) B^{*m\rho \lambda }+8\frac{\partial ^{2}W_{\left. pq\right] }}{%
\partial \varphi _{a}\partial \varphi _{c}}H_{c}^{*\mu }\eta ^{*m\nu \rho
\lambda }\right) \eta ^{n}\eta ^{p}\eta ^{q}  \nonumber \\
&&+gC_{c}^{*\mu \nu \rho \lambda }\left( \frac{\partial W_{ab}}{\partial
\varphi _{c}}\eta ^{a}C_{\mu \nu \rho \lambda }^{b}-\frac{1}{4}\frac{%
\partial ^{2}W_{ab}}{\partial \varphi _{c}\partial \varphi _{d}}\eta
^{a}\eta ^{b}\eta _{d\mu \nu \rho \lambda }\right.  \nonumber \\
&&\left. -\frac{1}{4\cdot \left( 4!\right) ^{2}}\varepsilon _{\mu \nu \rho
\lambda }f_{a\left[ mn\right. }\frac{\partial ^{2}W_{\left. pq\right] }}{%
\partial \varphi _{a}\partial \varphi _{c}}\eta ^{m}\eta ^{n}\eta ^{p}\eta
^{q}\right)  \nonumber \\
&&+g\left( \frac{\partial ^{2}W_{ab}}{\partial \varphi _{c}\partial \varphi
_{d}}\left( H_{c}^{*[\mu }C_{d}^{*\nu \rho \lambda ]}+C_{c}^{*[\mu \nu
}C_{d}^{*\rho \lambda ]}\right) \right.  \nonumber \\
&&\left. +\frac{\partial ^{3}W_{ab}}{\partial \varphi _{c}\partial \varphi
_{d}\partial \varphi _{e}}H_{c}^{*[\mu }H_{d}^{*\nu }C_{e}^{*\rho \lambda ]}+%
\frac{\partial ^{4}W_{ab}}{\partial \varphi _{c}\partial \varphi
_{d}\partial \varphi _{e}\partial \varphi _{f}}H_{c}^{*\mu }H_{d}^{*\nu
}H_{e}^{*\rho }H_{f}^{*\lambda }\right) \eta ^{a}C_{\mu \nu \rho \lambda
}^{b}  \nonumber \\
&&-\frac{1}{4}g\left( \frac{\partial ^{3}W_{ab}}{\partial \varphi
_{c}\partial \varphi _{d}\partial \varphi _{e}}\left( H_{d}^{*[\mu
}C_{e}^{*\nu \rho \lambda ]}+C_{d}^{*[\mu \nu }C_{e}^{*\rho \lambda
]}\right) \right.  \nonumber \\
&&+\frac{\partial ^{4}W_{ab}}{\partial \varphi _{c}\partial \varphi
_{d}\partial \varphi _{e}\partial \varphi _{f}}H_{d}^{*[\mu }H_{e}^{*\nu
}C_{f}^{*\rho \lambda ]}  \nonumber \\
&&\left. +\frac{\partial ^{5}W_{ab}}{\partial \varphi _{c}\partial \varphi
_{d}\partial \varphi _{e}\partial \varphi _{f}\partial \varphi _{g}}%
H_{d}^{*\mu }H_{e}^{*\nu }H_{f}^{*\rho }H_{g}^{*\lambda }\right) \eta
^{a}\eta ^{b}\eta _{c\mu \nu \rho \lambda }  \nonumber \\
&&-\frac{g}{4\cdot \left( 4!\right) ^{2}}\varepsilon _{\mu \nu \rho \lambda
}f_{a\left[ mn\right. }\left( \frac{\partial ^{3}W_{\left. pq\right] }}{%
\partial \varphi _{a}\partial \varphi _{c}\partial \varphi _{d}}\left(
H_{c}^{*[\mu }C_{d}^{*\nu \rho \lambda ]}+C_{c}^{*[\mu \nu }C_{d}^{*\rho
\lambda ]}\right) \right.  \nonumber \\
&&+\frac{\partial ^{4}W_{\left. pq\right] }}{\partial \varphi _{a}\partial
\varphi _{c}\partial \varphi _{d}\partial \varphi _{e}}H_{c}^{*[\mu
}H_{d}^{*\nu }C_{e}^{*\rho \lambda ]}  \nonumber \\
&&\left. \left. +\frac{\partial ^{5}W_{\left. pq\right] }}{\partial \varphi
_{a}\partial \varphi _{c}\partial \varphi _{d}\partial \varphi _{e}\partial
\varphi _{f}}H_{c}^{*\mu }H_{d}^{*\nu }H_{e}^{*\rho }H_{f}^{*\lambda
}\right) \eta ^{m}\eta ^{n}\eta ^{p}\eta ^{q}\right) .  \label{bfny3}
\end{eqnarray}
where we used the notations
\begin{eqnarray}
\hat{D}^{\mu }\varphi _{a} &=&\partial ^{\mu }\varphi _{a}+gW_{ab}A^{b\mu },
\label{bfny4} \\
\hat{F}_{\mu \nu }^{a} &=&\partial _{[\mu }A_{\nu ]}^{a}+g\frac{\partial
W_{bc}}{\partial \varphi _{a}}A_{\mu }^{b}A_{\nu }^{c},  \label{bfny5} \\
\left( \hat{D}_{\mu }\right) _{\;\;b}^{a} &=&\delta _{b}^{a}\partial _{\mu
}-g\frac{\partial W_{bc}}{\partial \varphi _{a}}A_{\mu }^{c},  \label{bfny6}
\\
\left( \hat{D}_{\mu }\right) _{b}^{\;\;a} &=&\delta _{b}^{a}\partial _{\mu
}+g\frac{\partial W_{bc}}{\partial \varphi _{a}}A_{\mu }^{c}.  \label{bfny7}
\end{eqnarray}
From the full solution (\ref{bfny3}) we can extract all the information on
the resulting interacting model.

Indeed, the pieces with antighost number zero from (\ref{bfny3}) produce the
\textit{Lagrangian action of the coupled theory}
\begin{eqnarray}
\bar{S}_{0}^{\left( \mathrm{I.a}\right) } &=&\int d^{4}x\left( H_{\mu }^{a}%
\hat{D}^{\mu }\varphi _{a}+\frac{1}{2}B_{a}^{\mu \nu }\hat{F}_{\mu \nu }^{a}-%
\frac{g}{4\cdot 4!}\varepsilon ^{\mu \nu \rho \lambda }f_{m\left[ ab\right. }%
\frac{\partial W_{\left. cd\right] }}{\partial \varphi _{m}}A_{\mu
}^{a}A_{\nu }^{b}A_{\rho }^{c}A_{\lambda }^{d}\right.  \nonumber \\
&&\left. +\mathcal{L}_{0}\left( \left[ y^{i}\right] \right) +gj_{a}^{\mu
}\left( y^{i}\right) k^{ae}W_{eb}\left( \varphi \right) A_{\mu }^{b}\right) .
\label{bfny8}
\end{eqnarray}
The first three types of terms describe the self-interactions of the BF
fields. They were partially obtained by us in~\cite{BFhamnoPT} under the
supplementary assumption that the interactions do not break the PT
invariance (which is the same with setting $f_{mab}=0$ in (\ref{bfny8})).
Here, we dropped this requirement and consequently allowed the appearance of
a new vertex that is quartic in the one-forms $A_{\mu }^{a}$. The last term
provides the couplings between the BF fields and the matter fields and can
be written in the compact form $g\bar{j}_{b}^{\mu }A_{\mu }^{b}$, where the
bar current is given in (\ref{barunbarcurr}), with the functions $U_{b}^{a}$
like in (\ref{sol14}). It is interesting to remark that we have a
generalized minimal coupling, in the sense that even if it is formally
expressed like `vector fields times currents', however the currents are not
the conserved, purely matter ones from (\ref{bfn39}), but the matter
currents in some `background' potentials $k^{ae}W_{eb}$ of scalar fields.

With the help of the elements of antighost number equal to one present in (%
\ref{bfny3}) we read the \textit{generating set of deformed gauge
transformations} for the action (\ref{bfny8}), which can be obtained by
detaching the antifields from these terms and by replacing the ghosts with
the corresponding gauge parameters
\begin{equation}
\hat{\delta}_{\epsilon }\varphi _{a}=-gW_{ab}\epsilon ^{b},\;\hat{\delta}%
_{\epsilon }A_{\mu }^{a}=\left( \hat{D}_{\mu }\right) _{\;\;b}^{a}\epsilon
^{b},  \label{bfny9}
\end{equation}
\begin{eqnarray}
\hat{\delta}_{\epsilon }H_{\mu }^{a} &=&2\left( \hat{D}^{\nu }\right)
_{\;\;b}^{a}\epsilon _{\mu \nu }^{b}+g\left( j_{m\mu }k^{me}\frac{\partial
W_{eb}}{\partial \varphi _{a}}-\frac{\partial W_{bc}}{\partial \varphi _{a}}%
H_{\mu }^{c}\right.  \nonumber \\
&&\left. +\frac{\partial ^{2}W_{bd}}{\partial \varphi _{a}\partial \varphi
_{c}}A^{d\nu }B_{c\mu \nu }+\frac{1}{4!}\varepsilon _{\mu \nu \rho \lambda
}f_{c\left[ mn\right. }\frac{\partial ^{2}W_{\left. pb\right] }}{\partial
\varphi _{a}\partial \varphi _{c}}A^{m\nu }A^{n\rho }A^{p\lambda }\right)
\epsilon ^{b}  \nonumber \\
&&-\frac{3}{2}g\frac{\partial ^{2}W_{bd}}{\partial \varphi _{a}\partial
\varphi _{c}}A^{b\nu }A^{d\rho }\epsilon _{c\mu \nu \rho }  \label{bfny10}
\end{eqnarray}
\begin{eqnarray}
\hat{\delta}_{\epsilon }B_{a}^{\mu \nu } &=&-3\left( \hat{D}_{\rho }\right)
_{a}^{\;\;b}\epsilon _{b}^{\mu \nu \rho }+2gW_{ab}\epsilon ^{b\mu \nu
}-g\left( \frac{\partial W_{ab}}{\partial \varphi _{c}}B_{c}^{\mu \nu
}\right.  \nonumber \\
&&\left. -\frac{1}{8}\varepsilon ^{\mu \nu \rho \lambda }f_{d\left[
ac\right. }\frac{\partial W_{\left. pb\right] }}{\partial \varphi _{d}}%
A_{\rho }^{c}A_{\lambda }^{p}\right) \epsilon ^{b},  \label{bfny11}
\end{eqnarray}
\begin{equation}
\hat{\delta}_{\epsilon }y^{i}=g\left( T_{a}\right)
_{\;j}^{i}y^{j}k^{ae}W_{eb}\epsilon ^{b}.  \label{bfny12}
\end{equation}
All the gauge transformations are deformed with respect to the free ones, (%
\ref{bfn2a})--(\ref{bfn2}). The striking features of the new gauge
symmetries can be synthesized by: 1. the matter fields gain gauge
transformations, which can be viewed as being obtained by gauging the rigid
symmetries (\ref{bfn40}) with the choice (\ref{ec8}) and by further putting
them in the same `background' potentials $k^{ae}W_{eb}$ of scalar fields; 2.
the BF one-forms $H_{\mu }^{a}$ are endowed with gauge transformations that
are proportional with the matter currents and with the derivatives of the
`background' potentials. Since the deformed gauge generators depend on the
fields, in contrast to the `free' theory (\ref{bfn1}), we expect that the
corresponding gauge algebra is non-Abelian, although the matrices $T_{a}$
commute (the second relation in (\ref{sol14})). As it will be seen below,
this is indeed the case. Another interesting observation is that the
deformed field strengths and gauge transformations of the one-forms $A_{\mu
}^{a}$ take generalized Yang-Mills forms, with the usual structure constants
replaced by the derivatives of the `background' potentials (see (\ref{bfny5}%
) and the second relation in (\ref{bfny9}), with the corresponding
`covariant derivatives' like in (\ref{bfny6})).

The pieces of antighost number two are known to describe the \textit{%
deformed gauge algebra and new first-order reducibility relations}. They
belong to four distinct categories. Firstly, those linear in the antifields
of the ghosts and quadratic in the ghosts with the pure ghost number equal
to one contain the structure functions corresponding to the commutators
among the gauge transformations (\ref{bfny9})--(\ref{bfny12}). Secondly, the
presence of elements which are simultaneously quadratic in the antifields of
the original fields as well as in the ghosts with the pure ghost number
equal to one shows that the gauge transformations (\ref{bfny9})--(\ref
{bfny12}) only close on-shell. Here `on-shell' means on the stationary
surface of field equations for the deformed action (\ref{bfny8}). As a
consequence, it follows that \textit{the deformed gauge algebra for the type
I solutions is open}, i.e., only closes on-shell, unlike the initial one,
which is Abelian. It is interesting to mention that the matrices $T_{a}$
from the gauge transformations (\ref{bfny12}) of the matter fields are
commuting (see the second relation in (\ref{sol14})), but the matter gauge
transformations are not. This is essentially so because (\ref{bfny12}) also
involves the functions $W_{eb}$ that satisfy the identities (\ref{sist2}).
Thirdly, there appear terms which are linear both in the antifields of the
ghosts and in the ghosts with the pure ghost number equal to two; the
functions that `glue' these BRST generators are precisely the deformed
first-order reducibility functions of the coupled model. Fourthly, we notice
the existence of pieces that are quadratic in the antifields of the original
fields and also linear in the ghosts with the pure ghost number equal to two
--- they exhibit the on-shell closeness of the first-order reducibility
relations, in contrast to the initial theory, whose reducibility
takes place everywhere on the space of field histories. The terms
with the antighost number equal to three and four complete the
tensor gauge structure of the interacting model. Among others,
they lead to the conclusion that the reducibility relations of
order two also hold only on-shell.

\textbf{I.b.}\ In the opposite situation, where the conserved matter
currents are not invariant under the gauge version of (\ref{bfn40})
\begin{equation}
\frac{\delta ^{R}j_{a}^{\mu }}{\delta y^{i}}\left( T_{b}\right)
_{\;j}^{i}y^{j}\neq 0,  \label{bfny13}
\end{equation}
it follows that $\left( S_{1},S_{1}\right) $ is not vanishing, hence the
second-order deformation $S_{2}$ as solution to the equation (\ref{bfn2.6})
will also be so. This happens for instance if the matter currents contain
spacetime derivatives. Moreover, it is possible to obtain other non-trivial,
higher-order deformations when solving the remaining equations ((\ref{bfn2.7}%
), etc.). The expressions of these deformations strongly depend on the
structure of the matter theory and cannot be output in the general setting
considered here. What is always valid is that the complete deformed solution
to the master equation starts like
\begin{equation}
\bar{S}^{(\mathrm{I.b})}=\bar{S}^{(\mathrm{I.a})}+g^{2}S_{2}^{\left( \mathrm{%
I.b}\right) }+\mathcal{O}\left( g^{3}\right) .  \label{bfny14}
\end{equation}
Accordingly, the Lagrangian action of the coupled gauge theory will contain
interactions of order $g^{2}$ and possibly of higher orders.

\subsubsection{Type II solutions}

Inserting (\ref{sol21})--(\ref{sol24}) in (\ref{sistx3}) we deduce that $%
\left( S_{1},S_{1}\right) $ becomes
\begin{equation}
\left( S_{1},S_{1}\right) =\int d^{4}x\left( 2\left( j_{c}^{\mu }\bar{f}%
_{ab}^{c}+\frac{\delta ^{R}j_{a}^{\mu }}{\delta y^{i}}\left( T_{b}\right)
_{\;j}^{i}y^{j}\right) A_{\mu }^{a}\eta ^{b}+\Lambda _{\mathrm{II}}\right) ,
\label{rez}
\end{equation}
with $\Lambda _{\mathrm{II}}$ denoting the restriction of $\Lambda $ to the
type II solutions. Again, we consider two basic situations.

\textbf{II.a.} Assume that
\begin{equation}
j_{c}^{\mu }\bar{f}_{ab}^{c}+\frac{\delta ^{R}j_{a}^{\mu }}{\delta y^{i}}%
\left( T_{b}\right) _{\;j}^{i}y^{j}=0,  \label{case1}
\end{equation}
or, which is the same, that the conserved matter currents transform under
the gauge version of the rigid symmetries (\ref{bfn40}) according to the
adjoint representation of the Lie algebra $L\left( \mathcal{G}\right) $.
Similar to the subcase \textbf{I.a}, the relation (\ref{case1}) might hold
only if the currents $j_{a}^{\mu }$ have no derivatives. This is so since if
we take $\bar{f}_{ab}^{c}=0$ in (\ref{case1}), then we arrive precisely at (%
\ref{bfny2}). In consequence, in this case we have that $\Lambda _{\mathrm{II%
}}=0$, which further leads to $\left( S_{1},S_{1}\right) =0$, so we can set $%
S_{2}=0$. Moreover, all the higher-order equations (\ref{bfn2.7}), etc., are
satisfied with the choice $S_{3}=S_{4}=\cdots =0$. Consequently, the
deformed solution to the master equation \textit{consistent to all orders in
the coupling constant} is provided by (\ref{bfn33}) where we use (\ref{ec8}%
), (\ref{sist1}) and (\ref{sol21})--(\ref{sol24}), and reduces to
\begin{eqnarray}
&&\overline{S}^{(\mathrm{II.a})}=\int d^{4}x\left( H_{\mu }^{a}D^{\mu
}\varphi _{a}+\mathcal{L}_{0}\left( \left[ y^{i}\right] \right) +gj_{a}^{\mu
}A_{\mu }^{a}\right.  \nonumber \\
&&-\frac{g}{4\cdot 4!}\varepsilon ^{\mu \nu \rho \lambda }f_{e[ab}\bar{f}%
_{cd]}^{e}A_{\mu }^{a}A_{\nu }^{b}A_{\rho }^{c}A_{\lambda }^{d}+\frac{1}{2}%
B_{a}^{\mu \nu }\bar{F}_{\mu \nu }^{a}  \nonumber \\
&&-g\varphi ^{*a}\left( \bar{f}_{ab}^{c}\varphi _{c}+F_{ab}\right) \eta
^{b}+gy_{i}^{*}\left( T_{a}\right) _{\;j}^{i}y^{j}\eta ^{a}+A_{a}^{*\mu
}\left( D_{\mu }\right) _{\;\;b}^{a}\eta ^{b}  \nonumber \\
&&+H_{a}^{*\mu }\left( 2\left( D^{\nu }\right) _{\;\;b}^{a}C_{\mu \nu }^{b}-g%
\bar{f}_{bc}^{a}\eta ^{b}H_{\mu }^{c}\right) +\frac{g}{2}\eta _{c}^{*}\bar{f}%
_{ab}^{c}\eta ^{a}\eta ^{b}  \nonumber \\
&&+B_{\mu \nu }^{*a}\left( -3\left( D_{\rho }\right) _{a}^{\;\;b}\eta
_{b}^{\mu \nu \rho }-g\bar{f}_{ab}^{c}\eta ^{b}B_{c}^{\mu \nu }+2g\left(
\bar{f}_{ab}^{c}\varphi _{c}+F_{ab}\right) C^{b\mu \nu }\right.  \nonumber \\
&&\left. +\frac{g}{8}\varepsilon ^{\mu \nu \rho \lambda }f_{e[ab}\bar{f}%
_{cd]}^{e}A_{\rho }^{b}A_{\lambda }^{c}\eta ^{d}\right) -6g\bar{f}%
_{ab}^{c}\left( H_{c}^{*\mu }B^{*a\nu \rho }C_{\mu \nu \rho }^{b}+B_{\mu \nu
}^{*a}B_{\rho \lambda }^{*b}\eta _{c}^{\mu \nu \rho \lambda }\right)
\nonumber \\
&&-\frac{3g}{2\cdot 4!}\varepsilon ^{\mu \nu \rho \lambda }f_{e[ab}\bar{f}%
_{cd]}^{e}B_{\mu \nu }^{*a}B_{\rho \lambda }^{*b}\eta ^{c}\eta
^{d}+C_{a}^{*\mu \nu }\left( -3\left( D^{\rho }\right) _{\;\;b}^{a}C_{\mu
\nu \rho }^{b}+g\bar{f}_{bc}^{a}\eta ^{b}C_{\mu \nu }^{c}\right)  \nonumber
\\
&&+\eta _{\mu \nu \rho }^{*a}\left( 4\left( D_{\lambda }\right)
_{a}^{\;\;b}\eta _{b}^{\mu \nu \rho \lambda }+g\bar{f}_{ab}^{c}\eta ^{b}\eta
_{c}^{\mu \nu \rho }-2g\left( \bar{f}_{ab}^{c}\varphi _{c}+F_{ab}\right)
C^{b\mu \nu \rho }\right.  \nonumber \\
&&\left. +\frac{g}{4!}\varepsilon ^{\mu \nu \rho \lambda }f_{e[ab}\bar{f}%
_{cd]}^{e}A_{\lambda }^{b}\eta ^{c}\eta ^{d}\right) +C_{a}^{*\mu \nu \rho
}\left( 4\left( D^{\lambda }\right) _{\;\;b}^{a}C_{\mu \nu \rho \lambda
}^{b}-g\bar{f}_{bc}^{a}\eta ^{b}C_{\mu \nu \rho }^{c}\right)  \nonumber \\
&&+g\eta _{\mu \nu \rho \lambda }^{*a}\left( -\bar{f}_{ab}^{c}\eta ^{b}\eta
_{c}^{\mu \nu \rho \lambda }+2\left( \bar{f}_{ab}^{c}\varphi
_{c}+F_{ab}\right) C^{b\mu \nu \rho \lambda }\right.  \nonumber \\
&&\left. -\frac{2}{\left( 4!\right) ^{2}}\varepsilon ^{\mu \nu \rho \lambda
}f_{e[ab}\bar{f}_{cd]}^{e}\eta ^{b}\eta ^{c}\eta ^{d}\right) +g\bar{f}%
_{ab}^{c}\left( 4\left( 3C_{c}^{*\mu \nu }B^{*a\rho \lambda }+2H_{c}^{*\mu
}\eta ^{*a\nu \rho \lambda }\right) \right.  \nonumber \\
&&\left. \left. +C_{c}^{*\mu \nu \rho \lambda }\eta ^{a}\right) C_{\mu \nu
\rho \lambda }^{b}\right) ,  \label{mastdef1}
\end{eqnarray}
where we used the notations
\begin{eqnarray}
D^{\mu }\varphi _{a} &=&\partial ^{\mu }\varphi _{a}+g\left( \bar{f}%
_{ab}^{c}\varphi _{c}+F_{ab}\right) A^{b\mu },  \label{xbfn15} \\
\overline{F}_{\mu \nu }^{a} &=&\partial _{[\mu }A_{\nu ]}^{a}+g\bar{f}%
_{bc}^{a}A_{\mu }^{b}A_{\nu }^{c},  \label{xbfn16} \\
\left( D_{\mu }\right) _{\;\;b}^{a} &=&\delta _{b}^{a}\partial _{\mu }-g\bar{%
f}_{bc}^{a}A_{\mu }^{c},  \label{xbfn17} \\
\left( D_{\mu }\right) _{a}^{\;\;b} &=&\delta _{a}^{b}\partial _{\mu }+g\bar{%
f}_{ac}^{b}A_{\mu }^{c}.  \label{xbfn18}
\end{eqnarray}
The Lagrangian formulation of the interacting model is deduced from (\ref
{mastdef1}) following the same pattern like in the previous subsubsection,
i.e. analyzing its components with definite, increasing antighost numbers.

Thus, the \textit{Lagrangian action of the interacting theory} takes the
simpler form
\begin{eqnarray}
\bar{S}_{0}^{\left( \mathrm{II.a}\right) } &=&\int d^{4}x\left( H_{\mu
}^{a}D^{\mu }\varphi _{a}+\frac{1}{2}B_{a}^{\mu \nu }\overline{F}_{\mu \nu
}^{a}-\frac{g}{4\cdot 4!}\varepsilon ^{\mu \nu \rho \lambda }f_{e[ab}\bar{f}%
_{cd]}^{e}A_{\mu }^{a}A_{\nu }^{b}A_{\rho }^{c}A_{\lambda }^{d}\right.
\nonumber \\
&&\left. +\mathcal{L}_{0}\left( \left[ y^{i}\right] \right) +gj_{a}^{\mu
}\left( y^{i}\right) A_{\mu }^{a}\right) ,  \label{bfny19}
\end{eqnarray}
and it is invariant under the \textit{deformed (generating set of) gauge
transformations}
\begin{equation}
\bar{\delta}_{\epsilon }\varphi _{a}=-g\left( \bar{f}_{ab}^{c}\varphi
_{c}+F_{ab}\right) \epsilon ^{b},\;\bar{\delta}_{\epsilon }A_{\mu
}^{a}=\left( D_{\mu }\right) _{\;\;b}^{a}\epsilon ^{b},  \label{bfny20}
\end{equation}
\begin{equation}
\bar{\delta}_{\epsilon }H_{\mu }^{a}=2\left( D^{\nu }\right)
_{\;\;b}^{a}\epsilon _{\mu \nu }^{b}-g\bar{f}_{bc}^{a}\epsilon ^{b}H_{\mu
}^{c},  \label{bfny22}
\end{equation}
\begin{eqnarray}
\bar{\delta}_{\epsilon }B_{a}^{\mu \nu } &=&-3\left( D_{\rho }\right)
_{a}^{\;\;b}\epsilon _{b}^{\mu \nu \rho }-g\bar{f}_{ab}^{c}\epsilon
^{b}B_{c}^{\mu \nu }+2g\left( \bar{f}_{ab}^{c}\varphi _{c}+F_{ab}\right)
\epsilon ^{b\mu \nu }  \nonumber \\
&&+\frac{g}{8}\varepsilon ^{\mu \nu \rho \lambda }f_{e[ab}\bar{f}%
_{cd]}^{e}A_{\rho }^{b}A_{\lambda }^{c}\epsilon ^{d},  \label{bfny21}
\end{eqnarray}
\begin{equation}
\bar{\delta}_{\epsilon }y^{i}=g\left( T_{a}\right) _{\;j}^{i}y^{j}\epsilon
^{a}.  \label{bfny23}
\end{equation}
Let us briefly comment on the physical features of the coupled model in this
situation. The first three terms from (\ref{bfny19}) describe again the
self-interactions among the BF fields. The abelian field strengths of the
one-forms $\left\{ A_{\mu }^{a}\right\} $ are now deformed into the standard
Yang-Mills form (\ref{xbfn16}), instead of the more general expression (\ref
{bfny5}) from the previous case. Also, the coupling between the BF and the
matter sector is a minimal one, of the form `vector fields times currents',
where the currents are now precisely the conserved, purely matter ones from (%
\ref{bfn39}). Along the same line, we read that the new gauge
transformations of the one-forms $\left\{ A_{\mu }^{a}\right\} $ are
standard Yang-Mills, being given by the second relation in (\ref{bfny20}),
with the corresponding covariant derivative expressed by (\ref{xbfn17}). The
matter fields are endowed, as a consequence of the type II solutions, with
the deformed gauge transformations (\ref{bfny23}), which are nothing but the
gauge version of the rigid ones (\ref{bfn40}) with the generators like in (%
\ref{ec8}) and satisfying the relation (\ref{sol24}). We would expect that
the deformed gauge algebra be precisely the Lie algebra $L\left( \mathcal{G}%
\right) $ with the structure constants $-\bar{f}_{ab}^{c}$. However, this is
not true since (\ref{mastdef1}) contains a term proportional with $%
\varepsilon ^{\mu \nu \rho \lambda }f_{e[ab}\bar{f}_{cd]}^{e}B_{\mu \nu
}^{*a}B_{\rho \lambda }^{*b}\eta ^{c}\eta ^{d}$, which shows that the
\textit{deformed gauge algebra is in fact open}, so it closes only on-shell.
(More precisely, it closes when the field equations for the two-forms hold, $%
\delta \bar{S}_{0}^{\left( \mathrm{II.a}\right) }/\delta B_{a}^{\mu \nu
}\approx 0$.) This is a relatively rare example of gauge theory with an open
gauge algebra, whose structure functions reduce entirely to the structure
constants of a Lie algebra. If we however forbid the breaking of PT
invariance, namely take $f_{eab}=0$ in (\ref{mastdef1}), then the deformed
gauge algebra indeed becomes the Lie algebra $L\left( \mathcal{G}\right) $.
As for the redundancy of the new gauge transformations, it remains of order
two, with \textit{the corresponding deformed reducibility relations closing
on-shell}.

\textbf{II.b.} In the opposite situation, where the conserved matter
currents do not transform under the gauge version of the rigid symmetries (%
\ref{bfn40}) according to the adjoint representation of the Lie algebra $%
L\left( \mathcal{G}\right) $%
\begin{equation}
j_{c}^{\mu }\bar{f}_{ab}^{c}+\frac{\delta ^{R}j_{a}^{\mu }}{\delta y^{i}}%
\left( T_{b}\right) _{\;j}^{i}y^{j}\neq 0,  \label{case2}
\end{equation}
we find ourselves in the same framework like in the subcase I.b. Thus, $%
\left( S_{1},S_{1}\right) $ is not vanishing, such that the second-order
deformation $S_{2}$ involved with the equation (\ref{bfn2.6}) will also be
so. In principle, it is possible to infer other non-trivial higher-order
deformations as solutions to the equations ((\ref{bfn2.7}), etc.). The
concrete form of these deformations depends again on the structure of the
matter theory and cannot be prescribed here. We can only write that the
complete deformed solution to the master equation begins like
\begin{equation}
\bar{S}^{(\mathrm{II.b})}=\bar{S}^{(\mathrm{II.a})}+g^{2}S_{2}^{\left(
\mathrm{II.b}\right) }+\mathcal{O}\left( g^{3}\right) ,  \label{mastdef2}
\end{equation}
such that the interacting Lagrangian action includes vertices of order $%
g^{2} $, and possibly of higher orders.

Finally, a word of caution. Once the deformations related to a given matter
theory are computed, special attention should be paid to the elimination of
non-locality, as well as of triviality of the resulting deformations. This
completes our general deformation procedure, based on local BRST cohomology.

\section{Examples}

Next, we consider two examples of matter theories --- Dirac fields and real
scalar fields --- and determine their consistent interactions with the
four-dimensional BF model under discussion in the light of the analysis
performed in the previous sections.

\subsection{Couplings for a set of Dirac fields}

First, we examine the consistent couplings with a collection of massive
Dirac fields. In view of this, we start from the Lagrangian action of the
matter fields $S_{0}^{\mathrm{matt}}\left[ y^{i}\right] $ in (\ref{bfn1}) of
the form
\begin{equation}
S_{0}^{\mathrm{matt}}\left[ \psi _{A}^{\alpha },\bar{\psi}_{\alpha
}^{A}\right] =\int d^{4}x\left( \bar{\psi}_{\alpha }^{A}\left( \mathrm{i}
\left( \gamma ^{\mu }\right) _{\;\;\beta }^{\alpha }\partial _{\mu }-m\delta
_{\beta }^{\alpha }\right) \psi _{A}^{\beta }\right) ,  \label{5.20}
\end{equation}
where $\psi _{A}^{\alpha }$ and $\bar{\psi}_{\alpha }^{A}$ ($A=\overline{%
1,I^{\prime }}$, $\alpha =1,2,3,4$) denote the spinor components of the
complex Dirac spinors $\psi _{A}$ and $\bar{\psi}^{A}$ (the bar operation
signifies spinor conjugation). The actions of $\delta $ and $\gamma $ on the
matter generators from the free BRST complex are expressed by
\begin{eqnarray}
\delta \psi _{A}^{\alpha } &=&0,\;\delta \bar{\psi}_{\alpha }^{A}=0,
\label{5.21} \\
\delta \psi _{\alpha }^{*A} &=&-\left( \mathrm{i}\left( \gamma ^{\mu
}\right) _{\;\;\alpha }^{\beta }\partial _{\mu }+m\delta _{\alpha }^{\beta
}\right) \bar{\psi}_{\beta }^{A},  \label{5.22} \\
\delta \bar{\psi}_{A}^{*\alpha } &=&-\left( \mathrm{i}\left( \gamma ^{\mu
}\right) _{\;\;\beta }^{\alpha }\partial _{\mu }-m\delta _{\beta }^{\alpha
}\right) \psi _{A}^{\beta },  \label{5.23} \\
\gamma \psi _{A}^{\alpha } &=&0,\;\gamma \bar{\psi}_{\alpha }^{A}=0,\;\gamma
\psi _{\alpha }^{*A}=0,\;\gamma \bar{\psi}_{A}^{*\alpha }=0,  \label{5.24}
\end{eqnarray}
where the antighost number one antifields $\psi _{\alpha }^{*A}$ and $\bar{%
\psi}_{A}^{*\alpha }$ are bosonic. Their actions on the variables from the
BF sector are correctly defined by the appropriate relations in (\ref{bfn13}%
)--(\ref{bfn21}). Let us consider the rigid symmetries
\begin{equation}
\Delta _{\xi }\bar{\psi}_{\alpha }^{A}=\bar{\psi}_{\alpha }^{B}\left(
T_{a}\right) _{\;B}^{A}\xi ^{a},\;\Delta _{\xi }\psi _{A}^{\alpha }=-\left(
T_{a}\right) _{\;A}^{B}\psi _{B}^{\alpha }\xi ^{a},  \label{ex4}
\end{equation}
of the action (\ref{5.20}), such that the corresponding conserved currents
read as
\begin{equation}
j_{a}^{\mu }=\mathrm{i}\bar{\psi}_{\alpha }^{A}\left( \gamma ^{\mu }\right)
_{\;\beta }^{\alpha }\left( T_{a}\right) _{\;A}^{B}\psi _{B}^{\beta }.
\label{ex5}
\end{equation}
From (\ref{ex5}) we find that
\begin{equation}
\frac{\delta ^{R}j_{a}^{\mu }}{\delta y^{i}}\left( T_{b}\right)
_{\;j}^{i}y^{j}=\frac{\delta ^{R}j_{a}^{\mu }}{\delta \bar{\psi}_{\alpha
}^{A}}\left( T_{b}\right) _{\;B}^{A}\bar{\psi}_{\alpha }^{B}-\frac{\delta
^{R}j_{a}^{\mu }}{\delta \psi _{A}^{\alpha }}\left( T_{b}\right)
_{\;A}^{B}\psi _{\;B}^{\alpha },  \label{5.31a}
\end{equation}
and hence
\begin{equation}
\frac{\delta ^{R}j_{a}^{\mu }}{\delta \bar{\psi}_{\alpha }^{A}}\left(
T_{b}\right) _{\;B}^{A}\bar{\psi}_{\alpha }^{B}-\frac{\delta ^{R}j_{a}^{\mu }%
}{\delta \psi _{A}^{\alpha }}\left( T_{b}\right) _{\;A}^{B}\psi
_{\;B}^{\alpha }=\mathrm{i}\bar{\psi}_{\alpha }^{A}\left( \gamma ^{\mu
}\right) _{\;\beta }^{\alpha }\left[ T_{a},T_{b}\right] _{\;A}^{B}\psi
_{B}^{\beta }.  \label{bfnyw}
\end{equation}

In the case of type I solutions (see (\ref{sol14})), the relation (\ref
{bfnyw}) becomes
\begin{equation}
\frac{\delta ^{R}j_{a}^{\mu }}{\delta \bar{\psi}_{\alpha }^{A}}\left(
T_{b}\right) _{\;B}^{A}\bar{\psi}_{\alpha }^{B}-\frac{\delta ^{R}j_{a}^{\mu }%
}{\delta \psi _{A}^{\alpha }}\left( T_{b}\right) _{\;A}^{B}\psi
_{\;B}^{\alpha }=0,  \label{bfny24}
\end{equation}
such that we are under the conditions of the subcase \textbf{I.a}.
Consequently, (\ref{xyz}) takes now the form
\begin{eqnarray}
\bar{a}_{1} &=&\left( \bar{\psi}_{\alpha }^{A}\bar{\psi}_{B}^{*\alpha }-\psi
_{\alpha }^{*A}\psi _{B}^{\alpha }\right) \left( T_{a}\right)
_{\;A}^{B}k^{ac}W_{cb}\eta ^{b}  \nonumber \\
&&+\mathrm{i}H_{m}^{*\mu }k^{ac}\frac{\partial W_{cb}}{\partial \varphi _{m}}%
\bar{\psi}_{\alpha }^{A}\left( \gamma _{\mu }\right) _{\;\beta }^{\alpha
}\left( T_{a}\right) _{\;A}^{B}\psi _{B}^{\beta }\eta ^{b},  \label{ex6}
\end{eqnarray}
while $\bar{a}_{0}$ (the solution to the equation (\ref{bfnbar0})) can be
written as
\begin{equation}
\bar{a}_{0}=\mathrm{i}\bar{\psi}_{\alpha }^{A}\left( \gamma ^{\mu }\right)
_{\;\beta }^{\alpha }\left( T_{a}\right) _{\;A}^{B}\psi _{B}^{\beta
}k^{ac}W_{cb}A_{\mu }^{b}.  \label{ex7}
\end{equation}
According to the general theory, the consistency of the first-order
deformation leads to no higher-order deformations in this situation. The
deformed Lagrangian action and accompanying gauge transformations are given
by (\ref{bfny8}) and respectively (\ref{bfny9})--(\ref{bfny12}) where we set
$y^{i}\rightarrow \psi _{A}^{\alpha },\bar{\psi}_{\alpha }^{A}$ and use (\ref
{ex4})--(\ref{ex5}).

For the type II solutions (see (\ref{sol21})--(\ref{sol24})) it follows that
the relation (\ref{bfnyw}) leads to
\begin{equation}
\frac{\delta ^{R}j_{a}^{\mu }}{\delta \bar{\psi}_{\alpha }^{A}}\left(
T_{b}\right) _{\;B}^{A}\bar{\psi}_{\alpha }^{B}-\frac{\delta ^{R}j_{a}^{\mu }%
}{\delta \psi _{A}^{\alpha }}\left( T_{b}\right) _{\;A}^{B}\psi
_{\;B}^{\alpha }=-j_{c}^{\mu }\bar{f}_{ab}^{c},  \label{bfny25}
\end{equation}
so we are in the subcase \textbf{II.a}. Furthermore, we have that
\begin{eqnarray}
\bar{a}_{1} &=&\left( \bar{\psi}_{\alpha }^{A}\bar{\psi}_{B}^{*\alpha }-\psi
_{\alpha }^{*A}\psi _{B}^{\alpha }\right) \left( T_{a}\right) _{\;A}^{B}\eta
^{a},  \label{bfny26} \\
\bar{a}_{0} &=&\mathrm{i}\bar{\psi}_{\alpha }^{A}\left( \gamma ^{\mu
}\right) _{\;\beta }^{\alpha }\left( T_{a}\right) _{\;A}^{B}\psi _{B}^{\beta
}A_{\mu }^{a}.  \label{bfny27}
\end{eqnarray}
The consistency of the first-order deformation again produces no
higher-order deformations. Similarly, the deformed action and its gauge
transformations follow from the relations (\ref{bfny19}) and (\ref{bfny20}%
)--(\ref{bfny23}) where we replace $y^{i}$ by $\psi _{A}^{\alpha }$ and $%
\bar{\psi}_{\alpha }^{A}$, and take into account the relations (\ref{ex4})--(%
\ref{ex5}).

\subsection{Couplings for a collection of real scalar fields}

Second, we analyze the case where the role of the matter fields is played by
a collection of real scalar fields, $\left\{ \phi ^{A}\right\} _{A=\overline{%
1,I}}$. In this situation the Lagrangian action of the matter fields from (%
\ref{bfn1}) is
\begin{equation}
S_{0}^{\mathrm{matt}}\left[ \phi ^{A}\right] =\int d^{4}x\left( \frac{1}{2}%
K_{AB}\left( \partial _{\mu }\phi ^{A}\right) \left( \partial ^{\mu }\phi
^{B}\right) -V\left( \phi ^{A}\right) \right) ,  \label{5.1}
\end{equation}
where $K_{AB}$ is an invertible, symmetric, constant matrix. We assume that
the matter action (\ref{5.1}) is invariant under the bosonic rigid
symmetries
\begin{equation}
\Delta _{\xi }\phi ^{A}=-\left( T_{a}\right) _{\;B}^{A}\phi ^{B}\xi ^{a}.
\label{exp6}
\end{equation}
This is true if the constant matrices $T_{a}$, of elements $\left(
T_{a}\right) _{\;B}^{A}$, are such that the following relations are
satisfied
\begin{equation}
\frac{\partial V}{\partial \phi ^{A}}\left( T_{a}\right) _{\;B}^{A}\phi
^{B}=0,  \label{exp3}
\end{equation}
\begin{equation}
\left( \tilde{T}_{a}\right) _{AB}=-\left( \tilde{T}_{a}\right) _{BA},
\label{exp4}
\end{equation}
where
\begin{equation}
\left( \tilde{T}_{a}\right) _{AB}=K_{AE}\left( T_{a}\right) _{\;B}^{E}.
\label{exp5}
\end{equation}
Assuming that such matrices exist, it follows that the conserved currents
associated with the rigid symmetries (\ref{exp6}) read as
\begin{equation}
j_{a}^{\mu }=\left( \partial ^{\mu }\phi ^{A}\right) \phi ^{B}\left( \tilde{T%
}_{a}\right) _{AB}.  \label{exp7}
\end{equation}
For the model under consideration we then find that
\begin{equation}
\frac{\delta ^{R}j_{a}^{\mu }}{\delta y^{i}}\left( T_{b}\right)
_{\;j}^{i}y^{j}=-\frac{\delta ^{R}j_{a}^{\mu }}{\delta \phi ^{A}}\left(
T_{b}\right) _{\;B}^{A}\phi ^{B}.  \label{xp8}
\end{equation}
With the help of the expression (\ref{exp7}) we deduce that
\begin{eqnarray}
-\frac{\delta ^{R}j_{a}^{\mu }}{\delta \phi ^{A}}\left( T_{b}\right)
_{\;B}^{A}\phi ^{B} &=&-\left[ T_{a},T_{b}\right] _{\;B}^{C}K_{AC}\left(
\partial ^{\mu }\phi ^{A}\right) \phi ^{B}  \nonumber \\
&&-\partial ^{\mu }\left( \frac{1}{2}\left\{ T_{a},T_{b}\right\}
_{\;B}^{C}K_{AC}\phi ^{A}\phi ^{B}\right) ,  \label{xp9}
\end{eqnarray}
where $\left\{ T_{a},T_{b}\right\} =T_{a}T_{b}+T_{b}T_{a}$.

For the type I solutions (see (\ref{sol14})) the relation (\ref{xp9})
becomes
\begin{equation}
-\frac{\delta ^{R}j_{a}^{\mu }}{\delta \phi ^{A}}\left( T_{b}\right)
_{\;B}^{A}\phi ^{B}=-\partial ^{\mu }\left( \frac{1}{2}\left\{
T_{a},T_{b}\right\} _{\;B}^{C}K_{AC}\phi ^{A}\phi ^{B}\right) \neq 0,
\label{xp10}
\end{equation}
so we are in the subcase \textbf{I.b}. In this context we have that
\begin{equation}
\bar{a}_{1}=k^{ac}\left( -\phi _{A}^{*}\left( T_{a}\right) _{\;B}^{A}\phi
^{B}W_{cb}+H_{m}^{*\mu }\frac{\partial W_{cb}}{\partial \varphi _{m}}\left(
\partial _{\mu }\phi ^{A}\right) \left( \tilde{T}_{a}\right) _{AB}\phi
^{B}\right) \eta ^{b}  \label{xp11}
\end{equation}
and respectively
\begin{equation}
\bar{a}_{0}=k^{ac}\left( \partial ^{\mu }\phi ^{A}\right) \left( \tilde{T}%
_{a}\right) _{AB}\phi ^{B}W_{cb}A_{\mu }^{b}.  \label{xp12}
\end{equation}
The consistency of the first-order deformation of the solution to the master
equation leads to a non-trivial deformation at the second order, of the form
\begin{eqnarray}
S_{2}^{\left( \mathrm{I.b}\right) } &=&-\frac{1}{4}k^{ap}k^{de}K_{AC}\left\{
T_{a},T_{d}\right\} _{\;B}^{C}\int d^{4}x\,\phi ^{A}\phi ^{B}\times
\nonumber \\
&&\times \left( W_{pb}A^{b\mu }+H_{m}^{*\mu }\eta ^{b}\frac{\partial W_{pb}}{%
\partial \varphi _{m}}\right) \left( W_{ec}A_{\mu }^{c}+H_{n\mu }^{*}\eta
^{c}\frac{\partial W_{ec}}{\partial \varphi _{n}}\right) .  \label{xp13}
\end{eqnarray}
On behalf of (\ref{bfny3}) adapted to our model and of (\ref{xp13}) we
obtain that $S_{3}^{\left( \mathrm{I.b}\right) }=0$ and also $S_{4}^{\left(
\mathrm{I.b}\right) }=S_{5}^{\left( \mathrm{I.b}\right) }=\cdots =0$. In
this situation we get that the full deformed Lagrangian action is a
polynomial of order two in the coupling constant
\begin{equation}
\bar{S}_{0}^{\left( \mathrm{I.b}\right) }=\bar{S}_{0}^{\left( \mathrm{BFI}%
\right) }+\int d^{4}x\left[ \frac{1}{2}K_{AB}\left( \left( \hat{D}_{\mu
}\right) _{\;C}^{A}\phi ^{C}\right) \left( \left( \hat{D}^{\mu }\right)
_{\;D}^{B}\phi ^{D}\right) -V\left( \phi ^{A}\right) \right] ,  \label{scIb}
\end{equation}
where
\begin{equation}
\left( \hat{D}_{\mu }\right) _{\;C}^{A}=\delta _{C}^{A}\partial _{\mu
}+g\left( T_{a}\right) _{\;C}^{A}k^{ab}W_{bc}A_{\mu }^{c}  \label{covderIb}
\end{equation}
and $\bar{S}_{0}^{\left( \mathrm{BFI}\right) }$ denotes the action that
describes the self-interactions among the BF fields for the type I
solutions, and reduces to the first three terms from the right-hand side of (%
\ref{bfny8}). The deformed gauge transformations of (\ref{scIb}) are like in
(\ref{bfny9}) and (\ref{bfny11})--(\ref{bfny12}) where we set $y^{i}=\phi
^{A}$ and $\left( T_{a}\right) _{\;j}^{i}=-\left( T_{a}\right) _{\;B}^{A}$,
while the gauge transformations of the one-forms $\left\{ H_{\mu
}^{a}\right\} $ are enriched with terms of order two in the coupling
constant
\begin{equation}
\hat{\delta}_{\epsilon }^{\prime }H_{\mu }^{a}=\hat{\delta}_{\epsilon
}H_{\mu }^{a}+g^{2}K_{AB}\left( T_{b}\right) _{\;C}^{A}\phi ^{C}\left(
T_{e}\right) _{\;D}^{B}\phi ^{D}k^{ef}W_{fg}A_{\mu }^{g}k^{bc}\frac{\partial
W_{cd}}{\partial \varphi _{a}}\epsilon ^{d}.  \label{trhIb}
\end{equation}
In the above $\hat{\delta}_{\epsilon }H_{\mu }^{a}$ can be found in (\ref
{bfny10}) and $j_{a}^{\mu }$ must be taken like in (\ref{exp7}). The
commutators among the deformed gauge transformations are also modified with
terms of order two in the coupling constant, but the reducibility relations
stop at order one in the coupling constant, and hence take the same form
like in the subcase I.a.

For the type II solutions (see (\ref{sol21})--(\ref{sol24})) the relation (%
\ref{xp9}) gives
\begin{equation}
-\frac{\delta ^{R}j_{a}^{\mu }}{\delta \phi ^{A}}\left( T_{b}\right)
_{\;B}^{A}\phi ^{B}+\bar{f}_{ab}^{c}j_{c}^{\mu }=-\partial ^{\mu }\left(
\frac{1}{2}\left\{ T_{a},T_{b}\right\} _{\;B}^{C}K_{AC}\phi ^{A}\phi
^{B}\right) ,  \label{xp14}
\end{equation}
so we are in the subcase \textbf{II.b}. Consequently, we have that
\begin{equation}
\bar{a}_{1}=-\phi _{A}^{*}\left( T_{a}\right) _{\;B}^{A}\phi ^{B}\eta ^{b}
\label{xp15}
\end{equation}
and respectively
\begin{equation}
\bar{a}_{0}=\left( \partial ^{\mu }\phi ^{A}\right) \left( \tilde{T}%
_{a}\right) _{AB}\phi ^{B}A_{\mu }^{a}.  \label{xp16}
\end{equation}
The consistency of the first-order deformation leads to the second-order
deformation like
\begin{equation}
S_{2}^{\left( \mathrm{II.b}\right) }=-\frac{1}{4}K_{AC}\left\{
T_{a},T_{d}\right\} _{\;B}^{C}\int d^{4}x\phi ^{A}\phi ^{B}A^{a\mu }A_{\mu
}^{d}.  \label{xp17}
\end{equation}
Using (\ref{mastdef1}) adapted to the present model and (\ref{xp17}) we get
that$\;S_{3}^{\left( \mathrm{II.b}\right) }=0$ and, furthermore, $%
S_{4}^{\left( \mathrm{II.b}\right) }=S_{5}^{\left( \mathrm{II.b}\right)
}=\cdots =0$. Consequently, the complete Lagrangian action of the coupled
model becomes
\begin{equation}
\bar{S}_{0}^{\left( \mathrm{II.b}\right) }=\bar{S}_{0}^{\left( \mathrm{BFII}%
\right) }+\int d^{4}x\left[ \frac{1}{2}K_{AB}\left( \left( \bar{D}_{\mu
}\right) _{\;C}^{A}\phi ^{C}\right) \left( \left( \bar{D}^{\mu }\right)
_{\;D}^{B}\phi ^{D}\right) -V\left( \phi ^{A}\right) \right] ,  \label{xp30}
\end{equation}
where
\begin{equation}
\left( \bar{D}_{\mu }\right) _{\;C}^{A}=\delta _{C}^{A}\partial _{\mu
}+g\left( T_{a}\right) _{\;C}^{A}A_{\mu }^{a}  \label{xp31}
\end{equation}
and $\bar{S}_{0}^{\left( \mathrm{BFII}\right) }$ means the action
responsible for the self-interactions among the BF fields for the type II
solutions, being represented by the first three terms from the right-hand
side of (\ref{bfny19}). It is again a polynomial of order two in the
deformation parameter. In this situation the gauge transformations of the
action (\ref{xp30}) gain no new components, of order two or higher in the
coupling constant, and are expressed like in (\ref{bfny20})--(\ref{bfny23})
where we set $y^{i}=\phi ^{A}$ and $\left( T_{a}\right) _{\;j}^{i}=-\left(
T_{a}\right) _{\;B}^{A}$. Consequently, the gauge algebra and the
reducibility relations are the same like in the subcase II.a.

\section{Conclusion}

The main result of this paper is that we can indeed add consistent
Lagrangian interactions to a ``free'' theory describing a
collection of BF-like models and a matter theory in four
dimensions. Our treatment is based on the deformation of the
solution to the master equation. The first-order deformation is
computed by means of the local BRST cohomology in ghost number
zero. Its existence is due to the hypothesis that the matter
theory is invariant under some (non-trivial) bosonic global
symmetries, which produce some (non-trivially) conserved currents
$j_{a}^{\mu }$. The consistency of the first-order deformation
restricts the commutators of the constant matrices $T_{a}$ that
enter the global matter symmetries to either vanish (type I
solutions) or close according to a Lie algebra (type II
solutions). The deformation procedure stops at order one if the
matter currents $j_{a}^{\mu }$ include no derivatives and if they
either remain invariant under the gauge version of the rigid
symmetries in the first case or transform under the gauge version
according to the adjoint representation of $L\left(
\mathcal{G}\right) $ in the second case. Otherwise, there appear
deformations of order $g^{2}$ and possibly of higher orders.

The common features of the interacting gauge models resulting from the two
types of solutions at order $g$ are: the matter fields are primarily coupled
to the vector fields $A_{\mu }^{a}$; all the fields (BF and matter) gain
deformed gauge transformations; the gauge algebra of the deformed gauge
transformations closes on-shell (in spite of the Abelian and respectively
Lie character of the matrices $T_{a}$), in contrast to the ``free'', Abelian
one; the reducibility relations hold on-shell, i.e. on the stationary
surface of deformed field equations, unlike the initial ones, that held
off-shell. The main differences between the two cases are revealed by the
couplings of the matter fields to the BF sector and by the expressions of
the gauge transformations. Indeed, for the type I solutions we find a
generalized minimal coupling in the sense that even if it is formally
expressed like `vector fields $A_{\mu }^{a}$ times currents', however the
currents are not the conserved matter currents $j_{a}^{\mu }$, but $%
j_{a}^{\mu }$ in some `background' potentials of scalar fields proportional
with $W_{ab}$, while for type II solutions we recover a genuine minimal
coupling. The same observation holds for the gauge transformations of the
matter fields: in the former case they can be viewed as being obtained by
gauging the rigid matter symmetries and by further putting them in the same
`background' potentials of scalar fields, while in the latter they reduce to
gauging the rigid symmetries only. The deformed field strengths and gauge
transformations of the one-forms $A_{\mu }^{a}$ inherit a similar behaviour:
they are generalized Yang-Mills-like for type I solutions, with the usual
structure constants replaced by the derivatives of the `background'
potentials, and standard Yang-Mills corresponding to the Lie algebra $%
L\left( \mathcal{G}\right) $ for type II solutions. Finally, we note that
the exemplification of our results in the case of a set of Dirac fields
leads to no deformations of order two or higher for either type I or II
solutions, while for a system of real scalar fields we obtain second-order
deformations for both solutions.

\section*{Acknowledgment}

The authors are partially supported from the type A grant 304/2004
with the Romanian National Council for Academic Scientific
Research (C.N.C.S.I.S.) and the Romanian Ministry of Education and
Research (M.E.C.), and also by the European Commission FP6 program
MRTN-CT-2004-005104.

\appendix

\section{Notations used in the subsection 4.2 \label{notat}}

The various notations used within the formula (\ref{bfn36}) are listed
below:
\begin{eqnarray}
K^{abc} &=&\eta ^{a}\eta ^{b}\varphi ^{*c}+2\eta ^{a}A^{b\mu }H_{\mu
}^{c}+2\left( A^{a\mu }A^{b\nu }-2B^{*a\mu \nu }\eta ^{b}\right) C_{\mu \nu
}^{c}  \nonumber \\
&&+4\left( \eta ^{a}\eta ^{*b\mu \nu \rho }+3B^{*a\mu \nu }A^{b\rho }\right)
C_{\mu \nu \rho }^{c}  \nonumber \\
&&-4\left( \eta ^{a}\eta ^{*b\mu \nu \rho \lambda }+6B^{*a\mu \nu }B^{*b\rho
\lambda }-4\eta ^{*a\mu \nu \rho }A^{b\lambda }\right) C_{\mu \nu \rho
\lambda }^{c},  \label{xbfn39}
\end{eqnarray}
\begin{eqnarray}
K_{d}^{abc} &=&\left( 4H_{d}^{*\nu }A^{a\mu }\eta ^{b}-C_{d}^{*\mu \nu }\eta
^{a}\eta ^{b}\right) C_{\mu \nu }^{c}-H_{d}^{*\mu }\eta ^{a}\eta ^{b}H_{\mu
}^{c}  \nonumber \\
&&+\left( 6H_{d}^{*\rho }A^{a\mu }A^{b\nu }-12H_{d}^{*\rho }B^{*a\mu \nu
}\eta ^{b}+6C_{d}^{*\mu \nu }\eta ^{a}A^{b\rho }\right.  \nonumber \\
&&\left. -C_{d}^{*\mu \nu \rho }\eta ^{a}\eta ^{b}\right) C_{\mu \nu \rho
}^{c}+\left( -48H_{d}^{*\lambda }B^{*a\mu \nu }A^{b\rho }\right.  \nonumber
\\
&&+12C_{d}^{*\mu \nu }A^{a\rho }A^{b\lambda }+16H_{d}^{*\lambda }\eta
^{*a\mu \nu \rho }\eta ^{b}-24C_{d}^{*\mu \nu }B^{*a\rho \lambda }\eta ^{b}
\nonumber \\
&&\left. -8C_{d}^{*\mu \nu \rho }A^{a\lambda }\eta ^{b}-C_{d}^{*\mu \nu \rho
\lambda }\eta ^{a}\eta ^{b}\right) C_{\mu \nu \rho \lambda }^{c},
\label{xbfn40}
\end{eqnarray}
\begin{eqnarray}
K_{de}^{abc} &=&-3\left( C_{d}^{*\mu \nu }H_{e}^{*\rho }\eta
^{a}+2H_{d}^{*\mu }H_{e}^{*\nu }A^{a\rho }\right) \eta ^{b}C_{\mu \nu \rho
}^{c}  \nonumber \\
&&-H_{d}^{*\mu }H_{e}^{*\nu }\eta ^{a}\eta ^{b}C_{\mu \nu }^{c}+\left(
-24H_{d}^{*\mu }H_{e}^{*\nu }B^{*a\rho \lambda }\eta ^{b}\right.  \nonumber
\\
&&+12H_{d}^{*\mu }H_{e}^{*\nu }A^{a\rho }A^{b\lambda }-24C_{d}^{*\mu \nu
}H_{e}^{*\rho }A^{a\lambda }\eta ^{b}  \nonumber \\
&&\left. -3C_{d}^{*\mu \nu }C_{e}^{*\rho \lambda }\eta ^{a}\eta
^{b}+4C_{d}^{*\mu \nu \rho }H_{e}^{*\lambda }\eta ^{a}\eta ^{b}\right)
C_{\mu \nu \rho \lambda }^{c},  \label{xbfn41}
\end{eqnarray}
\begin{eqnarray}
K_{def}^{abc} &=&-2\left( 4H_{d}^{*\mu }H_{e}^{*\nu }H_{f}^{*\rho
}A^{a\lambda }+3C_{d}^{*\mu \nu }H_{e}^{*\rho }H_{f}^{*\lambda }\eta
^{a}\right) \eta ^{b}C_{\mu \nu \rho \lambda }^{c}  \nonumber \\
&&-H_{d}^{*\mu }H_{e}^{*\nu }H_{f}^{*\rho }\eta ^{a}\eta ^{b}C_{\mu \nu \rho
}^{c},  \label{xbfn42}
\end{eqnarray}
\begin{equation}
K_{defg}^{abc}=-H_{d}^{*\mu }H_{e}^{*\nu }H_{f}^{*\rho }H_{g}^{*\lambda
}\eta ^{a}\eta ^{b}C_{\mu \nu \rho \lambda }^{c},  \label{xbfn43}
\end{equation}
\begin{eqnarray}
U_{d}^{abc} &=&\left( -2\eta ^{a}A_{\mu }^{b}A_{\nu }^{c}+B_{\mu \nu
}^{*a}\eta ^{b}\eta ^{c}\right) B_{d}^{\mu \nu }-A_{\mu }^{a}\eta ^{b}\eta
^{c}A_{d}^{*\mu }  \nonumber \\
&&+\left( -A_{\mu }^{a}A_{\nu }^{b}A_{\rho }^{c}+6\eta ^{a}B_{\mu \nu
}^{*b}A_{\rho }^{c}+\eta ^{b}\eta ^{c}\eta _{\mu \nu \rho }^{*a}\right) \eta
_{d}^{\mu \nu \rho }  \nonumber \\
&&-\frac{1}{3}\eta ^{a}\eta ^{b}\eta ^{c}\eta _{d}^{*}+\left( -12A_{\mu
}^{a}A_{\nu }^{b}B_{\rho \lambda }^{*c}+12\eta ^{a}B_{\mu \nu }^{*b}B_{\rho
\lambda }^{*c}\right.  \nonumber \\
&&\left. -8\eta ^{a}\eta _{\mu \nu \rho }^{*b}A_{\lambda }^{c}+\eta _{\mu
\nu \rho \lambda }^{*c}\eta ^{a}\eta ^{b}\right) \eta _{d}^{\mu \nu \rho
\lambda },  \label{xbfn44}
\end{eqnarray}
\begin{eqnarray}
U_{d,e}^{abc} &=&\left( H_{e}^{*\mu }A^{a\nu }\eta ^{b}\eta ^{c}+\frac{1}{6}%
C_{e}^{*\mu \nu }\eta ^{a}\eta ^{b}\eta ^{c}\right) B_{d\mu \nu }  \nonumber
\\
&&-\frac{1}{3}H_{e}^{*\mu }\eta ^{a}\eta ^{b}\eta ^{c}A_{d\mu }^{*}+\left(
-3H_{e}^{*\rho }\eta ^{a}A^{b\mu }A^{c\nu }\right.  \nonumber \\
&&-3H_{e}^{*\rho }\eta ^{a}\eta ^{b}B^{*c\mu \nu }+\frac{3}{2}C_{e}^{*\mu
\nu }\eta ^{a}\eta ^{b}A^{c\rho }  \nonumber \\
&&\left. +\frac{1}{6}C_{e}^{*\mu \nu \rho }\eta ^{a}\eta ^{b}\eta
^{c}\right) \eta _{d\mu \nu \rho }+\left( 24A^{a\mu }H_{e}^{*\nu }\eta
^{b}B^{*c\rho \lambda }\right.  \nonumber \\
&&+4H_{e}^{*\lambda }A^{a\mu }A^{b\nu }A^{c\rho }-4H_{e}^{*\lambda }\eta
^{a}\eta ^{b}\eta ^{*c\mu \nu \rho }  \nonumber \\
&&+6C_{e}^{*\mu \nu }\eta ^{a}\eta ^{b}B^{*c\rho \lambda }-6C_{e}^{*\mu \nu
}\eta ^{a}A^{b\rho }A^{c\lambda }  \nonumber \\
&&\left. +8C_{e}^{*\mu \nu \rho }\eta ^{a}\eta ^{b}A^{c\lambda }+\frac{1}{6}%
C_{e}^{*\mu \nu \rho \lambda }\eta ^{a}\eta ^{b}\eta ^{c}\right) \eta _{d\mu
\nu \rho \lambda },  \label{xbfn45}
\end{eqnarray}
\begin{eqnarray}
U_{d,ef}^{abc} &=&\frac{1}{6}H_{e}^{*\mu }H_{f}^{*\nu }\eta ^{a}\eta
^{b}\eta ^{c}B_{d\mu \nu }+\frac{3}{2}H_{e}^{*\mu }H_{f}^{*\nu }\eta
^{a}\eta ^{b}A^{c\rho }\eta _{d\mu \nu \rho }  \nonumber \\
&&+\frac{1}{2}C_{e}^{*\mu \nu }H_{f}^{*\rho }\eta ^{a}\eta ^{b}\eta ^{c}\eta
_{d\mu \nu \rho }+\left( 6H_{e}^{*\mu }H_{f}^{*\nu }\eta ^{a}\eta
^{b}B^{*c\rho \lambda }\right.  \nonumber \\
&&-6H_{e}^{*\mu }H_{f}^{*\nu }\eta ^{a}A^{b\rho }A^{c\lambda }+6C_{e}^{*\mu
\nu }H_{f}^{*\rho }\eta ^{a}\eta ^{b}A^{c\lambda }  \nonumber \\
&&\left. +\frac{2}{3}C_{e}^{*\mu \nu \rho }H_{f}^{*\lambda }\eta ^{a}\eta
^{b}\eta ^{c}+\frac{1}{2}C_{e}^{*\mu \nu }C_{f}^{*\rho \lambda }\eta
^{a}\eta ^{b}\eta ^{c}\right) \eta _{d\mu \nu \rho \lambda },  \label{xbfn46}
\end{eqnarray}
\begin{eqnarray}
U_{d,efg}^{abc} &=&\left( 2H_{e}^{*\mu }H_{f}^{*\nu }H_{g}^{*\rho }\eta
^{a}\eta ^{b}A^{c\lambda }+C_{e}^{*\mu \nu }H_{f}^{*\rho }H_{g}^{*\lambda
}\eta ^{a}\eta ^{b}\eta ^{c}\right) \eta _{d\mu \nu \rho \lambda }  \nonumber
\\
&&+\frac{1}{6}H_{e}^{*\mu }H_{f}^{*\nu }H_{g}^{*\rho }\eta ^{a}\eta ^{b}\eta
^{c}\eta _{d\mu \nu \rho },  \label{xbfn47}
\end{eqnarray}
\begin{equation}
U_{d,efgh}^{abc}=\frac{1}{6}H_{e}^{*\mu }H_{f}^{*\nu }H_{g}^{*\rho
}H_{h}^{*\lambda }\eta ^{a}\eta ^{b}\eta ^{c}\eta _{d\mu \nu \rho \lambda },
\label{xbfn48}
\end{equation}
\begin{eqnarray}
K^{abcdf} &=&\frac{1}{8}\varepsilon ^{\mu \nu \rho \lambda }\left[ \left(
\frac{1}{3!}A_{\mu }^{a}A_{\nu }^{b}-B_{\mu \nu }^{*a}\eta ^{b}\right)
A_{\nu }^{c}A_{\rho }^{d}\right. +\frac{1}{3}\left( B_{\mu \nu }^{*a}B_{\rho
\lambda }^{*b}\right.  \nonumber \\
&&\left. \left. -\frac{2}{3}\eta _{\mu \nu \rho }^{*a}A_{\lambda }^{b}+\frac{%
1}{4!}\eta _{\mu \nu \rho \lambda }^{*a}\eta ^{b}\right) \eta ^{c}\eta
^{d}\right] \eta ^{f},  \label{bfk1}
\end{eqnarray}
\begin{eqnarray}
K_{e}^{abcdf} &=&\frac{1}{4!}\varepsilon ^{\mu \nu \rho \lambda }\left[
\frac{1}{2}\left( \frac{1}{5!}C_{e\mu \nu \rho \lambda }^{*}\eta ^{a}+\frac{1%
}{3!}C_{e\mu \nu \rho }^{*}A_{\lambda }^{a}+\frac{1}{2}C_{e\mu \nu
}^{*}B_{\rho \lambda }^{*a}\right. \right.  \nonumber \\
&&\left. +\frac{1}{3}H_{e\mu }^{*}\eta _{\nu \rho \lambda }^{*a}\right) \eta
^{b}\eta ^{c}-H_{e\mu }^{*}\left( A_{\nu }^{a}A_{\rho }^{b}-2B_{\nu \rho
}^{*a}\eta ^{b}\right) A_{\lambda }^{c}  \nonumber \\
&&\left. -\frac{1}{2}C_{e\mu \nu }^{*}A_{\rho }^{a}A_{\lambda }^{b}\eta
^{c}\right] \eta ^{d}\eta ^{f},  \label{bfk2}
\end{eqnarray}
\begin{eqnarray}
K_{eg}^{abcdf} &=&\frac{1}{2\cdot 4!}\varepsilon ^{\mu \nu \rho \lambda
}\left[ \frac{1}{2}\left( \frac{1}{15}H_{e\mu }^{*}C_{g\nu \rho \lambda
}^{*}\eta ^{a}+\frac{1}{20}C_{e\mu \nu }^{*}C_{g\rho \lambda }^{*}\eta
^{a}\right. \right.  \nonumber \\
&&\left. \left. +H_{e\mu }^{*}C_{g\nu \rho }^{*}A_{\lambda }^{a}\right) \eta
^{b}-H_{e\mu }^{*}H_{g\nu }^{*}\left( A_{\rho }^{a}A_{\lambda }^{b}-2B_{\rho
\lambda }^{*a}\eta ^{b}\right) \right] \eta ^{c}\eta ^{d}\eta ^{f},
\label{bfk3}
\end{eqnarray}
\begin{equation}
K_{egh}^{abcdf}=\frac{1}{4\cdot 4!}\varepsilon ^{\mu \nu \rho \lambda
}H_{e\mu }^{*}H_{g\nu }^{*}\left( \frac{1}{10}C_{h\rho \lambda }^{*}\eta
^{a}+\frac{1}{3}H_{h\rho }^{*}A_{\lambda }^{a}\right) \eta ^{b}\eta ^{c}\eta
^{d}\eta ^{f},  \label{bfk4}
\end{equation}
\begin{equation}
K_{eghl}^{abcdf}=\frac{1}{2\cdot 4!\cdot 5!}\varepsilon ^{\mu \nu \rho
\lambda }H_{e\mu }^{*}H_{g\nu }^{*}H_{h\rho }^{*}H_{l\lambda }^{*}\eta
^{a}\eta ^{b}\eta ^{c}\eta ^{d}\eta ^{f},  \label{bfk5}
\end{equation}
\begin{eqnarray}
K_{b}^{a} &=&4\varepsilon ^{\mu \nu \rho \lambda }\left[ 2\left( -C_{\mu \nu
\rho \lambda }^{a}\eta _{b}^{*}+C_{\mu \nu \rho }^{a}A_{b\lambda
}^{*}\right) +C_{\mu \nu }^{a}B_{b\rho \lambda }\right.  \nonumber \\
&&\left. -\left( \varphi ^{*a}\eta _{b\mu \nu \rho \lambda }-H_{\mu
}^{a}\eta _{b\nu \rho \lambda }\right) \right] ,  \label{bfk7}
\end{eqnarray}
\begin{eqnarray}
K_{b,c}^{a} &=&4\varepsilon ^{\mu \nu \rho \lambda }\left[ \eta _{b\mu \nu
\rho \lambda }\left( C_{\sigma \tau \kappa \varsigma }^{a}C_{c}^{*\sigma
\tau \kappa \varsigma }+C_{\sigma \tau \kappa }^{a}C_{c}^{*\sigma \tau
\kappa }+C_{\sigma \tau }^{a}C_{c}^{*\sigma \tau }\right. \right.  \nonumber
\\
&&\left. +H_{\sigma }^{a}H_{c}^{*\sigma }\right) +C_{\mu \nu \rho \lambda
}^{a}\left( \eta _{b\sigma \tau \kappa }C_{c}^{*\sigma \tau \kappa
}+B_{b\sigma \tau }C_{c}^{*\sigma \tau }-2A_{b\sigma }^{*}H_{c}^{*\sigma
}\right)  \nonumber \\
&&\left. +\eta _{b\nu \rho \lambda }\left( 3C_{\mu \sigma \tau
}^{a}C_{c}^{*\sigma \tau }-2C_{\mu \sigma }^{a}H_{c}^{*\sigma }\right)
+3B_{b\rho \lambda }C_{\mu \nu \sigma }^{a}H_{c}^{*\sigma }\right] ,
\label{bfk8}
\end{eqnarray}
\begin{eqnarray}
K_{b,cd}^{a} &=&4\varepsilon ^{\mu \nu \rho \lambda }\left[ \eta _{b\mu \nu
\rho \lambda }\left( C_{\sigma \tau \kappa \varsigma }^{a}\left(
4H_{c}^{*\sigma }C_{d}^{*\tau \kappa \varsigma }+3C_{c}^{*\sigma \tau
}C_{d}^{*\kappa \varsigma }\right) \right. \right.  \nonumber \\
&&\left. +3C_{\sigma \tau \kappa }^{a}H_{c}^{*\sigma }C_{d}^{*\tau \kappa
}+C_{\sigma \tau }^{a}H_{c}^{*\sigma }H_{d}^{*\tau }\right)  \nonumber \\
&&\left. +C_{\mu \nu \rho \lambda }^{a}\left( 3\eta _{b\sigma \tau \kappa
}H_{c}^{*\sigma }C_{d}^{*\tau \kappa }+B_{b\sigma \tau }H_{c}^{*\sigma
}H_{d}^{*\tau }\right) \right] ,  \label{bfk9}
\end{eqnarray}
\begin{eqnarray}
K_{b,cde}^{a} &=&4\varepsilon ^{\mu \nu \rho \lambda }\left[ \eta _{b\mu \nu
\rho \lambda }\left( 6C_{\sigma \tau \kappa \varsigma }^{a}H_{c}^{*\sigma
}H_{d}^{*\tau }C_{e}^{*\kappa \varsigma }+C_{\sigma \tau \kappa
}^{a}H_{c}^{*\sigma }H_{d}^{*\tau }H_{e}^{*\kappa }\right) \right.  \nonumber
\\
&&\left. +C_{\mu \nu \rho \lambda }^{a}\eta _{b\sigma \tau \kappa
}H_{c}^{*\sigma }H_{d}^{*\tau }H_{e}^{*\kappa }\right] ,  \label{bfk10}
\end{eqnarray}
\begin{equation}
K_{b,cdef}^{a}=4\varepsilon _{\mu \nu \rho \lambda }\eta _{b}^{\mu \nu \rho
\lambda }C_{\sigma \tau \kappa \varsigma }^{a}H_{c}^{*\sigma }H_{d}^{*\tau
}H_{e}^{*\kappa }H_{f}^{*\varsigma },  \label{bfk11}
\end{equation}
\begin{eqnarray}
K_{ab}^{c} &=&\varepsilon _{\mu \nu \rho \lambda }\left[ -6\left( \eta
_{a}^{\mu \nu \sigma }B_{b}^{\rho \lambda }A_{\sigma }^{c}+3\eta _{a}^{\mu
\sigma \tau }\eta _{b\sigma \tau }^{\nu }B^{*c\rho \lambda }\right) \right.
\nonumber \\
&&-2\eta _{a}^{\mu \nu \rho \lambda }\left( \eta _{b}^{\sigma \tau \kappa
\varsigma }\eta _{\sigma \tau \kappa \varsigma }^{*c}+2\eta _{b}^{\sigma
\tau \kappa }\eta _{\sigma \tau \kappa }^{*c}+2B_{b}^{\sigma \tau }B_{\sigma
\tau }^{*c}\right.  \nonumber \\
&&\left. \left. -2A_{b}^{*\sigma }A_{\sigma }^{c}-2\eta _{b}^{*}\eta
^{c}\right) +4\eta _{a}^{\mu \nu \rho }A_{b}^{*\lambda }\eta ^{c}-B_{a}^{\mu
\nu }B_{b}^{\rho \lambda }\eta ^{c}\right] ,  \label{bfk12}
\end{eqnarray}
\begin{eqnarray}
K_{ab,d}^{c} &=&\varepsilon _{\mu \nu \rho \lambda }\left[ -9\eta _{a}^{\mu
\sigma \tau }\eta _{b\sigma \tau }^{\nu }\left( \eta ^{c}C_{d}^{*\rho
\lambda }-2A^{c\rho }H_{d}^{*\lambda }\right) \right.  \nonumber \\
&&-\eta _{a}^{\sigma \tau \kappa \varsigma }\eta _{b\sigma \tau \kappa
\varsigma }\left( \eta ^{c}C_{d}^{*\mu \nu \rho \lambda }+4C_{d}^{*\mu \nu
\rho }A^{c\lambda }\right.  \nonumber \\
&&\left. +12C_{d}^{*\mu \nu }B^{*c\rho \lambda }+8H_{d}^{*\mu }\eta ^{*c\nu
\rho \lambda }\right) +6\eta _{a}^{\mu \nu \sigma }B_{b}^{\rho \lambda }\eta
^{c}H_{d\sigma }^{*}  \nonumber \\
&&-2\eta _{a}^{\mu \nu \rho \lambda }\left( \eta _{b}^{\sigma \tau \kappa
}\left( \eta ^{c}C_{d\sigma \tau \kappa }^{*}+3A_{\kappa }^{c}C_{d\sigma
\tau }^{*}-6B_{\tau \kappa }^{*c}H_{d\sigma }^{*}\right) \right.  \nonumber
\\
&&\left. \left. +2A_{b}^{*\sigma }\eta ^{c}H_{d\sigma }^{*}+B_{b}^{\sigma
\tau }\left( \eta ^{c}C_{d\sigma \tau }^{*}+2A_{\tau }^{c}H_{d\sigma
}^{*}\right) \right) \right] ,  \label{bfk13}
\end{eqnarray}
\begin{eqnarray}
K_{ab,de}^{c} &=&-\varepsilon _{\mu \nu \rho \lambda }\left[ 2\eta _{a}^{\mu
\nu \rho \lambda }\left( 3\eta _{b}^{\sigma \tau \kappa }\left( H_{d\sigma
}^{*}C_{e\tau \kappa }^{*}\eta ^{c}+H_{d\sigma }^{*}H_{e\tau }^{*}A_{\kappa
}^{c}\right) \right. \right.  \nonumber \\
&&\left. +B_{b}^{\sigma \tau }H_{d\sigma }^{*}H_{e\tau }^{*}\eta ^{c}\right)
+\eta _{a}^{\sigma \tau \kappa \varsigma }\eta _{b\sigma \tau \kappa
\varsigma }\left( \left( 4H_{d}^{*\mu }C_{e}^{*\nu \rho \lambda }\right.
\right.  \nonumber \\
&&\left. \left. +3C_{d}^{*\mu \nu }C_{e}^{*\rho \lambda }\right) \eta
^{c}+12H_{d}^{*\mu }C_{e}^{*\nu \rho }A^{c\lambda }+12H_{d}^{*\mu
}H_{e}^{*\nu }B^{*c\rho \lambda }\right)  \nonumber \\
&&\left. +9\eta _{a}^{\mu \sigma \tau }\eta _{b\sigma \tau }^{\nu
}H_{d}^{*\rho }H_{e}^{*\lambda }\eta ^{c}\right] ,  \label{bfk14}
\end{eqnarray}
\begin{eqnarray}
K_{ab,def}^{c} &=&-2\varepsilon _{\mu \nu \rho \lambda }\left[ \eta
_{a}^{\sigma \tau \kappa \varsigma }\eta _{b\sigma \tau \kappa \varsigma
}\left( 3H_{d}^{*\mu }H_{e}^{*\nu }C_{f}^{*\rho \lambda }\eta
^{c}+2H_{d}^{*\mu }H_{e}^{*\nu }H_{f}^{*\rho }A^{c\lambda }\right) \right.
\nonumber \\
&&\left. +\eta _{a}^{\mu \nu \rho \lambda }\eta _{b}^{\sigma \tau \kappa
}H_{d\sigma }^{*}H_{e\tau }^{*}H_{f\kappa }^{*}\eta ^{c}\right] ,
\label{bfk15}
\end{eqnarray}
\begin{equation}
K_{ab,defg}^{c}=-\varepsilon _{\mu \nu \rho \lambda }\eta _{a}^{\sigma \tau
\kappa \varsigma }\eta _{b\sigma \tau \kappa \varsigma }H_{d}^{*\mu
}H_{e}^{*\nu }H_{f}^{*\rho }H_{g}^{*\lambda }\eta ^{c},  \label{bfk16}
\end{equation}
\begin{eqnarray}
K_{d}^{abc} &=&\eta _{d}^{\mu \nu \rho }\left( \eta _{\mu \nu \rho
}^{*a}\eta ^{b}\eta ^{c}-6B_{\mu \nu }^{*a}A_{\rho }^{b}\eta ^{c}-A_{\mu
}^{a}A_{\nu }^{b}A_{\rho }^{c}\right)  \nonumber \\
&&-A_{d}^{*\mu }A_{\mu }^{a}\eta ^{b}\eta ^{c}+B_{d}^{\mu \nu }\left( B_{\mu
\nu }^{*a}\eta ^{b}\eta ^{c}-A_{\mu }^{a}A_{\nu }^{b}\eta ^{c}\right)
\nonumber \\
&&+\eta _{d}^{\mu \nu \rho \lambda }\left( \eta _{\mu \nu \rho \lambda
}^{*a}\eta ^{b}\eta ^{c}-4\eta _{\mu \nu \rho }^{*a}A_{\lambda }^{b}\eta
^{c}\right.  \label{bfk17} \\
&&\left. +12B_{\mu \nu }^{*a}\left( B_{\rho \lambda }^{*b}\eta ^{c}-A_{\rho
}^{b}A_{\lambda }^{c}\right) \right) -\frac{1}{6}\eta _{d}^{*}\eta ^{a}\eta
^{b}\eta ^{c},  \nonumber
\end{eqnarray}
\begin{eqnarray}
K_{d,e}^{abc} &=&-\eta _{d}^{\mu \nu \rho }\left( \left( \frac{1}{6}\eta
^{a}C_{e\mu \nu \rho }^{*}-\frac{3}{2}A_{\rho }^{a}C_{e\mu \nu }^{*}+3B_{\nu
\rho }^{*a}H_{e\mu }^{*}\right) \eta ^{b}\eta ^{c}\right.  \nonumber \\
&&\left. +3A_{\nu }^{a}A_{\rho }^{b}\eta ^{c}H_{e\mu }^{*}\right) +\eta
_{d}^{\mu \nu \rho \lambda }\left( \left( \frac{1}{6}\eta ^{a}C_{e\mu \nu
\rho \lambda }^{*}+2A_{\lambda }^{a}C_{e\mu \nu \rho }^{*}\right. \right.
\nonumber \\
&&\left. +6B_{\rho \lambda }^{*a}C_{e\mu \nu }^{*}+4\eta _{\nu \rho \lambda
}^{*a}H_{e\mu }^{*}\right) \eta ^{b}\eta ^{c}-2\left( 3A_{\rho
}^{a}A_{\lambda }^{b}\eta ^{c}C_{e\mu \nu }^{*}\right.  \nonumber \\
&&\left. \left. +12B_{\nu \rho }^{*a}A_{\lambda }^{b}\eta ^{c}H_{e\mu
}^{*}+2A_{\nu }^{a}A_{\rho }^{b}A_{\lambda }^{c}H_{e\mu }^{*}\right) \right)
+\left( -\frac{1}{3}A_{d}^{*\mu }\eta ^{a}H_{e\mu }^{*}\right.  \nonumber \\
&&\left. +B_{d}^{\mu \nu }\left( \frac{1}{6}\eta ^{a}C_{e\mu \nu
}^{*}+A_{\nu }^{a}H_{e\mu }^{*}\right) \right) \eta ^{b}\eta ^{c},
\label{bfk18}
\end{eqnarray}
\begin{eqnarray}
K_{d,ef}^{abc} &=&-\frac{1}{2}\eta _{d}^{\mu \nu \rho }\left( \eta
^{a}H_{e\mu }^{*}C_{f\nu \rho }^{*}-3A_{\rho }^{a}H_{e\mu }^{*}C_{f\nu
}^{*}\right) \eta ^{b}\eta ^{c}  \nonumber \\
&&+\eta _{d}^{\mu \nu \rho \lambda }\left( \left( \frac{2}{3}\eta
^{a}H_{e\mu }^{*}C_{f\nu \rho \lambda }^{*}+\frac{1}{2}\eta ^{a}C_{e\mu \nu
}^{*}C_{f\rho \lambda }^{*}\right. \right.  \nonumber \\
&&\left. +6A_{\lambda }^{a}H_{e\mu }^{*}C_{f\nu \rho }^{*}+\frac{1}{6}%
B_{\rho \lambda }^{*a}H_{e\mu }^{*}H_{f\nu }^{*}\right) \eta ^{b}\eta ^{c}
\nonumber \\
&&\left. -6A_{\rho }^{a}A_{\lambda }^{b}\eta ^{c}H_{e\mu }^{*}H_{f\nu
}^{*}\right) +\frac{1}{6}B_{d}^{\mu \nu }\eta ^{a}\eta ^{b}\eta ^{c}H_{e\mu
}^{*}H_{f\nu }^{*},  \label{bfk19}
\end{eqnarray}
\begin{equation}
K_{d,efg}^{abc}=\left( \eta _{d}^{\mu \nu \rho \lambda }\left( \eta
^{a}C_{g\rho \lambda }^{*}-2A_{\lambda }^{a}H_{g\rho }^{*}\right) -\frac{1}{6%
}\eta _{d}^{\mu \nu \rho }\eta ^{a}H_{g\rho }^{*}\right) \eta ^{b}\eta
^{c}H_{e\mu }^{*}H_{f\nu }^{*},  \label{bfk20}
\end{equation}
\begin{equation}
K_{d,efgh}^{abc}=\frac{1}{6}\eta _{d}^{\mu \nu \rho \lambda }\eta ^{a}\eta
^{b}\eta ^{c}H_{e\mu }^{*}H_{f\nu }^{*}H_{g\rho }^{*}H_{h\lambda }^{*}.
\label{bfk21}
\end{equation}

\end{document}